\documentclass[onecolumn,showpacs,preprintnumbers,amsmath,amssymb,nofootinbib]{revtex4}
\usepackage{graphicx}
\usepackage{epsfig}
\usepackage{dcolumn}
\usepackage{bm}
\usepackage[english]{babel}
\usepackage{amsmath}
\usepackage{amsfonts}
\usepackage{amssymb}
\usepackage{delarray}
\numberwithin{equation}{section}
\newcommand{\sI}{{\scriptscriptstyle I}}
\newcommand{\sII}{{\scriptscriptstyle II}}
\newcommand{\sIII}{{\scriptscriptstyle III}}

\begin{document}

\title{MACROSCOPIC EFFECT OF QUANTUM GRAVITY:\\ GRAVITON, GHOST AND INSTANTON CONDENSATION\\ ON HORIZON SCALE OF THE UNIVERSE}

\author{Leonid Marochnik}
\email{lmarochnik@gmail.com}
 \affiliation{Physics Department, University of Maryland, College Park, MD 20742, USA}
 \author{Daniel Usikov}%
 \email{dusikov@gmail.com}
\affiliation{Physics Department, University of Maryland, College Park, MD 20742, USA}
\author{Grigory Vereshkov}
\email{gveresh@gmail.com} \affiliation{Research Institute of
Physics, Southern Federal University, Rostov-on-Don 344090,
Russia, \protect\\
Institute for Nuclear Research of the Russian Academy of Sciences, Moscow 117312, Russia‡}

\begin{abstract}

We discuss a special class of quantum gravity phenomena that occur on the scale of the Universe as a whole at any stage of
its evolution, including the contemporary Universe.  These phenomena are
a direct consequence of the zero rest mass of gravitons, conformal
non-invariance of the graviton field, and one-loop finiteness of
quantum gravity, i.e. it is a direct consequence of first
principles only.
The effects are due to graviton--ghost condensates arising from the
interference of quantum coherent states. Each of coherent states
is a state of gravitons and ghosts of a wavelength of the order of
the horizon scale and of different occupation numbers. The state
vector of the Universe is a coherent superposition of vectors of
different occupation numbers. One--loop approximation of quantum
gravity is believed to be applicable to the contemporary Universe
because of its remoteness from the Planck epoch.  To substantiate
the reliability of macroscopic quantum effects, the formalism of
one--loop quantum gravity is discussed in detail. The theory is
constructed as follows:  Faddeev -- Popov path
integral in Hamilton gauge $\longrightarrow$ factorization of classical and quantum
variables, allowing the existence of a self--consistent system of
equations for gravitons, ghosts and macroscopic geometry
$\longrightarrow$ transition to the one--loop approximation,
taking into account that contributions of ghost fields to
observables cannot be eliminated in any way.
The ghost sector corresponding to the Hamilton gauge automatically ensures of one--loop
finiteness of the theory off the mass shell. The
Bogolyubov--Born--Green--Kirckwood--Yvon (BBGKY) chain for the
spectral function of gravitons renormalized by ghosts is used to
build a self--consistent theory of gravitons in the isotropic
Universe. It is the first use of this technique in quantum gravity
calculations. We found three exact solutions of the equations,
consisting of BBGKY chain and macroscopic Einstein's equations. It
was found that these solutions describe virtual graviton and ghost
condensates as well as condensates of instanton fluctuations. All
exact solutions, originally found by the BBGKY formalism, are
reproduced at the level of exact solutions for field operators and
state vectors. It was found that exact solutions correspond to
various condensates with different graviton--ghost compositions.
Each exact solution corresponds to a certain phase state of
graviton--ghost substratum. We establish conditions under which a continuous quantum--gravity phase transitions occur between different phases of the graviton--ghost condensate.

\end{abstract}


\maketitle

\newpage

\tableofcontents

\newpage

\section{\ Introduction}\label{Intr}

Macroscopic quantum effects are quantum phenomena that occur on a
macroscopic scale. To date, there are two known macroscopic
quantum effects: superfluidity at the scale of liquid helium
vessel and superconductivity at the scale of superconducting
circuits of electrical current. These effects have been thoroughly
studied experimentally and theoretically understood. A key role in
these effects is played by coherent quantum condensates of
micro-objects with the De Broglie wavelength of the order of
macroscopic size of the system. The third macroscopic quantum
effect under discussion in this paper is condensation of gravitons
and ghosts in the self--consistent field of the expanding
Universe. A description of this effect by an adequate mathematical formalism
is the problem at the present time.

We show that condensation of gravitons and ghosts is a consequence
of quantum interference of states forming the coherent
superposition. In this superposition, quantum fields have a
certain wavelength, and with different amplitudes of probability
they are in states corresponding to different occupation numbers
of gravitons and ghosts. Intrinsic properties of the theory
automatically lead to a characteristic wavelength of gravitons and
ghosts in the condensate. This wavelength  is always of the order
of a distance to the horizon of events\footnote{Everywhere in this
paper we discuss quantum states of gravitons and ghosts that are
self--consistent with the evolution of macroscopic geometry of the
Universe. In the mathematical formalism of the theory, the ghosts
play a role of a second physical subsystem, the average
contributions of which to the macroscopic Einstein equations
appear on an equal basis with the average contribution of
gravitons. At first glance, it may seem that the status of the
ghosts as the second subsystem is in a contradiction with the
well--known fact that the Faddeev--Popov ghosts are not physical
particles. However the paradox, is in the fact that we have no
contradiction with the standard concepts of quantum theory of
gauge fields but rather full agreement with these. The
Faddeev--Popov ghosts are indeed not physical particles in a
quantum--field sense, that is, they are not particles that are in
the asymptotic states whose energy and momentum are connected by a
definite relation. Such ghosts are nowhere to be found on the
pages of our work. We discuss only virtual gravitons and virtual
ghosts that exist in the area of interaction. As to virtual
ghosts, they cannot be eliminated in principle due to lack of
ghost--free gauges in quantum gravity. In the strict mathematical
sense, the non--stationary Universe as a whole is a region of
interaction, and, formally speaking, there are no real gravitons
and ghosts in it. Approximate representations of  real particles,
of course, can be introduced for  shortwave quantum modes.  In our
work,  quantum states of shortwave ghosts are not introduced and
consequently are not discussed. Furthermore,  macroscopic quantum
effects,  which are discussed in our work, are formed by the most
virtual modes of all virtual modes. These modes are selected by
the equality $\lambda H = 1$, where $\lambda$ is the wavelength,
$H$ is the Hubble function. The same equality also characterizes
the intensity of interaction of the virtual modes with the
classical gravitational field, i.e. it reflects the essentially
non--perturbative nature of the effects. An approximate transition
to real, weakly interacting particles, situated on the mass shell
is impossible for these modes, in principle (see also the footnote
$^2$ on p. 5).}.

In this fact, a common feature of macroscopic quantum effects is
manifested: such effects are always formed by quantum
micro--objects, whose wavelengths are of the order of macroscopic
values. With this in mind, we can say that macroscopic quantum
gravity effects exist across the Universe as a whole. The existence of the effects of this type was first discussed in \cite{MUV}.

Quantum theory of gravity is a non--renormalized theory \cite{36} and for
this reason it is impossible to calculate effects with an
arbitrary accuracy in any order of the theory of perturbations.
The program combining gravity with other physical interactions
within the framework of supergravity or superstrings theory
assumes the ultimate formulation of the theory containing no
divergences. Today we do not have such a theory;  nevertheless, we
can hope to obtain physically meaningful results. Here are the
reasons for this assumption.

First, in all discussed options for the future theory, Einstein's
theory of gravity is contained as a low energy limit. Second, from
all physical fields, which will appear in a future theory
(according to present understanding), only the quantum component
of gravitational field (graviton field) has a unique combination
of zero rest mass and conformal non--invariance properties. Third,
physically meaningful effects of quantum gravity can be identified
and quantified in one--loop approximation. Fourth, as was been
shown by t'Hooft and Veltman \cite{3}, the one--loop quantum
gravity {\it with ghost sector} and without fields of matter is
finite. For the property of one--loop finiteness, proven in
\cite{3} on the graviton mass shell, we add the following key
assertion. {\it All one--loop calculations in quantum gravity must
be done in such a way that the feature of one--loop finiteness
(lack of divergences in terms of observables) must automatically
be implemented not only on the graviton mass shell but also
outside it.}

Let us emphasize the following important fact. Because of
conformal non--invariance and zero rest mass of gravitons, no
conditions exist in the Universe to place gravitons on the mass
shell precisely. Therefore, in the absence of one--loop
finiteness, divergences arise in observables. To eliminate them,
the Lagrangian of Einstein's theory must be modified, by amending
the definition of gravitons. In other words, in the absence of
one--loop finiteness, gravitons generate divergences, contrary to
their own definition. Such a situation does not make any sense, so
{\it the one--loop finiteness off the mass shell is a prerequisite
for internal consistency of the theory.}

These four conditions provide for the reliability of theory
predictions. Indeed, the existence of quantum component of the
gravitational field leaves no doubt. Zero rest mass of this
component means no threshold for quantum processes of graviton
vacuum polarization and graviton creation by external or
self--consistent macroscopic gravitational field. The combination
of zero rest mass and conformal non--invariance of graviton field
leads to the fact that these processes are occurring even in the
isotropic Universe at any stage of its evolution, including the
contemporary Universe. Vacuum polarization and particle creation
belong to effects  predicted by the theory already in one--loop
approximation. In this approximation, calculations of quantum
gravitational processes involving gravitons are not accompanied by
the emergence of divergences. Thus, the one--loop finiteness of
quantum gravity allows uniquely describe mathematically graviton
contributions to the macroscopic observables. Other one--loop
effects in the isotropic Universe are suppressed either because of
conformal invariance of non--gravitational quantum fields, or (in
the modern Universe) by non--zero rest mass particles, forming
effective thresholds for quantum gravitational processes in the
macroscopic self--consistent field.

Effects of vacuum polarization and particle creation in the sector
of matter fields of $J=0,\ 1/2,\ 1$ spin were well studied in the
1970's by many authors (see \cite{52} and references therein). The
theory of classic gravitational waves in the isotropic Universe
was formulated by Lifshitz in 1946 \cite{13}. Grishchuk \cite{40}
considered a number of cosmological applications of this theory
that are result of conformal non--invariance of gravitational
waves. Isaacson \cite{41, 41a} has formulated the task of
self--consistent description of gravitational waves and background
geometry. The model of Universe consisting of short gravitational
waves was described for the first time in \cite{20, 20a}. The
energy--momentum tensor of classic gravitational waves of super
long wavelengths was constructed in \cite{42, 43}. The canonic
quantization of gravitational field was done in \cite{11,
44,HuParker1977}. The local speed of creation of shortwave
gravitons was calculated in \cite{45}.  In all papers listed
above, the ghost sector of graviton theory was not taken into
account. One--loop quantum gravity in the form of the theory of
gravitons defined on the background spacetime was described by De
Witt \cite{7}. Calculating methods of this theory were discussed
by Hawking \cite{37}.

The exact equations of self-consistent theory of gravitons in the Heisenberg representation with the ghost sector automatically providing a one-loop finiteness off the mass shell are obtained in our work \cite{VM}. In \cite{VM}, it is shown that {\it the Heisenberg representation of quantum gravity (as well as the Heisenberg representation of quantum Yang-Mills theory \cite{FS}) exists only in the Hamilton gauge. The ghost sector corresponding to this gauge represented by the complex scalar field with minimal coupling to gravity}.

One--loop finiteness provides the simplicity and elegance of a
mathematical theory that allows, in turn, discovering a number of
new approximate and exact solutions of its equations. This paper
is focused on three exact solutions corresponding to three
different quantum states of graviton--ghost subsystem in the space
of the non--stationary isotropic Universe with self--consistent
geometry. The first of these solutions describes a coherent
condensate of virtual gravitons and ghosts; the second solution
describes a coherent condensate of instanton fluctuations. The
third solution describes the self--polarized condensate in the De
Sitter space. This solution allows interpretation in terms of
virtual particles as well as in terms of instanton fluctuations.

The principal nature of macroscopic quantum gravity effects, the
need for strict proof of their inevitability and reliability
impose stringent requirements for constructing a mathematical
algorithm of the theory.
Sections \ref{qg} and \ref{scgt} are devoted to the derivation of
the equations of the theory with a discussion of all the
mathematical details. In Section \ref{qg}, we start with exact
quantum theory of gravity, presented in terms of path integral of
Faddeev--Popov \cite{5} and De Witt \cite{4, 4a}. Key ideas of this
Section are the following. (i) The necessity  to gauge the full
metric (before its separation into the background and
fluctuations) and the inevitability of appearance of a ghost
sector in the exact path integral and operator Einstein's
equations (Sections \ref{pi} and \ref{OEE}); (ii) The principal
necessity to use normal coordinates (exponential parameterization)
in a mathematically rigorous procedure for the separation of
classical and quantum variables is discussed in Sections \ref{fac}
and \ref{var}; (iii) The derivation of differential identities,
providing the consistency of classical and quantum equations
performed jointly in any order of the theory of perturbations is
given Section \ref{equiv}. Rigorously derived equations of gauged
one--loop quantum gravity are presented in Section \ref{1loop}.

The status of properties of ghost sector generated by gauge is
crucial to properly assess the structure of the theory and its
physical content. Let us immediately emphasize that the standard
presentation on the ghost status in the theory of  $S$--matrix can
not be exported to the theory of gravitons in the macroscopic
spacetime with self--consistent geometry. Two internal
mathematical properties of the quantum theory of gravity make such
export fundamentally impossible. First, there are no gauges that
completely eliminate the diffeomorphism group degeneracy in the
theory of gravity. This means that among the objects of the
quantum theory of fields inevitably arise ghosts interacting with
macroscopic gravity. Secondly, gravitons and ghosts cannot be in
principle situated precisely on the mass shell because of their
conformal non--invariance and zero rest mass. This is because
there are no asymptotic states, in which interaction of quantum
fields with macroscopic gravity could be neglected. Restructuring
of vacuum graviton and ghost modes with a wavelength of the order
of the distance to the horizon of events takes place at all stages
of cosmological evolution, including the contemporary Universe.
Ghost trivial vacuum, understood as the quantum state with zero
occupation numbers for all modes, simply is absent from physically
realizable states. Therefore, direct participation of ghosts in
the formation of macroscopic observables is
inevitable\footnote{Once again, we emphasize that the equal
participation of virtual gravitons and ghosts in the formation of
macroscopic observables in the non--stationary Universe does not
contradict the generally accepted concepts of the quantum theory
of gauge fields.  On the contrary it follows directly from the
mathematical structure of this theory. In order to clear up this
issue once and for all, recall some details of the theory of $S-$
matrix. In constructing this theory, all space--time is divided
into regions of asymptotic states and the region of effective
interaction. Note that this decomposition is carried out by means
of, generally speaking, an artificial procedure of  turning on and
off the interaction adiabatically. (For obvious reasons, the
problem of self--consistent description of gravitons and ghosts in
the non-stationary Universe with $\lambda H = 1$ by means of an
analogue of such procedure cannot be considered a priori.) Then,
after splitting the space--time into two regions, it is assumed
that the asymptotic states are ghost--free. In the most elegant
way, this selection rule is implemented in the BRST formalism,
which shows that the BRST invariant states turn out to be
gauge--invariant automatically. The virtual ghosts, however,
remain in the area of interaction, and this points to the fact
that virtual gravitons and ghosts are  parts of the Feynman
diagrams on an equal footing. In the self--consistent theory of
gravitons in the non--stationary Universe,  virtual ghosts of
equal weight as the gravitons, appear at the same place where they
appear in the theory of $ S-$matrix, i.e. at the same place as
they were introduced by Feynman, i.e. in the region of
interaction. Of course, the fact that in the real non--stationary
Universe, both the observer and virtual particles with $\lambda H
= 1$ are in the area of interaction, is  highly nontrivial. It is
quite possible that this property of the real world is manifested
in the effect of dark energy. An active and irremovable
participation of virtual ghosts in the formation of macroscopic
properties of the real Universe poses the question of their
physical nature. Today, we can only say with certainty that the
mathematical inevitability of ghosts provides the one--loop
finiteness off the mass shell, i.e. the mathematical consistency
of one-loop quantum gravity without fields of matter. Some
hypothetical ideas about the nature of the ghosts are briefly
discussed in the final Section \ref{con}.}.

Section \ref{scgt} is devoted to general discussion of equations
of the theory of gravitons in the isotropic Universe. It focuses
on three issues: (i) Canonical quantization of gravitons and ghosts (Sections \ref{3vs}, \ref{quant}); (ii) Construction of the state vector
of a general form as a product of normalized superpositions
(Section \ref{stvec}); (iii) The proof of the
one--loop finiteness of macroscopic observables (Section
\ref{fin}). The main conclusion is that {\it the
quantum ghost fields are inevitable and unavoidable components of
the quantum gravitational field.}  As noted above, one--loop
finiteness is seen by us as a universal property of quantum
gravity, which extends off the mass shell. The requirement of
compensation of divergences in terms of macroscopic observables,
resulting from one--loop finiteness, uniquely captures the dynamic
properties of quantum ghost fields in the isotropic Universe.
The existence of Quantum Gravity in the Heisenberg representation in the Hamilton gauge is a nontrivial
property of the theory. Exactly this property automatically provides a one--loop finiteness of the theory off mass shell.

Sections \ref{swg} and \ref{lgw} contain approximate solutions to
obtain quantum ensembles of short and long gravitational waves. In
Section \ref{sce} it is shown that approximate solutions obtained
can be used to construct scenarios for the evolution of the early
Universe. In one such scenario, the Universe is filled with
ultra--relativistic gas of short--wave gravitons and with a
condensate of super--long wavelengths, which is dominated by
ghosts. The evolution of this Universe is oscillating in nature.

At the heart of cosmological applications of one--loop quantum
gravity is the Bogolyubov--Born--Green--Kirckwood--Yvon (BBGKY)
chain (or hierarchy) for the spectral function of gravitons,
renormalized by ghosts. We present the first use of this technique
in quantum gravity calculations. Each equation of the BBGKY chain
connects the expressions for neighboring moments of the spectral
function. In Section \ref{chain} the BBGKY chain is derived by
identical mathematical procedures from graviton and ghost operator
equations. Among these procedures is averaging of bilinear forms
of field operators over the state vector of the general form,
whose mathematical structure is given in Section \ref{stvec}. The
need to work with state vectors of the general form is dictated by
the instability of the trivial graviton--ghost vacuum (see \cite{MUV2}, Section
III F). Evaluation of mathematical correctness of procedures
for BBGKY structure is entirely a question of the existence of
moments of the spectral function as mathematical objects. A
positive answer to this question is guaranteed by one--loop
finiteness (Section \ref{fin}). The set of moments of the spectral
function contains information on the dynamics of operators as well
as on the properties of the quantum state over which the averaging
is done. The set of solutions of BBGKY chain contains all possible
self--consistent solutions of operator equation, averaged over all
possible quantum ensembles.

{\it A nontrivial fact is that in the one--loop quantum gravity
BBGKY chain can formally be introduced at an axiomatic level.}
Theory of gravitons provided by BBGKY chain, conceptually and
mathematically corresponds to the axiomatic quantum field theory
in the Wightman formulation (see Chapter 8 in the monograph
\cite{6}). Here, as in Wightman, the full information on the
quantum field is contained in an infinite sequence of averaged
correlation functions. Definitions of these functions clearly
relate to the symmetry properties of manifold on one  this field
is  defined. {\it Once the BBGKY chain is set up, the existence of
finite solutions for the observables is provided by inherent
mathematical properties of equations of the chain.} This means
that the phenomenology of BBGKY chain is more general than field
operators, state vectors and graviton--ghost compensation of
divergences that were used in its derivation.

Exact solutions of the equations, consisting of BBGKY chain and
macroscopic Einstein's equations are obtained in Sections
\ref{ggh_cond} and \ref{Sitt}. Two solutions given in
\ref{ggh_cond}, describe heterogeneous graviton--ghost
condensates, consisting of three subsystems. Two of these are
condensates of spatially homogeneous modes with the equations of
state $p=-\varepsilon/3$ and $p=\varepsilon$. The third subsystem
is a condensate of quasi--resonant modes with a constant conformal
wavelength corresponding to the variable physical wavelength of
the order of the distance to the horizon of events. The equations
of state of condensates of quasi--resonant modes differ from
$p\sim-\varepsilon/3$  by logarithmic terms, through which the
first solution is $p\gtrsim -\varepsilon/3$, while the second is
$p\lesssim-\varepsilon/3$. Furthermore, the solutions differ by
the sign of the energy density of condensates of spatially
homogenous modes. The third solution describes a homogeneous
condensate of quasi--resonant modes with a constant physical
wavelength. The equation of state of this condensate is
$p=-\varepsilon$  and its self--consistent geometry is the De
Sitter space. The three solutions are interpreted as three
different phase states of graviton--ghost system. The problem of
quantum--gravity phase transitions is discussed in
Section\ref{pt}.

Solutions obtained in Section \ref{Bog} in terms of moments of the
spectral function, are reproduced in Sections \ref{exact} and
\ref{inst} at the level of dynamics of operators and state
vectors. A microscopic theory provides details to clarify the
structure of graviton--ghost condensates and clearly demonstrates
the effects of quantum interference of coherent states. In Section
\ref{K} it is shown that the condensate of quasi--resonant modes
with the equation of state  $p\gtrsim -\varepsilon/3$ consists of
virtual gravitons and ghosts. In Section \ref{S} a similar
interpretation is proposed for the condensate in the De Sitter
space, but it became necessary to extend  the mathematical
definition of the moments of the spectral function.

New properties of the theory, whose existence was not anticipated
in advance, are studied in Section \ref{inst}. In Section
\ref{git} we find that the self--consistent theory of gravitons
and ghosts is invariant with respect to the Wick turn. In this
Section we also construct the formalism of quantum theory in the
imaginary time and discuss the physical interpretation of this
theory. The subjects of the study are correlated fluctuations
arising in the process of tunnelling between degenerate states of
graviton--ghost systems, divided by classically impenetrable
barriers. The level of these fluctuations is evaluated by
instanton solutions (as in Quantum Chromodynamics).  In Section
\ref{ins1} it is shown that the condensate of quasi--resonant
modes with the equation of state $p\lesssim-\varepsilon/3$  is of
purely instanton nature. In Section \ref{ins2} the instanton
condensate theory is formulated for the De Sitter space.

Potential use of the results obtained to construct scenarios of
cosmological evolution was briefly discussed in Sections
\ref{approx} --- \ref{inst} to obtain approximate and exact
solutions. Future issues of the theory of the theory are briefly discussed in
the Conclusion (Section \ref{con}).

A system of units is used, in which the speed of light is  $c=1$,
Planck constant is $\hbar=197.327$ MeV$\cdot$fm; Einstein's
gravity constant is    $\varkappa\equiv 8\pi G=8\pi\cdot
1.324\cdot 10^{-42}$ MeV$^{-1}\cdot$fm.

\section{Basic Equations}\label{qg}

According to De Witt \cite{7}, one of formulations of one--loop
quantum gravity (with no fields of matter) is reduced to the zero
rest mass quantum field theory with spin $J=2$, defined for the
background spacetime with classic metric. The graviton dynamics is
defined by the interaction between quantum field and classic
gravity, and the background space geometry, in turn, is formed by
the energy--momentum tensor (EMT) of gravitons.

In the current Section we describe how to get the self--consistent
system of equations, consisting of quantum operator equations for
gravitons and ghosts and classic $C$--number Einstein equations
for macroscopic metrics with averaged EMT of gravitons and ghosts
on the right hand side. The theory is formulated without any
constrains on the graviton wavelength that allows the use of the
theory for the description of quantum gravity effects at the long
wavelength region of the specter. The equations of the theory
(except the gauge condition) are represented in 4D form which is
general covariant with respect to the transformation of the
macroscopic metric.

The mathematically consistent system of 4D quantum and classic
equations with no restrictions with respect to graviton
wavelengths is obtained by a regular method for the first time.
{\it The case of a gauged path integral with ghost sector} is seen
as a source object of the theory. Important elements of the method
are {\it exponential parameterization of the operator of the
density of the contravariant metric; factorization of path
integral measure; consequent integration over quantum and classic
components of the gravitational field.} Mutual compliance of
quantum and classic equations, expressed in terms of fulfilling of
the conservation of averaged EMT at the operator equations of
motion is provided by the virtue of the theory construction
method.

\subsection{Path Integral and Faddeev--Popov Ghosts}\label{pi}

Formally, the exact scheme of quantum gravity is based on the
amplitude of transition, represented by path integral \cite{4, 5}:
\begin{equation}
\begin{array}{c}
 \displaystyle
 \langle \mbox{\rm out}|\mbox{\rm in}\rangle=
\int \exp\left(\frac{i}{\hbar}\int
(L_{grav}+L_\Lambda)d^4x\right)\left(\mbox{\rm det}\, \hat M^i_{\;
 k}\right)
 \times\prod_x\left(\prod_i\delta(\hat A_k\sqrt{-\hat g}\hat
 g^{ik}-B^i)\right)d\hat\mu=
 \\[5mm] \displaystyle
 =\int \exp\left(\frac{i}{\hbar}\int (L_{grav}+L_\Lambda+L_{ghost})d^4x\right)
 \times\prod_x\left(\prod_i\delta(\hat A_k\sqrt{-\hat g}\hat
 g^{ik}-B^i)\right)d\hat\mu d\mu_\theta\ ,
 \end{array}
 \label{2.1}
 \end{equation}
where
\[
\displaystyle L_{grav}+L_\Lambda=-\frac{1}{2\varkappa}\sqrt{-\hat g}\hat
g^{ik}\hat R_{ik}-\sqrt{-\hat g}\Lambda
\]
is the density of gravitational Lagrangian, with cosmological
constant included;  $L_{ghost}$  is the density of ghost
Lagrangian, explicit form of which is defined by localization of
$\mbox{\rm det}\, \hat M^i_{\; k}$;  $\hat A_k$ is gauge operator,
$B^i(x)$ is the given field;   $\hat M^i_{\; k}$ is an operator of
equation for infinitesimal parameters of transformations for the
residual degeneracy $\eta^i=\delta x^i$;
\begin{equation}
 \displaystyle
  d\hat\mu=\prod_x\left\{(-\hat g)^{5/2}\prod_{i\leqslant
 k}d\hat g^{ik}\right\}
 \label{2.2}
 \end{equation}
is the gauge invariant measure of path integration over
gravitational variables; $d\mu_\theta$ is the measure of
integration over ghost variables. Operator $\hat M^i_{\; k}$ is of
standard definition:
\begin{equation}
 \displaystyle
 \hat M^i_{\;  k}\eta^k\equiv \hat A_k(\delta\sqrt{-\hat g}\hat
 g^{ik})=0,
 \label{2.3}
 \end{equation}
where
\begin{equation}
 \displaystyle
\delta\sqrt{-\hat g}\hat  g^{ik}=-\partial_l(\sqrt{-\hat g}\hat
g^{ik}\eta^l)+\sqrt{-\hat g}\hat  g^{il}\partial_l\eta^k+
\sqrt{-\hat g}\hat  g^{kl}\partial_l\eta^i
 \label{2.4}
 \end{equation}
is variation of metrics under the action of infinitesimal
transformations of the group of diffeomorphisms. According to
(\ref{2.1}), the allowed gauges are constrained by the condition
of existence of the inverse operator $(\hat M^i_{\;  k})^{-1}$.

The equation (\ref{2.1}) explicitly manifests the fact that the
source path integral is defined as a mathematical object only
after the gauge has been imposed. In the theory of gravity, there
are no gauges completely eliminating the degeneracy with respect
to the transformations (\ref{2.4}). Therefore, the sector of
nontrivial ghost fields, interacting with gravity, is necessarily
present in the path integral. {\it This aspect of the quantum
gravity is important for understanding of its mathematical
structure, which is fixed before any approximations are
introduced.} By that reason, in this Section we discuss the
equations of the theory, by explicitly defining the concrete
gauge.

The mathematical procedure of transition from path integral (\ref{2.1}) to the equations of Quantum Gravity in the Heisenberg representation (with the canonical quantization of gravitons and ghosts) is described by us in detail \cite{VM}. The first step of this procedure is to represent the integral (\ref{2.1}) as  a path integral over the canonical variables. Such an integral was proposed by Faddeev \cite{F9} on the basis of the general theory of Hamilton systems with explicitly unsolvable constraints \cite{F16}. The second step is to introduce the normal coordinates of the gravitational field using the exponential parameterization of metric. The Hamilton gauge  of the normal coordinates specifies the Faddeev path integral in such a way that
the ghost sector (corresponding to it) allows to introduce canonical variables of ghost fields and to
represent the ghost Lagrangian in the Hamilton form. In the third step of this procedure, the transition from the gauged path integral to the canonical Hamilton formalism in the Heisenberg representation is made (using the standard definition of the operator of evolution). The results of the \cite{VM}  are rigorous basis of the simplified procedure for obtaining gauged equations of quantum gravity with ghosts, which is described below.

Hamilton gauge is that of synchronous type:
\begin{equation}
 \displaystyle
 \sqrt{-\hat g}\hat g^{00}=\sqrt{\bar\gamma}\ , \qquad \sqrt{-\hat g}\hat
 g^{0\alpha}=0\ .
 \label{2.5}
 \end{equation}
For that gauge
\begin{equation}
 \displaystyle
 \hat A_k=(1,\, 0,\, 0,\, 0)\ , \qquad B^i=(\sqrt{\bar\gamma},\, 0,\, 0,\, 0),
 \label{2.6}
 \end{equation}
where  $\bar\gamma=\bar\gamma({\bf x})$ is the metric determinant
of the basic 3D space of constant curvature (for the plane
cosmological model $\bar\gamma=1$).

The construction of the ghost sector, i.e. finding of the
Lagrangian density $L_{ghost}$, is reduced to two operations.
First, $\mbox{\rm det}\, \hat M^i_{\; k}$  is represented in the
form, factorized over independent degrees of freedom for ghosts,
and then the localization of the obtained expression is conducted.
Substitution of (\ref{2.6}) and (\ref{2.4}) to (\ref{2.3}) gives
the following system of equations
\begin{equation}
\begin{array}{c}
 \displaystyle
 -\partial_\alpha\sqrt{\bar\gamma}\eta^\alpha+
  \sqrt{\bar\gamma}\frac{\partial\eta^0}{\partial t}=0,
 \qquad
 \sqrt{-\hat g}\hat  g^{\alpha\beta}\partial_\beta\eta^0+
 \frac{\partial\sqrt{\bar\gamma}\eta^\alpha}{\partial t}=0.
 \end{array}
 \label{2.7}
 \end{equation}
According to (\ref{2.7}), with respect to variables $\eta^0$,
$\sqrt{\bar\gamma}\eta^\alpha$  the operator--matrix $\hat M^i_{\;
k}$  reads
\begin{equation}
 \displaystyle
   \hat M^i_{\; k}= \left( {{\displaystyle \sqrt{\bar\gamma}\frac{\partial}{\partial t}
   \qquad\qquad\qquad
   -\partial_\alpha}
 \atop {\displaystyle  \sqrt{-\hat g}\hat  g^{\alpha\beta}\partial_\beta
 \qquad\qquad
\displaystyle \delta_\beta^\alpha\frac{\partial}{\partial t}}}
 \right)
\label{2.8}
\end{equation}
(Note matrix--operator is obtained in the form (\ref{2.8}) without
the substitution of transformation parameters if Leutwiller
measure $d\hat \mu_L=\hat g\hat g^{00}d\hat \mu$ is used. The
measure discussion see, e.g. \cite{8}.) Functional determinant of
matrix--operator  $\mbox{\rm det}\, \hat M^i_{\; k}$ is
represented in the form of the determinant of matrix $\hat M^i_{\;
k}$, every element of which is a functional determinant of
differential operator. As it is follows from (\ref{2.8}),
  \begin{equation}
 \displaystyle \mbox{\rm det}\, \hat M^i_{\; k}=
\left(\mbox{\rm det}\; \partial_i\sqrt{\hat g}\hat
g^{ik}\partial_k\right)\times \left(\mbox{\rm det}\;
\frac{\partial}{\partial t}\right)\times \left(\mbox{\rm det}\;
\frac{\partial}{\partial t}\right).
 \label{2.9}
\end{equation}
One can see that the first multiplier in (\ref{2.9}) is
4--invariant determinant of the operator of the zero rest mass
Klein--Gordon--Fock equation, and two other multipliers do not
depend on gravitational variables.

Localization of determinant (\ref{2.9}) by representing it in a
form of path integral over the ghost fields is a trivial
operation. As it follows from (\ref{2.9}), the class of
synchronous gauges contains three dynamically independent ghost
fields $\theta,\, \varphi,\, \chi$, two of each --- $ \varphi,\,
\chi$ do not interact with gravity. For the obvious reason, the
trivial ghosts  $ \varphi,\, \chi$ are excluded from the theory.
The Lagrangian density of nontrivial ghosts coincides exactly with
Lagrangian density of complex Klein--Gordon--Fock fields (taking
into account the Grassman character of  fields $\bar\theta,\,
\theta$):
\begin{equation}
 \displaystyle L_{ghost}=-\frac{1}{4\varkappa}\sqrt{-\hat g}\hat
g^{ik}\partial_i\bar \theta\cdot \partial_k\theta.
 \label{2.10}
\end{equation}
The normalization multiplier $-1/4\varkappa$  in (\ref{2.10}) is
chosen for the convenience. The integral measure over ghost fields
has a simple form:
\[
\displaystyle d\mu_\theta=\prod_xd\bar\theta d\theta\ .
\]

The calculations above comply with both general requirements to
the construction of ghost sector. First, path integration should
be carried out only over the dynamically independent ghost fields.
Second, in the ghost sector, it is necessary to extract and then
to take into account only the nontrivial ghost fields, i.e. those
interacting with gravity.

\subsection{Einstein Operator Equations}\label{OEE}

Let us take into account the fact that the calculation of gauged
path integral should be mathematically equivalent to the solution
of dynamical operator equations in the Heisenberg representation.
It is also clear that operator equations of quantum theory should
have a definite relationship with Einstein equations. In the
classic theory, it is possible to use any form of representation
of Einstein equations, e.g.
\begin{equation}
\begin{array}{c}
 \displaystyle
 (-\hat g)^n\left(\hat g^{il}\hat g^{km}\hat R_{lm}-
 \frac12\hat g^{ik}\hat g^{lm}\hat R_{lm}-\hat g^{ik}\varkappa\Lambda\right)=0\ ,\qquad(a)
 \\[3mm]
 \displaystyle (-\hat g)^n\left( \hat g^{km}\hat R_{im}-
 \frac12 \delta_i^k\hat g^{lm}\hat R_{lm}-\delta_i^k\varkappa\Lambda\right)=0\ ,\qquad\qquad (b)
 \\[3mm]
 \displaystyle (-\hat g)^n\left(\hat R_{ik}-
\frac12\hat g_{ik}\hat g^{lm}\hat R_{lm}-\hat
g^{ik}\varkappa\Lambda\right)=0\ ,\qquad\qquad\qquad (c)
 \end{array}
  \label{2.14}
 \end{equation}
where, for example,  $n=0,\; 1/2,\; 1$. Transition from one to
another is reduced to the multiplication by metric tensor and its
determinant, which are  trivial operations in case when the metric
is a $C$--number function. If the metric is an operator, then the
analogous operations will, at least, change renormalization
procedures of quantum non--polynomial theory. Thus, the question
about the form of notation for Einstein's operator equations has
first--hand relation to the calculation procedure. Now we show
that in the quantum theory one should use operator equations
(\ref{2.14}$b$) with $n=1/2$, supplemented by the energy--momentum
pseudo--tensor of ghosts.

In the path integral formalism, the renormalization procedures are
defined by the dependence of Lagrangian of interactions and the
measure of integration of the field operator in terms of which the
polynomial expansion of non--polynomial theory is defined
\cite{9}. The introduction of such an operator, i.e. {\it the
parameterization of the metric}, is, generally speaking, not
simple. Nevertheless, it is possible to find a special
parameterization for which the algorithms of renormalization
procedures are defined only by Lagrangian of interactions.
Obviously, in such a parameterization the measure of integration
should be trivial. It reads:
\begin{equation}
\displaystyle
 d\hat\mu=\prod_x\prod_{i\leqslant k}d\hat\Psi_i^k\ ,
\label{2.15}
\end{equation}
where $\hat\Psi_i^k$  is a dynamic variable. The metric is
expressed via this variable. It is shown in \cite{9} that the
trivialization of measure (\ref{2.15}) takes place for the
exponential parameterization that reads
\begin{equation}
\begin{array}{c}
 \displaystyle \sqrt{-\hat{g}}\hat{g}^{ik} = \sqrt{-\bar g}\bar g^{il}(\exp{\hat\Psi})_l^k=
 \sqrt{-\bar g}\bar g^{il}
  \left(
         \delta_l^k + \hat\Psi_l^k + \frac{1}{2}\hat\Psi_l^m\hat\Psi_m^k+\ldots
  \right)\ ,
\end{array}
\label{2.16}
\end{equation}
where $\bar g^{ik}$  is the defined metric of an auxiliary basic
space. In that class of our interest, the metric is defined by the
interval
\[
\displaystyle d\bar s^2=dt^2-\bar\gamma_{\alpha\beta}dx^\alpha dx^\beta\ ,
\]
where  $\bar\gamma_{\alpha\beta}$ is the metric of 3D space with a
constant curvature. (For the flat Universe
$\bar\gamma_{\alpha\beta}$ is the Euclid metric.)

The exponential parameterization is singled out among all other
parameterizations by the property that $\hat\Psi_i^k$  are the
normal coordinates of gravitational fields \cite{10}. In that
respect, the gauge conditions (\ref{2.5}) are identical to $\hat
\Psi_0^i=0$. The fact that the "gauged"\ coordinates are the
normal coordinates, leads to a simple and elegant ghost sector
(\ref{2.10}). The status of $\hat\Psi_i^k$, as normal coordinates,
is of principal value for the mathematical correctness while
separating the classic and quantum variables (see Section
\ref{var}). Besides, in the framework of perturbation theory the
normal coordinates allow to organize a calculation procedure,
which is based on a simple classification of nonlinearity of
quantum gravity field. It is important that this procedure is
mathematically non--contradictive at every order of perturbation
theory over amplitude of quantum fields (see Section \ref{equiv},
\ref{1loop}).

Operator Einstein equations that are mathematically equivalent to
the path integral of a trivial measure are derived by the
variation of gauged action by variables $\hat\Psi_i^k$. The
principal point is that the gauged action necessarily includes the
ghost sector because there are no gauges that are able to
completely eliminate the degeneracy. According to (\ref{2.10}), in
the Hamilton gauge we get
\begin{equation}
\displaystyle
 S=-\int d^4x\left\{\frac{1}{2\varkappa}\sqrt{-\hat g}\hat g^{ik}\left(\hat R_{ik}+
 \frac12\partial_i\bar\theta\cdot\partial_k\theta\right)+\sqrt{-\hat g}\Lambda\right\}\ .
\label{2.17}
\end{equation}
In accordance with definition (\ref{2.16}), the variation is done
by the rule
\[
\displaystyle \delta\sqrt{-\hat g}\hat g^{ik} =\sqrt{-\hat g}\hat g^{il}\delta\hat \Psi_l^k\ .
\]
Thus, from (\ref{2.17}) it follows
\begin{equation}
\begin{array}{c}
\displaystyle \hat{\mathcal{G}}_i^k\equiv \sqrt{-\hat g}\hat
g^{kl}\hat R_{il}+\varkappa\sqrt{-\hat g}\delta_i^k\Lambda-
\varkappa\left(\sqrt{-\hat g}\hat g^{kl}\hat
T^{(ghost)}_{il}-\frac12\delta_i^k\sqrt{-\hat g}\hat g^{ml}\hat
T^{(ghost)}_{ml}\right)=0\ .
\end{array}
 \label{2.18}
\end{equation}
After subtraction of semi--contraction from (\ref{2.18}) we obtain
a mathematically equivalent equation
\begin{equation}
\begin{array}{c}
\displaystyle
\hat{\mathcal{E}}_i^k=\hat{\mathcal{G}}_i^k-\frac12\delta_i^k\hat{\mathcal{G}}_l^l\equiv
 \sqrt{-\hat g}\hat g^{kl}\hat
R_{il}-\frac12\delta_i^k\sqrt{-\hat g}\hat g^{ml}\hat
R_{ml}-
\varkappa\left(\sqrt{-\hat g}\hat g^{kl}\hat
T^{(ghost)}_{il}+\sqrt{-\hat g}\delta_i^k\Lambda\right)=0\ .
\end{array}
\label{2.19}
\end{equation}
In (\ref{2.18}), (\ref{2.19}) there is an object
\begin{equation}
\displaystyle \hat
T^{(ghost)}_{ik}=-\frac{1}{4\varkappa}\left(\partial_i\bar\theta\cdot\partial_k\theta+
\partial_k\bar\theta\cdot\partial_i\theta-
\hat g_{ik}\hat g^{lm}\partial_l\bar\theta\cdot\partial_m\theta\right)\ ,
\label{2.20}
\end{equation}
which has the status of the energy--momentum pseudo--tensor of
ghosts.

In accordance with the general properties of Einstein's theory,
six spatial components of equations (\ref{2.18}) are considered as
quantum equations of motion:
\begin{equation}
\begin{array}{c}
\displaystyle \sqrt{-\hat g}\hat g^{\beta l}\hat R_{\alpha
l}+\varkappa\sqrt{-\hat g}\delta_\alpha^\beta\Lambda=
 \varkappa\left(\sqrt{-\hat g}\hat g^{\beta l}\hat
T^{(ghost)}_{\alpha l}-\frac12\delta_\alpha^\beta\sqrt{-\hat
g}\hat g^{ml}\hat T^{(ghost)}_{ml}\right)\ .
\end{array}
\label{2.21}
\end{equation}
(Everywhere in this work the Greek metric indexes stand for
$\alpha,\ \beta=1,\ 2,\ 3$.) In the classic theory,  equations of
constraints $\hat{\mathcal{E}}_0^0=0$ and
$\hat{\mathcal{E}}_0^\alpha=0$ are the first integrals of
equations of motion (\ref{2.21}). Therefore, in the quantum theory
formulated in the Heisenberg representation four primary
constraints from (\ref{2.19}), have the status of the {\it
initial} conditions for the Heisenberg state vector. They read:
\begin{equation}
\begin{array}{c}
\displaystyle \left\{\sqrt{-\hat g}\hat g^{0l}\hat
R_{0l}-\frac12\sqrt{-\hat g}\hat g^{ml}\hat
R_{ml}-\varkappa\left(\sqrt{-\hat g}\hat g^{0l}\hat
T^{(ghost)}_{0l}+\sqrt{-\hat g}\Lambda\right)\right\}|\Psi\rangle=0\ ,
 \\[3mm]
\displaystyle \left\{\sqrt{-\hat g}\hat g^{\alpha l}\hat
R_{0l}-\varkappa\sqrt{-\hat g}\hat g^{\alpha l}\hat
T^{(ghost)}_{0l}\right\}|\Psi\rangle=0\ .
 \end{array}
 \label{2.22}
\end{equation}
If conditions (\ref{2.22}) are valid from the start, then the
internal properties of the theory must provide their validity at
any subsequent moment of time. Four secondary relations, defined
by the gauge non containing the higher order derivatives, also
have the same status:
\begin{equation}
\displaystyle \left\{\hat A_k(\sqrt{-\hat g}\hat
g^{ik})-B^i\right\}|\Psi\rangle=0\ .
 \label{2.23}
\end{equation}

The system of equations of quantum gravity is closed by the
ghosts' equations of motion, obtained by the variation of action
(\ref{2.17}) over ghost variables:
\begin{equation}
\begin{array}{c}
 \displaystyle \partial_i\sqrt{-\hat g}\hat g^{ik}\partial_k\theta=0\ ,
 \qquad \partial_i\sqrt{-\hat g}\hat g^{ik}\partial_k\bar\theta=0\ .
\end{array}
\label{2.24}
\end{equation}
Ghost fields  $\bar \theta$ and  $\theta$  are not defined by
Grassman scalars, therefore  $T^{(ghost)}_{ik}$ is not a tensor.
Nevertheless, all mathematical properties of equations
(\ref{2.24}) and expressions (\ref{2.20}) coincide with the
respected properties of equations and EMT of complex scalar
fields. This fact is of great importance when concrete
calculations are done (see Section \ref{scgt}).

 \subsection{Factorization of the Path Integral}\label{fac}

Transition from the formally exact scheme (\ref{2.21}) ---
(\ref{2.24})) to the semi--quantum theory of gravity can be done
after some additional hypotheses are included in the theory. The
physical content of these hypotheses consists of the assertion of
existence of classical spacetime with metric $g_{ik}$,
connectivity $\Gamma^i_{kl}$  and curvature $R_{ik}$. The first
hypothesis is formulated at the level of operators. Assume that
{\it operator of metric $\hat g^{ik}$  is a functional of
$C$--number function  $g^{ik}$  and the quantum operator $\hat
\psi_i^k$.} The second hypothesis is related to the state vector.
{\it Each state vector that is involved in the scalar product
$\langle \mbox{\rm out}|\mbox{\rm in}\rangle$, is represented in a
factorized form $|\Psi\rangle=|\Phi\rangle|\psi\rangle$, where
$|\psi\rangle$ are the vectors of quantum states of gravitons;
$|\Phi\rangle$ are the vectors of quasi--classic states of
macroscopic metric}. In the framework of these hypotheses the
transitional amplitude is reduced to the product of amplitudes:
\begin{equation}
\displaystyle
 \langle \mbox{\rm out}|\mbox{\rm in}\rangle=
 \langle \Phi_{out}|\Phi_{in}\rangle\langle
 \psi_{out}|\psi_{in}\rangle\ .
 \label{2.25}
\end{equation}
Thus, {\it the physical assumption about existence of classic
spacetime formally (mathematically) means that the path integral
must be calculated first by exact integration over quantum
variables, and then by approximate integration over the classic
metric.}

Mathematical definition of classic and quantum variables with
subsequent integrations are possible only after the trivialization
and factorization of integral measure are done. As already noted,
trivial measure (\ref{2.15}) takes place in exponential
parameterization (\ref{2.16}). The existence of $|\mbox{\rm
in}\rangle=|\Psi\rangle$ vector allows the introduction of classic
$C$--number variables as follows
\[
\displaystyle \Phi_i^k=\langle\Psi|\hat\Psi_i^k|\Psi\rangle\
,\qquad \sqrt{-g}g^{ik}=\sqrt{-\bar g}\bar g^{il}(\exp{\Phi})_l^k\
.
\]
Quantum graviton operators are defined as the difference
$\hat\psi_i^k=\hat\Psi_i^k-\Phi_i^k$. Factorized amplitude
(\ref{2.25}) is calculated via the factorized measure
 \begin{equation}
\begin{array}{c}
\displaystyle
  d\hat \mu=d\mu_g\times d\mu_{\psi}\ ,
\\[3mm]\displaystyle
 d\mu_g=\prod_x\left\{(-g)^{5/2}\prod_{i\leqslant
 k}dg^{ik}\right\}\ , \qquad
 d\mu_{\psi}=\prod_x\prod_{i\leqslant k}d\hat\psi_i^k\ .
\end{array}
\label{2.26}
\end{equation}

Factorization of the measure allows the subsequent integration,
first by $d\mu_{\psi}$, $d\mu_\theta$, then by approximate
integration over  $d\mu_g$. In the operator formalism, such
consecutive integrations correspond to the solution of
self--consistent system of classic and quantum equations.
Classical equations are obtained by averaging of operator
equations (\ref{2.19}). They read:
\begin{equation}
\begin{array}{c}
\displaystyle
  \langle \Psi|\hat{\mathcal{E}}_i^k|\Psi\rangle=0\ .
\end{array}
\label{2.27}
\end{equation}
Subtraction of (\ref{2.27}) from (\ref{2.19}) gives the quantum dynamic equations
\begin{equation}
\begin{array}{c}
\displaystyle
 \hat{\mathcal{E}}_i^k-\langle \Psi|\hat{\mathcal{E}}_i^k|\Psi\rangle=0\ .
\end{array}
\label{2.28}
\end{equation}
Synchronous gauge (\ref{2.23}) is converted to the gauge of
classical metric and to conditions imposed on the state vector:
\begin{equation}
\begin{array}{c}
\displaystyle
\sqrt{-g}g^{00}=\sqrt{\bar \gamma},\qquad  \sqrt{-g}g^{0\alpha}=0\ ,
\\[5mm]
\displaystyle
  \hat\psi_0^i|\Psi\rangle=0\ .
\end{array}
\label{2.29}
\end{equation}
Quantum equations (\ref{2.24}) of ghosts' dynamics are added to
equations (\ref{2.27}) --- (\ref{2.29}).

Theory of gravitons in the macroscopic spacetime with
self--consistent geometry is without doubt an approximate theory.
Formally, the approximation is in the fact that the single
mathematical object $\sqrt{-\hat g}\hat g^{ik}$  is replaced by
two objects --- classical metric and quantum field, having
essentially different physical interpretations. That "coercion"\
of the theory can lead to a controversy, i.e. to the system of
equations having no solutions, if an inaccurate mathematics of the
adopted hypotheses is used. The scheme described above does not
have such a controversy. The most important element of the scheme
is the exponential parameterization (\ref{2.16}), which separates
the classical and quantum variables, as can be seen from
(\ref{2.26}). After the background and quantum fluctuations are
introduced, this parameterization looks as follows:
\begin{equation}
\begin{array}{c}
\displaystyle \sqrt{-\hat{g}}\hat{g}^{ik} = \sqrt{-\bar g}\bar
g^{il}\left(\exp{(\Phi+\hat\psi})\right)_l^k=
  \sqrt{-g} g^{il}(\exp\hat\psi)_l^k\ ,
\end{array}
\label{2.30}
\end{equation}
Note that the auxiliary basic space vanishes from the theory, and
instead the macroscopic (physical) spacetime with self--consistent
geometry takes its place.

If the geometry of macroscopic spacetime satisfies symmetry
constrains, the factorization of the measure (\ref{2.26}) becomes
not a formal procedure but strictly mathematical in its nature.
These restrictions must ensure the existence of an algorithm
solving the equations of constraints in the framework of the
perturbation theory (over the amplitude of quantum fields). The
theory of gravity is non--polynomial, so after the separation of
single field into classical and quantum components, the use of the
perturbation theory in the quantum sector becomes unavoidable. The
classical sector remains non--perturbative. In the general case,
when quantum field is defined in an arbitrary Riemann space, the
equations of constraints is not explicitly solvable. The problem
can be solved in the framework of perturbation theory if
background $g_{ik}$   and the free (linear) tensor field  $\hat
\psi_i^k$ belong to different irreducible representations of the
symmetry group of the background spacetime. In that case at the
level of linear field we obtain (\ref{2.26}), because the full
measure is represented as a product of measure of integration over
independent irreducible representations. At the next order,
factorization is done over coordinates, because the classical
background and the induced quantum fluctuations have essentially
different spacetime dynamics. Note, to factorize the measure by
symmetry criterion we do not need to go to the short--wave
approximation.

Background metric of isotropic cosmological models and classical
spherically symmetric non--stationary gravitational field meet the
constrains described above. These two cases are covering all
important applications of semi--quantum theory of gravity which
are quantum effects of vacuum polarization and creation of
gravitons in the non--stationary Universe and in the neighborhood
of black holes.

\subsection{Variational Principle for Classic and Quantum Equations}\label{var}

Geometrical variables can be identically transformed to the form
of functionals of classical and quantum variables. At the first
step of transformation there is no need to fix the
parameterization. Let us introduce the notations:
\begin{equation}
\begin{array}{c}
\displaystyle \sqrt{-\hat g}\hat g^{ik}=\sqrt{-g}\hat X^{ik},\qquad
 \frac{1}{\sqrt{-\hat g}}\hat g_{ik}=\frac{1}{\sqrt{-g}}\hat Y_{ik}\ ,
 \qquad \hat Y_{il}\hat X^{lk}=\delta_i^k\ .
\end{array}
\label{2.31}
\end{equation}
According to (\ref{2.30}), formalism of the theory allows
definition of quantum field $\hat\psi_i^k$  as symmetric tensor in
physical space, $g_{kl}\hat\psi^l_k=\hat\psi_{ik}=\hat\psi_{ki}$.
Objects, introduced in (\ref{2.31}), have the same status. With
any parameterization the following relationships take place:
\[
\displaystyle \lim_{\hat \psi_l^m\to 0}\hat X^{ik}=g^{ik},\qquad
\lim_{\hat \psi_l^m\to 0}\hat Y_{ik}=g_{ik}\ .
\]
We should also remember that the mixed components of tensors
$\hat X_i^k,\ \hat Y_i^k$ do not contain the background metric as
functional parameters. For any parameterization, these tensors are
only functionals of quantum fields $\hat \psi_i^k$ which are also
defined in mixed indexes. For the exponential parameterization:
\begin{equation}
\begin{array}{c}
\displaystyle \hat X_i^k=\delta_i^k+\hat\psi_i^k+\frac12\hat\psi_i^l\hat\psi_l^k+...\ ,\qquad
\hat Y_i^k=\delta_i^k-\hat\psi_i^k+\frac12\hat\psi_i^l\hat\psi_l^k+...\ ,
\qquad \hat g=g\cdot \hat d =ge^{\hat\psi}\ ,
\end{array}
\label{2.32}
\end{equation}
where  $d=\mbox{\rm det}\,|| \hat X^i_k||$. One can seen from
(\ref{2.32}), that the determinant of the full metric contains
only the trace of the quantum field.

Regardless of parameterization, the connectivity and curvature of
the macroscopic space $\Gamma_{ik}^l$, $R^i_{\, klm}$ are
extracted from full connectivity and curvature as additive terms:
\[
\displaystyle \hat \Gamma_{ik}^l=\Gamma_{ik}^l+ \hat {\mathcal{T}}_{ik}^l,\qquad
\hat R^i_{\, klm}=R^i_{\, klm}+\hat {\mathcal{R}}^i_{\, klm}\ .
\]
Quantum contribution to the curvature tensor,
\[
\displaystyle \hat {\mathcal{R}}^i_{\, klm}=\hat
{\mathcal{T}}_{km\ ;l}^i-\hat {\mathcal{T}}_{kl\ ;m}^i+\hat
{\mathcal{T}}_{nl}^i\hat {\mathcal{T}}_{km}^n-\hat
{\mathcal{T}}_{nm}^i\hat {\mathcal{T}}_{kl}^n\ ,
\]
is expressed via the quantum contribution to the full
connectivity:
\begin{equation}
\begin{array}{c}
\displaystyle \hat{\mathcal{T}}_{ik}^l=\frac12\left(-\hat
Y_{im}\hat X^{ml}_{\;\;\; ;k} -\hat Y_{km}\hat X^{ml}_{\;\;\;
;i}+\hat Y_{ij}\hat Y_{kn}\hat X^{ml}\hat X^{jn}_{\;\;\;
;m}\right)
+\frac14Y_{jn}\left(\delta_i^lX^{jn}_{\;\;\;
;k}+\delta_k^lX^{jn}_{\;\;\; ;i}-Y_{ik}X^{ml}X^{jn}_{\;\;\;
;m}\right)\ .
\end{array}
\label{2.33}
\end{equation}
The density of Ricci tensor in mixed indexes reads
\begin{equation}
\begin{array}{c}
\displaystyle \displaystyle \sqrt{-\hat g}\hat g^{kl}\hat
R_{il}=\sqrt{-g}\left\{\hat X^{kl}R_{il}+ \frac12\left[\hat
Y_{in}\left(\hat X^{ml}\hat X^{nk}_{\;\;\; ;m}-\hat X^{mk}\hat
X^{nl}_{\;\;\; ;m}\right)-\hat X^{lk}_{\;\;\;\; ;\ i} -\frac12
\delta_i^k\hat Y_{nj}\hat X^{ml}\hat X^{jn}_{\;\;\;
;m}\right]_{;l}\right.-
\\[5mm]
\displaystyle \left.-\frac14\left(\hat Y_{jn}\hat
Y_{sm}-\frac12\hat Y_{jm}\hat Y_{sn}\right)\hat X^{kl}\hat
X^{jm}_{\;\;\; ;i}\hat X^{ns}_{\;\;\; ;l}+\frac12\hat Y_{ml}\hat
X^{km}_{\;\;\; ;n}\hat X^{nl}_{\;\;\; ;i}\right\}\ .
\end{array}
\label{2.34}
\end{equation}
Symbol ";"\ in (\ref{2.33}), (\ref{2.34}) and in  what follows
stands for the covariant derivatives in background space. The
density of gauged gravitational Lagrangian is represented in a
form which is characteristic for the theory of quantum fields in
the classical background spacetime:
\begin{equation}
\begin{array}{c}
\displaystyle
S=\int d^4x\sqrt{-g}\left({\mathcal{L}}_{grav}-\sqrt{\hat d}\Lambda-
\frac{1}{4\varkappa}\hat X^{ik}\bar\theta_{ ,i}\theta_{ , k}\right)\ ,
\\[5mm]
\displaystyle
{\mathcal{L}}_{grav}=-\frac{1}{2\varkappa}\hat X^{ik}R_{ik}
+\frac{1}{8\varkappa}\left[\hat X^{kl}\left(\hat Y_{jn}\hat Y_{sm}-
\frac12\hat Y_{jm}\hat Y_{sn}\right)
\hat X^{jm}_{\;\;\; ;k}\hat X^{sn}_{\;\;\; ;l}-
2\hat Y_{ik}\hat X^{il}_{\;\;\; ;m}\hat X^{km}_{\;\;\;\; ;l}\right]\ .
\end{array}
\label{2.35}
\end{equation}
When the expression for ${\mathcal{L}}_{grav}$  was obtained from
contraction of tensor (\ref{2.34}), the full covariant divergence
in the background space have been excluded. Formulas (\ref{2.34}),
(\ref{2.35}) apply for at any parameterization.

Let us discuss the variation method. In the exact quantum theory
of gravity with the trivial measure (\ref{2.15}), the variation of
the action over variables  $\hat \Psi_i^k$ leads to the Einstein
equations in mixed indexes (\ref{2.18}) and (\ref{2.19}). In the
exact theory, the exponential parameterization is convenient, but,
generally speaking, is not necessary. A principally different
situation takes place in the approximate self--consistent theory
of gravitons in the macroscopic spacetime. In that theory the
number of variables doubles, and with this, the classical and
quantum components of gravitational fields have to have the status
of the dynamically independent variables due to the doubling of
the number of equations. The variation should be done separately
over each type of variables. The formalism of the path integration
suggests a rigid criterion of dynamic independence: {\it the full
measure of integration, by definition, must be factorized with
respect to the dynamically independent variables.} Obviously, only
the exponential parameterization (\ref{2.30}), leading to the
factorized measure (\ref{2.26}), meets the criterion.

The variation of the action over the classic variables is done
together with the operation of averaging over the quantum
ensemble. In the result, equations for metric of the macroscopic
spacetime are obtained:
\begin{equation}
\displaystyle \langle\Psi|\frac{\delta S}{\delta g^{in}}|\Psi\rangle=
-2\varkappa \sqrt{-g}g_{nk}\langle\Psi|\hat G_i^k-\frac12\delta_i^k\hat G_l^l|\Psi\rangle=0\ ,
\label{2.36}
\end{equation}
where $\hat G_i^k=\hat {\mathcal{G}}_i^k/\sqrt{-g}$. Variation of
the action over background variables, defined as
$\Phi_i^k=\langle\Psi|\hat\Psi^k_i|\Psi\rangle$, yields the
equations:
\begin{equation}
\displaystyle \langle\Psi|\frac{\delta S}{\delta\Phi_k^i}|\Psi\rangle=
-2\varkappa \sqrt{-g}\langle\Psi|\hat G_i^k|\Psi\rangle=0\ .
\label{2.37}
\end{equation}
Equations (\ref{2.36}) and (\ref{2.37}) are mathematically
identical. We should also mention that if the variations over the
background metric are done with the fixed {\it mixed} components
of the quantum field, these equations are valid for any
parameterization.

Exponential parameterization (\ref{2.30}) has a unique property:
{\it the variations over classic  $\Phi_i^k$ (before averaging)
and quantum $\hat\psi_i^k$  (without averaging) variables lead to
the same equations.} That fact is a direct consequence of the
relations, showing that variations $\delta\Phi_l^k$  and
$\delta\hat\psi_l^k$  are multiplied by the same operator
multiplier:
\[
\begin{array}{c}
\displaystyle
\delta\sqrt{-\hat g}\hat g^{ik}=\sqrt{-\hat g}\hat g^{il}\delta\Phi_l^k\ ,\qquad
\hat\psi_i^k=const\ ,
\\[5mm]
\displaystyle
\delta\sqrt{-\hat g}\hat g^{ik}=\sqrt{-\hat g}\hat g^{il}\delta\hat\psi_l^k\ ,\qquad
\Phi_i^k=const\ .
\end{array}
\]
By a simple operation of subtraction, the identity allows the
extraction of pure background terms from the equation of quantum
field. The equations of graviton theory in the macroscopic space
with self--consistent geometry are written as follows:
\begin{equation}
\displaystyle  \langle\Psi|\hat E_i^k|\Psi\rangle\equiv
\langle\Psi|\hat G_i^k-\frac12\delta_i^k\hat G_l^l|\Psi\rangle=0\ ,
\label{2.38}
\end{equation}
\begin{equation}
\displaystyle \hat L_i^k\equiv
\hat G_i^k-\frac12\delta_i^k\hat G_l^l-\langle\Psi|\hat G_i^k-
\frac12\delta_i^k\hat G_l^l|\Psi \rangle=0\ .
\label{2.39}
\end{equation}

With the exponential parameterization, the formalism of the theory
can be expressed in an elegant form. Let us go to the rules of
differentiation of exponential matrix functions
\begin{equation}
\displaystyle \hat Y_{im}\hat X^{mk}_{\;\;\; ;\; l}=\hat
\psi_{i;\; l}^k\ ,\qquad \hat X^{ik}_{\;\;\; ;\; l}=\hat X^{im}\hat
\psi_{m\;\; ;\; l}^k\ . \label{2.40}
\end{equation}
Taking into account (\ref{2.40}), we get the quantum contribution
to the full connectivity  (\ref{2.33}) as follows
\begin{equation}
\begin{array}{c}
\displaystyle \hat{\mathcal{T}}_{ik}^l=\frac12\left(-\hat
\psi_{i;k}^l -\hat \psi_{k;i}^l+\hat Y_{kn}\hat
X^{lm}\hat\psi^n_{i;m}\right)+
\frac14\left(\delta_i^l\hat\psi_{;k}+\delta_k^l\hat\psi_{;i}-
\hat Y_{ik}\hat X^{lm}\hat\psi_{;m}\right)\ .
\end{array}
\label{2.41}
\end{equation}
Formulas (\ref{2.35}) could be rewritten as follows:
\begin{equation}
\begin{array}{c}
\displaystyle  \hat X^l_k=(\exp \hat\psi)_k^l,\qquad \sqrt{\hat d}=e^{\hat\psi/2}\ ,
\\[5mm]
\displaystyle S=\int
d^4x\sqrt{-g}\left({\mathcal{L}}_{grav}-\Lambda
e^{\hat\psi/2}-\frac{1}{4\varkappa}\hat
X_k^l\bar\theta^{;k}\theta_{; l}\right)\ ,
\\[5mm]
\displaystyle {\mathcal{L}}_{grav}=-\frac{1}{2\varkappa}\hat
X_k^lR^k_l+
\frac{1}{8\varkappa}\hat X_k^l\left(\hat
\psi_n^{m;k}\hat \psi_{m;l}^n-\frac12\hat\psi^{;k}\hat\psi_{;l}
-2\hat\psi^{k;m}_n\hat\psi^n_{m;l}\right)\ .
\end{array}
\label{2.42}
\end{equation}
As is seen from (\ref{2.42}), for the exponential
parameterization, the non--polynomial structures of quantum theory
of gravity have been completely reduced to the factorized
exponents\footnote{We are using the standard definitions. Matrix
functions are defined by their expansion into power series as any
operator functions:
\[
   \hat U(\hat V)=\sum_nc_n{\hat V}^n\ .
\]
The derivative of  $n$--th degrees of matrix by the same matrix is
defined as
 \[
\displaystyle \frac{\partial {\hat V}^n}{\partial \hat V}=n{\hat V}^{n-1}\ .
  \]
The derivative by numerical (non matrix) parameter $z$  is
\[
\displaystyle \frac{\partial {\hat V}^n}{\partial z}=n{\hat
V}^{n-1}\cdot\frac{\partial {\hat V}}{\partial z}\ .
\]
If matrix function  $ \hat U(\hat V)$ and its derivative  $\hat
W=\partial {\hat U}^n/\partial \hat V$ are elementary functions,
then
\[
\displaystyle \frac{\partial {\hat U}}{\partial z}=\hat
W\cdot\frac{\partial {\hat V}}{\partial z}\ .
\]
Formulas (\ref{2.40}) --- (\ref{2.42}) are the consequence of
these definitions. It worth to mention, that in matrix analysis in
all intermediate formulas one should be careful with the index
ordering.}.

The explicit form of the tensor, in the terms of which the
self--consistent system of equations could be written is as
follows
\begin{equation}
\begin{array}{c}
\displaystyle \hat E_i^k\equiv \hat G_i^k-\frac12\delta_i^k\hat
G_l^l=
\hat X^{kl}{R_{li} -\frac12\delta_i^k\hat
X^{lm}}R_{ml}-\delta_i^k\varkappa\Lambda e^{\hat\psi/2}+
\\[5mm]
\displaystyle + \frac12\left[\hat X^{lm}\left(\hat\psi^k_{i;m}-
\hat\psi^k_{m;i}\right) -\hat X^{km} \hat
\psi^l_{i;m}+ \frac12\delta_i^k\left( \hat X^{mn} \hat
\psi_{n;m}^l+ \hat X^{lm} \hat\psi_{m;n}^n\right)\right]_{;\ l}-
\\[5mm]
\displaystyle -\frac14\hat X^{kl}\left(\hat\psi^n_{m;i}
\hat\psi_{n;l}^m- \frac12\hat\psi_{;i}
\hat\psi_{;l}-2\hat\psi_{l;m}^n\hat
\psi_{n;i}^m\right) +
\frac18\delta_i^k\hat
X^{rl}\left(\hat\psi^n_{m;r} \hat\psi_{n;l}^m-
\frac12\hat\psi_{;r} \hat\psi_{;l}-2\hat\psi_{l;m}^n\hat
\psi_{n;r}^m\right)-
\\[5mm]
\displaystyle
-\frac{1}{4}\left[\hat X^{kl}\left(\bar
\theta_{;l}\theta_{;i}+\bar
\theta_{;i}\theta_{;l}\right)-\delta_i^k\hat
X^{ml}\bar\theta_{;m}\theta_{;l}\right]\ .
\end{array}
\label{2.43}
\end{equation}
Let us introduce the following notations:
\begin{equation}
\begin{array}{c}
\displaystyle \hat X_{(1)}^{ik}=\hat X^{ik}-
g^{ik}=\hat\psi^{ik}+\frac12\hat\psi^{il}\hat\psi_l^k+...\ ,
\\[3mm]
\displaystyle \hat X_{(2)}^{ik}=\hat X^{ik}-
g^{ik}-\hat\psi^{ik}=\frac12\hat\psi^{il}\hat\psi_l^k+...\ .
\end{array}
\label{2.44}
\end{equation}
With use of (\ref{2.44}), let us extract from (\ref{2.43}) the
terms not containing the quantum field, and the terms linear over
the quantum field:
\begin{equation}
\begin{array}{c}
\displaystyle \hat
E_i^k=R_i^k-\frac12\delta_i^kR-\delta_i^k\varkappa\Lambda+
\frac12\left(\hat\psi^{k\; ;l}_{i;l}-\hat\psi^{k\; ;l}_{l;i}
-\hat\psi^{l;k}_{i\;
;l}+\delta_i^k\hat\psi_{m;l}^{l;m}\right)+
\hat\psi^k_lR_i^l-\frac12\delta_i^k\hat\psi^m_lR_m^l-
 \frac12\delta_i^k\varkappa\Lambda\hat\psi-\varkappa\hat T_i^k\ ,
\\[5mm]
\displaystyle \hat T_i^k=\hat T_{i(grav)}^k+\hat T_{i(ghost)}^k\ ,
\end{array}
\label{2.45}
\end{equation}
where
\begin{equation}
\begin{array}{c}
\displaystyle \varkappa\hat T_{i(grav)}^k=\frac14\hat
X^{kl}\left(\hat\psi^n_{m;i} \hat\psi_{n;l}^m-
\frac12\hat\psi_{;i} \hat\psi_{;l}-2\hat\psi_{l;m}^n\hat
\psi_{n;i}^m\right)
-\frac18\delta_i^k\hat
X^{rl}\left(\hat\psi^n_{m;r} \hat\psi_{n;l}^m-
\frac12\hat\psi_{;r} \hat\psi_{;l}-2\hat\psi_{l;m}^n\hat
\psi_{n;r}^m\right)-
\\[5mm]
\displaystyle -\frac12\left[\hat
X_{(1)}^{lm}\left(\hat\psi^k_{i;m}- \hat\psi^k_{m;i}\right) - \hat
X_{(1)}^{km} \hat\psi^l_{i;m}+\frac12\delta_i^k\left( \hat
X_{(1)}^{mn} \hat\psi_{n;m}^l+ \hat X_{(1)}^{lm}
\hat\psi_{m;n}^n\right)\right]_{;\ l}-
\\[5mm]
\displaystyle -\hat X_{(2)}^{kl}{R_{li} +\frac12\delta_i^k\hat
X_{(2)}^{lm}}R_{ml}+\delta_i^k\varkappa\Lambda
\left(e^{\hat\psi/2}-1-\frac12\hat\psi\right)
\end{array}
\label{2.46}
\end{equation}
is the EMT of gravitons;
\begin{equation}
\begin{array}{c}
\displaystyle \varkappa\hat T_{i(ghost)}^k=-\frac{1}{4}\left[\hat
X^{kl}\left(\bar \theta_{;l}\theta_{;i}+\bar
\theta_{;i}\theta_{;l}\right)-\delta_i^k\hat
X^{ml}\bar\theta_{;m}\theta_{;l}\right]
\end{array}
\label{2.47}
\end{equation}
is the EMT of ghosts. In the averaging of (\ref{2.45}), it was
taken into account that  $\langle\Psi|\hat
\psi_i^k|\Psi\rangle\equiv 0$ by definition of the quantum field.
Averaged equations for the classic fields (\ref{2.38}) take form
of the standard Einstein equations containing averaged EMT of
gravitons, renormalized by ghosts:
\begin{equation}
\begin{array}{c}
\displaystyle \langle\Psi|\hat E_i^k|\Psi\rangle\equiv
R_i^k-\frac12\delta_i^kR-\delta_i^k\varkappa\Lambda
 -\varkappa\langle\Psi|\hat T_i^k|\Psi\rangle=0\ .
\end{array}
\label{2.48}
\end{equation}
Quantum dynamic equations for gravitons (\ref{2.39}) could be
rewritten as follows:
\begin{equation}
\begin{array}{c}
\displaystyle \hat L_i^k\equiv\frac12\left(\hat\psi^{k\;
;l}_{i;l}-\hat\psi^{k\; ;l}_{l;i} -\hat\psi^{l;k}_{i\;
;l}+\delta_i^k\hat\psi_{m;l}^{l;m}\right)-
\frac12\delta_i^k\varkappa\Lambda\hat\psi
+\hat\psi^k_lR_i^l-\frac12\delta_i^k\hat\psi^m_lR_m^l-\varkappa\left(\hat
T_i^k-\langle\Psi|\hat T_i^k|\Psi\rangle\right)=0\ .
\end{array}
\label{2.49}
\end{equation}
As is seen in the equations (\ref{2.49}), in the theory of
gravitons all nonlinear effects are in the difference between the
EMT operator and its average value. System of equations
(\ref{2.48}), (\ref{2.49}) is closed by the quantum dynamic
equations for ghosts, which could be also written in 4D covariant
form:
\begin{equation}
\displaystyle (\hat X^{ik}\theta_{;k})_{;i}=0,\qquad (\hat X^{ik}\bar\theta_{;k})_{;i}=0
\label{2.50}
\end{equation}
Equations (\ref{2.50}) provide the realization of the conservative
nature of the ghosts' EMT:
\begin{equation}
\displaystyle \langle\Psi|\hat T_{i(ghost)}^k|\Psi\rangle_{;k}=0\ .
\label{2.51}
\end{equation}

\subsection{Differential Identities}\label{equiv}

In the exact theory, which is dealing with the full metric, there is an identity:
\begin{equation}
\begin{array}{c}
\displaystyle \hat D_k\left\{\hat g^{kl}\hat
R_{li}-\frac12\delta_i^k\hat g^{ml}\hat
R_{lm}-\delta_i^k\varkappa\Lambda+\frac14\left[\hat
g^{kl}\left(\bar \theta_{;l}\theta_{;i}+\bar
\theta_{;i}\theta_{;l}\right)-\delta_i^k\hat
g^{ml}\bar\theta_{;m}\theta_{;l}\right]\right\}=0\ ,
\end{array}
\label{2.52}
\end{equation}
where $\hat D_k$ is the covariant derivative in the space with
metric  $\hat g_{ik}$. This identity is satisfied  by Bianchi
identity and by the ghost equations of motion. In terms of
covariant derivative in the background space, identity
(\ref{2.52}) could be rewritten as follows:
\begin{equation}
\displaystyle \hat E_{i;k}^k-\frac12(\ln\hat d)_{;k}\hat
E_i^k+\hat{\mathcal{T}}_{kl}^k\hat
E_i^l-\hat{\mathcal{T}}_{ik}^l\hat E_l^k\equiv \hat
E_{i;k}^k-\hat{\mathcal{T}}_{ik}^l\hat E_l^k=0\ . \label{2.53}
\end{equation}
For the exponential parameterization, taking into account
(\ref{2.41}), the expression (\ref{2.53}) can be  transformed to
the following form
\begin{equation}
\displaystyle \hat E_{i;k}^k+\frac12\hat\psi^l_{k;i}\left(\hat
E_l^k-\frac12\delta_l^k\hat E_l^l\right)=0\ . \label{2.54}
\end{equation}
Identity transformation  $\hat E_i^k\equiv \langle\Psi|\hat
E_i^k|\Psi\rangle+\hat L_i^k$ and the subsequent averaging of
(\ref{2.54}) yields:
\begin{equation}
\displaystyle \langle\Psi|\hat
E_i^k|\Psi\rangle_{;k}+\frac12\langle\Psi|\hat\psi^l_{k;i}\left(\hat
L_l^k-\frac12\delta_l^k\hat L_m^m\right)|\Psi\rangle=0\ .
\label{2.55}
\end{equation}
Here we have used explicitly the fact that $\langle\Psi|\hat
\psi_i^k|\Psi\rangle\equiv 0$, $\langle\Psi|\hat
L_i^k|\Psi\rangle\equiv 0$,   by definition. Next, expression
(\ref{2.48}) is substituted into (\ref{2.55}). Taking into account
the Bianchi identity and the conservation of the ghost EMT, we
obtain:
\begin{equation}
\displaystyle \langle\Psi|\hat
T_{i(grav)}^k|\Psi\rangle_{;k}=\frac12\langle\Psi|\hat\psi^l_{k;i}\left(\hat
L_l^k-\frac12\delta_l^k\hat L_m^m\right)|\Psi\rangle\ .
\label{2.56}
\end{equation}
As is seen from (\ref{2.56}), quantum equations of motion
(\ref{2.49}) provide the conservation of the averaged EMT of
gravitons:
\begin{equation}
\displaystyle \langle\Psi|\hat T_{i(grav)}^k|\Psi\rangle_{;k}=0\ .
\label{2.57}
\end{equation}

Take notice, that tensors $\hat E_i^k$  and $\hat L_i^k$ in
(\ref{2.54}), (\ref{2.56}) are multiplied by the linear forms of
graviton field operators only. Such a structure of identities is
only valid for the exponential parameterization. This fact is of
key value for the computations in the framework of perturbation
theory. The order $n$   of the perturbation theory is defined by
the highest degree of the field operator in the quantum dynamic
equations for gravitons (\ref{2.49}). The EMT of gravitons which
is consistent with the quantum equation of order $n$ contains
averaged products of field operators of the order $n+1$ (e.g., the
quadratic EMT is consistent with the linear operator equation). We
see that by defining the order of the perturbation theory, we have
identity (\ref{2.56}), in which all terms are of the same maximal
order of the quantum field amplitude:
\begin{equation}
\displaystyle \langle\Psi|\hat
T_{i(grav)}^{k(n+1)}|\Psi\rangle_{;k}=\frac12\langle\Psi|\hat\psi^l_{k;i}\left(\hat
L^{k(n)}_l-\frac12\delta_l^k\hat L^{m(n)}_m\right)|\Psi\rangle\ .
\label{2.58}
\end{equation}
Such a structure of the identity automatically provides the
conservation condition (\ref{2.57}) at any order of perturbation
theory\footnote{In the framework of the perturbation theory, any
parameterization, except the exponential one, creates
mathematically contradictory models, in which the perturbative EMT
of gravitons  $\langle\Psi|\hat T_{i(grav)}^{k(n+1)}|\Psi\rangle$
is not conserved. In our opinion, a discussion of artificial
methods of solutions of this problem, appeared, for example, if
linear parameterization $g_{ik}=g_{ik}+\hat\psi_{ik}$ is used,
makes no sense. The algorithm we have suggested here is well
defined because it is based on the exact procedure of separation
between the classical and quantum variables in terms of normal
coordinates. We believe there is no other mathematically
non--contradictive scheme. }.

\subsection{One--Loop Approximation}\label{1loop}

In the framework of one--loop approximation, quantum fields
interact only with the classic gravitational field. Accordingly,
equations (\ref{2.49}) are being converted into linear operator
equations:
\begin{equation}
\begin{array}{c}
\displaystyle
\hat L_i^k=\frac12\left( \hat\psi_{i;l}^{k;l}-\hat\psi_{i;\; l}^{l;k}
-\hat\psi_{l;i}^{k;\; l}         +\delta_i^k\hat\psi_{l;m}^{m;l}\right)+
\hat\psi_l^kR_i^l-\frac12\delta_i^k\hat\psi_m^lR_l^m-\frac12\delta_i^k\varkappa\Lambda\hat\psi
 =0\ .
\end{array}
\label{2.59}
\end{equation}
Of course, these equations are separated into the equations of
constraints (initial conditions):
\begin{equation}
\displaystyle \hat L_0^0|\Psi\rangle=0,\qquad \hat
L_0^\alpha|\Psi\rangle=0,\qquad \hat L_\alpha^0|\Psi\rangle=0\ ,
\label{2.60}
\end{equation}
and the equations of motion:
\begin{equation}
\displaystyle \hat L_\alpha^\beta-\frac12\delta_\alpha^\beta\hat L_l^l=0\ .
\label{2.61}
\end{equation}
The equations for ghosts (\ref{2.50}) are also transformed into the linear operator equations:
\begin{equation}
\begin{array}{c}
\displaystyle
         \theta_{;i}^{;i}=0\ ,\qquad \bar\theta_{;i}^{;i}=0\ .
\end{array}
\label{2.62}
\end{equation}

In the one--loop approximation, the state vector is represented as
a product of normalized state vectors of gravitons and ghosts:
\begin{equation}
\displaystyle |\Psi\rangle=|\Psi_g\rangle|\Psi_{gh}\rangle\ .
 \label{2.63}
 \end{equation}
Equations for macroscopic metric (\ref{2.48}) take the form:
\begin{equation}
\begin{array}{c}
\displaystyle
  R_i^k -\frac{1}{2}\delta_i^k R- \varkappa\delta_i^k\Lambda =
\varkappa\left(\langle \Psi_g|\hat{T}_{i(grav)}^k|\Psi_g\rangle+
  \langle \Psi_{gh}|\hat{T}_{i(ghost)}^k|\Psi_{gh}\rangle\right)\ .
  \end{array}
\label{2.64}
\end{equation}
The averaged EMTs of gravitons and ghosts in equations
(\ref{2.64}) are the quadratic forms of the quantum fields.
Assuming that $\hat X^{ik}=g^{ik}$, $\hat X^{ik}_{(1)}=\hat
\psi^{ik}$, $\hat X^{ik}_{(2)}=\hat \psi^{il}\hat \psi_l^k/2$  in
(\ref{2.46}), (\ref{2.47}), we obtain:
\begin{equation}
\begin{array}{c}
\displaystyle
  \hat{T}_{i(grav)}^k =
\frac{1}{4\varkappa}\left\{
 \hat\psi_{m;i}^l\hat\psi_l^{m;k}
-\frac{1}{2}\hat\psi_{;i}\hat\psi^{;k}
 -\hat\psi_{i;m}^l\hat\psi_l^{m;k}  -\hat\psi_l^{k;m}\hat\psi_{m;i}^l-\frac{1}{2}\delta_i^k\left(\hat\psi_{m;n}^l\hat\psi_l^{m;n}
  -\frac{1}{2}\hat\psi_{;n}\hat\psi^{;n}-2\hat\psi^l_{n;m}\hat\psi_l^{m;n}
  \right)-\right.
\\[3mm] \\ \displaystyle
 \left. -
  2\left[\hat\psi_m^l\hat\psi_i^{k;m}
        -\hat\psi_m^k\hat\psi_i^{l;m}
        -\hat\psi^{lm}\hat\psi_{m;i}^k
         +\frac12\delta_i^k\left(\hat\psi_m^n\hat\psi_n^l\right)^{;m}
        \right]_{;\ l}-2\hat\psi_m^k\hat\psi_l^mR_i^l
        +\delta_i^k\hat\psi_l^n\hat\psi_n^mR_m^l+\frac12\delta_i^k\varkappa\Lambda\hat\psi^2
  \right\}\ ,
  \end{array}
\label{2.65}
\end{equation}

\begin{equation}
\displaystyle \hat
T_{i(ghost)}^{k}=-\frac{1}{4\varkappa}\left(\bar\theta_{;i}\theta^{;k}+\bar\theta^{;k}\theta_{;i}-
\delta_i^k\bar\theta^{;l}\theta_{;l}\right)\ . \label{2.66}
\end{equation}
Quantum equations (\ref{2.59}), (\ref{2.62}) provide the
conservation of tensors (\ref{2.65}), (\ref{2.66}) in the
background space:
\begin{equation}
\displaystyle \langle \Psi_g|\hat{T}_{i(grav)}^k|\Psi_g\rangle_{
;\ k}=0, \qquad
  \langle \Psi_{gh}|\hat{T}_{i(ghost)}^k|\Psi_{gh}\rangle_{;\ k}=0\ .
 \label{2.67}
 \end{equation}

The ghost sector of the theory (\ref{2.59}) --- (\ref{2.67})
corresponds to the gauge (\ref{2.29}). Note, however, that all
equations of the theory, except gauges, are formally general
covariant in the background space. That provides a way of
expanding the class of gauges for classic fields. Obviously, we
can move from the initial 4--coordinates, corresponding to the
classic sector of gauges (\ref{2.29}), to any other coordinates,
conserving quantum gauge condition
\begin{equation}
\displaystyle \hat\psi_0^i|\Psi\rangle=0\ .
 \label{2.68}
\end{equation}
It is not difficult to see, that in the classic sector any gauges
of synchronous type are allowed:
\begin{equation}
\displaystyle g_{00}=N^2(t), \qquad g_{0\alpha}=0\ .
 \label{2.69}
\end{equation}
where $N(t)$ is an arbitrary function of time.

An important technical detail is that in the perturbation theory
the graviton field should be consistent with an additional
identity. In one--loop approximation that identity is obtained
from the covariant differentiation of equation (\ref{2.59}):
 \begin{equation}
\displaystyle  \hat Q_i\equiv\left(R_l^k+\varkappa\Lambda\delta_l^k\right)\hat\psi_{k;i}^l=0\ .
 \label{2.70}
\end{equation}
The appearance of conditions (\ref{2.70}) reflects the fact that
we are dealing with an approximate theory. As it was already
mentioned in Section \ref{fac}, the partition of the metric into
classic and quantum components, and, respectively, the
factorization of the path integral, can be only done under the
condition that additional constrains are applied to the geometry
of background space. These constrains are manifested through the
structure of the Ricci tensor of the background space which should
provide the identity (\ref{2.70}) for the solutions of dynamic
equations for gravitons. In the Heisenberg form of quantum theory
the additional identity can be written as conditions on the state
vector:
\begin{equation}
\displaystyle  \hat
Q_i|\Psi\rangle\equiv\left(R_l^k+\varkappa\Lambda\delta_l^k\right)\hat\psi_{k;i}^l|\Psi\rangle=0\
.
 \label{2.71}
\end{equation}
Status of all constrains for the state vectors are the same and
are as follows. If (\ref{2.60}), (\ref{2.68}), (\ref{2.71}) exist
at the initial moment of time, the internal properties of the
theory should provide their existence at any following instance of
time.

While one is conducting a concrete one--loop calculation, there is
a problem of gauge invariance of the total EMT of gravitons and
ghosts. As was mentioned by De Witt \cite{7}, after the separation
of the metric into background and graviton components, the
transformations of the diffeomorphism group (\ref{2.4}) can be
represented as transformations of the internal gauge symmetry of
graviton field. In the framework of one--loop approximation, these
transformations are as follows:
\begin{equation}
 \displaystyle
\delta\hat\psi_i^k=-\delta_i^k\eta^l_{;l}+\eta^k_{;i}+\eta_i^{;k} \ .
 \label{2.72}
 \end{equation}
The problem of gauge non--invariance is twofold. First, the EMT of
gravitons (\ref{2.65}) is not invariant with respect to
transformations in (\ref{2.72}). Second, the ghost sector (the
ghost EMT), inevitably presented in the theory, depends on the
gauge. Concerning the first problem, it is known that the
operation removing gauge non--invariant terms from the EMT of
gravitons belongs to the operation of averaging over a quantum
ensemble. In the general case of arbitrary background geometry and
arbitrary graviton wavelengths we encounter a number of problems
(when conducting this operation), which should be discussed
separately.

In the particular case of the theory of gravitons in a homogeneous
and isotropic Universe, the averaging problem has a consistent
mathematical solution. It was shown in Section \ref{3vs} that {\it
removing the gauge non--invariant contributions from the EMT of
gravitons from the quantum ensemble has been set
gauge--invariantly.} To address the second aspect of the problem,
we should take into account that the theory of gravitons in the
macroscopic space with the self--consistent geometry operates with
macroscopic observables. Therefore, in this theory one--loop
finiteness, as the general property of one--loop quantum gravity,
should have a specific embodiment: {\it by their mathematical
definition, macroscopic observables must be the finite values.}
This requirement on the theory in the Heisenberg representation is realized in Hamilton gauge
(\ref{2.29}) only.

\section{Self--Consistent Theory of Gravitons in the Isotropic Universe}\label{scgt}

\subsection{Elimination of 3--Vector and 3--Scalar Modes
by Conditions Imposed on the State Vector}\label{3vs}

We consider the quantum theory of gravitons in the spacetime with
the following background metric
\begin{equation}
 \displaystyle
 ds^2=g_{ik}dx^idx^k=N^2(t)dt^2-a^2(t)(dx^2+dy^2+dz^2)\ .
 \label{3.1}
 \end{equation}
In this space the graviton field is expanded over the irreducible
representations of the group of three--dimensional rotations, i.e.
over 3--tensor $\hat\psi_{\alpha(t)}^\beta$, 3--vector
$\hat\psi_{i(v)}^k=(\hat\psi_{0(v)}^\alpha,\,
\hat\psi_{\alpha(v)}^\beta)$  and 3--scalar
$\hat\psi_{i(s)}^k=(\hat\psi_{0(s)}^0,\, \hat\psi_{0(s)}^\alpha,\,
\hat\psi_{\alpha(s)}^\beta)$ modes. Equations (\ref{2.59}) are
split  into three independent systems of equations, so that each
of such systems represents each mode separately. The state vector
of gravitons is of multiplicative form that reads
 \[
 \displaystyle
|\Psi_g\rangle=|\Psi_t\rangle|\Psi_v\rangle|\Psi_s\rangle\ .
 \]
The averaged EMT (\ref{2.63}) is presented by an additive form that reads:
\begin{equation}
\begin{array}{c}
 \displaystyle
\langle \Psi_g|\hat T_{i(grav)}^k|\Psi_g\rangle=
\langle
\Psi_t|\hat T_{i(t)}^k|\Psi_t\rangle +
 \langle \Psi_v|\hat T_{i(v)}^k|\Psi_v\rangle+\langle \Psi_s|\hat T_{i(s)}^k|\Psi_s\rangle \ .
 \end{array}
 \label{3.2}
 \end{equation}
The averaged EMT contains no products of modes that belong to
different irreducible representations. This is because the
equality $\langle \Psi_g|\hat\psi_i^k|\Psi_g\rangle=0$ is divided
into three following three independent equalities
\begin{equation}
\begin{array}{c}
\displaystyle
\langle \Psi_s|\hat\psi_{i(s)}^k|\Psi_s\rangle=0,\qquad
\langle \Psi_v|\hat\psi_{i(v)}^k|\Psi_v\rangle=0,
\qquad
\langle \Psi_t|\hat\psi_{i(t)}^k|\Psi_t\rangle=0\ .
\end{array}
\label{3.3}
\end{equation}
Equalities (\ref{3.3}) are conditions that provide the consistency
of properties of quantum ensemble of gravitons with the properties
of homogeneity and isotropy of the background. In the homogeneous
and isotropic space, the same equalities hold for Fourier images
of the graviton field. Therefore, the satisfaction of these
equalities is provided by the isotropy of graviton spectrum in the
${\bf k}$--space and by the equivalence of different
polarizations.

3--tensor modes $\hat\psi_{\alpha(t)}^\beta$ and their EMT
$\langle\Psi_t|\hat T_{i(t)}^k|\Psi_t\rangle$, respectively, are
gauge invariant objects. Gauge non--invariant modes
$\hat\psi_{i(v)}^k,\, \hat\psi_{i(s)}^k$ are eliminated by
conditions that, imposed on the state vector, read
 \begin{equation}
 \displaystyle
\hat\psi_{i(v)}^k|\Psi_v\rangle=\hat\psi_{i(s)}^k|\Psi_s\rangle=0\ .
 \label{3.4}
 \end{equation}
Note that the conditions (\ref{3.4}) automatically follow from
equations (\ref{2.59}) and conditions (\ref{2.66}). As a result of
this, a gauge non--invariant EMT of 3--scalar and 3--vector modes
is eliminated from the macroscopic Einstein equations, and we get
\begin{equation}
\displaystyle \langle \Psi_v|\hat T_{i(v)}^k|\Psi_v\rangle=0\ ,\qquad
\langle \Psi_s|\hat T_{i(s)}^k|\Psi_s\rangle=0\ .
 \label{3.5}
 \end{equation}

The important fact is that {\it in the isotropic Universe, the
separation of gauge invariant EMT of 3--tensor gravitons is
accomplished without the use of short--wave approximation}. In
connection with this, note the following fact. In the theory,
which formally operates with waves of arbitrary lengths, the
problem of existence of a quantum ensemble of waves with
wavelengths greater than the distance from horizon is open
\cite{11}. In cosmology, the existence of such an ensemble is
provided by the following experimental fact. In the real Universe
(whose properties are controlled by observational data beginning
from the instant of recombination), the characteristic scale of
casually--connected regions is much greater (many orders of
magnitude) than the formal horizon of events. The standard
explanation of this fact is based on the hypothesis of early
inflation. Taking into account these circumstances, we do not
impose any additional restrictions on the quantum ensemble.

The procedure described above is based on the existence of
independent irreducible representations of graviton modes only.
But in this procedure, gauge--non--invariant modes are eliminated
by using of a gauge, i.e. they are eliminated by using of
gauge--non--invariant procedures. The gauge--invariant procedure
of getting the same results is presented in \cite{MUV2}, Section III A.

\subsection{Canonical Quantization of 3--Tensor Gravitons and Ghosts}\label{quant}

The parameters of gauge transformations do not contain terms of
expansion over transverse 3--tensor plane waves. Therefore,
Fourier images of tensor fluctuations are gauge--invariant by
definition. We have
\begin{equation}
\begin{array}{c}
\displaystyle  \hat \psi_0^0({\bf k})=0,\qquad \hat \psi^0_\alpha({\bf
k})=-\hat \psi_0^\alpha({\bf k})=0\ ,
\qquad
\hat\psi_\alpha^\beta({\bf
k}\sigma)=Q_\alpha^\beta({\bf k}\sigma)\hat\psi_{{\bf k}\sigma}\
,
  \\[5mm]
\displaystyle k_\alpha Q^\alpha_\beta({\bf k}\sigma)\equiv 0\
,\qquad Q^\alpha_\alpha({\bf k}\sigma)\equiv 0\ ,
\end{array}
 \label{3.29}
 \end{equation}
where  $\sigma$ is the index of transverse polarizations. The
operator equation for 3--tensor gravitons is
 \begin{equation}
 \begin{array}{c}
 \displaystyle
\psi_{\alpha(t)}^\beta(t,{\bf x})=\sum_{{\bf
k}\sigma}Q_\alpha^\beta({\bf k}\sigma)\psi_{{\bf
k}\sigma}(t)e^{i{\bf kx}}\ ,
\qquad \ddot \psi_{{\bf
k}\sigma}+3H\dot\psi_{{\bf
k}\sigma}+\frac{k^2}{a^2}\psi_{{\bf k}\sigma}=0\ ,
\end{array}
 \label{3.30}
 \end{equation}
where $H=\dot a/a$  is Hubble function and dots mean derivatives
with respect to the physical time $t$.

The special property of the gauge used is the following. The
differential equation for ghosts is obtained from the equation for
gravitons by exchange of graviton operator with the ghost
operator. It reads
\begin{equation}
 \displaystyle
\theta(t,{\bf x})=\sum_{{\bf k}}\theta_{{\bf k}}(t)e^{i{\bf kx}}\
, \qquad \ddot \theta_{{\bf k}}+3H\dot\theta_{{\bf
k}}+\frac{k^2}{a^2}\theta_{{\bf k}}=0\ .
 \label{3.31}
\end{equation}

Macroscopic Einstein equations (\ref{2.64}) read
\begin{equation}
\displaystyle
3H^2=\varkappa\left(\varepsilon_g+\Lambda\right)\ ,
\label{3.32}
\end{equation}
\begin{equation}
\displaystyle
2\dot H+3H^2=\varkappa\left(\Lambda-p_g\right)\ ,
\label{3.33}
\end{equation}
where
\begin{equation}
\begin{array}{c}
\displaystyle \varepsilon_g=\frac{1}{8\varkappa}\sum_{{{\bf
k}\sigma}}\langle\Psi_g|\dot{\hat \psi}_{{\bf k}\sigma}^+\dot{\hat
\psi}_{{\bf k}\sigma}+\frac{k^2}{a^2}\hat \psi_{{\bf
k}\sigma}^+\hat \psi_{{\bf
k}\sigma}|\Psi_g\rangle-
\frac{1}{4\varkappa}\sum_{\bf
k}\langle\Psi_{gh}|\dot{\bar\theta}_{\bf k}\dot\theta_{\bf
k}+\frac{k^2}{a^2}\bar\theta_{\bf k}\theta_{\bf
k}|\Psi_{gh}\rangle\ ,
 \\[5mm]
\displaystyle p_g=\frac{1}{8\varkappa}\sum_{{{\bf
k}\sigma}}\langle\Psi_g|\dot{\hat \psi}_{{\bf k}\sigma}^+\dot{\hat
\psi}_{{\bf k}\sigma}-\frac{k^2}{3a^2}\hat \psi_{{\bf
k}\sigma}^+\hat \psi_{{\bf
k}\sigma}|\Psi_g\rangle-
\frac{1}{4\varkappa}\sum_{\bf
k}\langle\Psi_{gh}|\dot{\bar\theta}_{\bf k}\dot\theta_{\bf
k}-\frac{k^2}{3a^2}\bar\theta_{\bf k}\theta_{\bf
k}|\Psi_{gh}\rangle
\end{array}
\label{3.34}
 \end{equation}
are the energy density and pressure of gravitons that are
renormalized by ghosts. Formulas (\ref{3.34})  were obtained after
elimination of 3--scalar and 3--vector modes from equations
(\ref{2.65}) and (\ref{2.66}). We also took into account the
following definitions
\[
 \displaystyle \langle\Psi|\hat T_0^0|\Psi\rangle= \varepsilon_g\ ,
\qquad \langle\Psi|\hat
T_\alpha^\beta|\Psi\rangle=\frac{\delta_\alpha^\beta}{3}\langle\Psi|\hat
T_\gamma^\gamma|\Psi\rangle=-\delta_\alpha^\beta p_g .
 \]
Also we have the following rules of averaging of bilinear forms
that are the consequence of homogeneity and isotropy of the
background
\[
\begin{array}{c}
\displaystyle \langle\Psi_g|\hat\psi_{{\bf
k}\sigma}^+\hat\psi_{{\bf
k'}\sigma'}|\Psi_g\rangle=\langle\Psi_g|\hat\psi_{{\bf
k}\sigma}^+\hat\psi_{{\bf k}\sigma}|\Psi_g\rangle\delta_{{\bf
kk'}}\delta_{\sigma\sigma'}\ ,
 \qquad
\langle\Psi_{gh}|\bar\theta_{{\bf k}}\theta_{{\bf
k'}}|\Psi_{gh}\rangle=\langle\Psi_{gh}|\bar\theta_{{\bf
k}}\theta_{{\bf k}}|\Psi_{gh}\rangle\delta_{{\bf kk'}}\ .
\end{array}
\]

The self--consistent system of equations (\ref{3.30}) ---
(\ref{3.33}) is a particular case of general equations of
one--loop quantum gravity (\ref{2.59}), (\ref{2.62}), (\ref{2.64})
--- (\ref{2.66}). In turn, these general equations are the result
of the transition to the one--loop approximation from exact
equations (\ref{2.46}) --- (\ref{2.50}) that were obtained by
variation of gauged action over classic and quantum variables. To
canonically quantize 3--tensor gravitons and ghosts, one needs to
make sure that the variational procedure takes place for equations
(\ref{3.30}) --- (\ref{3.33}) directly. To do so, in the action
(\ref{2.42}) we keep only background terms and terms that are
quadratic over 3--tensor fluctuations and ghosts. Then, we exclude
the full derivative from the background sector and make the
transition to Fourier images in the quantum sector. As a result of
these operations, we obtain the following
\begin{equation}
\begin{array}{c}
\displaystyle S=\int dt\left(-\frac{3{\dot a}^2a}{\varkappa
N}-\Lambda a^3N+L_{grav}+L_{ghost}\right)\ ,
 \\[5mm]
 \displaystyle
L_{grav}+L_{ghost}=\frac{1}{8\varkappa}\sum_{{{\bf
k}\sigma}}\left(\frac{a^3}{N}\dot{\hat \psi}_{{\bf
k}\sigma}^+\dot{\hat \psi}_{{\bf k}\sigma}-Nak^2\hat \psi_{{\bf
k}\sigma}^+\hat \psi_{{\bf
k}\sigma}\right)
-\frac{1}{4\varkappa}\sum_{\bf
k}\left(\frac{a^3}{N}\dot{\bar\theta}_{\bf k}\dot\theta_{\bf
k}-Nak^2\bar\theta_{\bf k}\theta_{\bf k}\right)\ .
\end{array}
\label{3.35}
 \end{equation}
In (\ref{3.35}), the background metric is taken to be in the form
of (\ref{3.1}), and the $N$ function is taken to be a variation
variable (the choice of this function, e.g. $N=1$, to be made
after variation of action). Here and further on, the normalized
volume is supposed to be unity, so $V=\int d^3x =1$. The terms
which are linear over the graviton field are eliminated from
(\ref{3.35}) because of zero trace of 3--tensor fluctuations.
Variations of action over $N$ and $a$ are done with the following
averaging. These procedures lead to equations (\ref{3.32}),
(\ref{3.33}) and expressions (\ref{3.34}). Variation of action
over quantum variables leads to the quantum equations of motion
(\ref{3.30}) and (\ref{3.31}).

In accordance with the standard procedure of canonical
quantization of gravitons, one introduces generalized momenta
\begin{equation}
\displaystyle \hat\pi_{{\bf k}\sigma}=\frac{\partial L}{\partial
\dot{\hat\psi}_{{\bf
k}\sigma}}=\frac{a^3}{4\varkappa}\dot{\hat\psi}_{{\bf k}\sigma}^+
\ . \label{3.36}
 \end{equation}
Then, commutation relations between operators that are defined at
the same instant of time read
 \begin{equation}
\begin{array}{c}
\displaystyle \left[\hat\pi_{{\bf k}\sigma}\, ,\ \hat\psi_{{\bf
k'}\sigma'}\right]_{-}\equiv
\frac{a^3}{4\varkappa}\left[\dot{\hat\psi}^+_{{\bf k}\sigma}\ ,\
\hat\psi_{{\bf k'}\sigma'}\right]_{-}=-i\hbar \delta_{{\bf k}{\bf
k'}}\delta_{\sigma\sigma'}\ .
\end{array}
\label{3.37}
 \end{equation}
Formulas (\ref{3.36}) and (\ref{3.37}) are presented for the $N=1$
case. Note also that the derivative in (\ref{3.36}) should be
calculated taking into account the  $\psi_{{\bf
k}\sigma}^+=\psi_{-{\bf k}-\sigma}$ condition.

The ghost quantization contains three specific issues. First,
there is the following technical detail that must be taken into
account for the definition of generalized momenta of ghost fields.
The argument in respect to which the differentiation is conducted
needs to be considered as a left co--multiplier of quadratic form.
Executing the appropriate requirement and taking into account
Grassman's character of ghost fields, we obtain
 \begin{equation}
\begin{array}{c}
 \displaystyle
{\mathcal{P}}_{\bf k}=\frac{\partial L}{\partial \dot\theta_{\bf
k}}=\frac{a^3}{4\varkappa}\dot{\bar\theta}_{\bf k}\ , \qquad
{\bar{\mathcal{P}}}_{\bf k}=\frac{\partial L}{\partial
\dot{\bar\theta}_{\bf k}}=-\frac{a^3}{4\varkappa}\dot{\theta}_{\bf
k}\ .
 \end{array}
\label{3.38}
\end{equation}
Second, the quantization of Grassman's fields is carried out by
setting the following anti--commutation relations
\begin{equation}
\begin{array}{c}
 \displaystyle
\left[{\mathcal{P}}_{\bf k}\ ,\ \theta_{\bf k}\right]_+\equiv
\frac{a^3}{4\varkappa}\left[\dot{\bar\theta}_{{\bf k}}\ ,\
\theta_{{\bf k'}}\right]_+=-i\hbar \delta_{{\bf k}{\bf k'}}\ ,
\qquad 
\left[{\bar{\mathcal{P}}}_{\bf k}\ ,\ \bar\theta_{\bf k}\right]_+\equiv
-\frac{a^3}{4\varkappa}\left[\dot{\theta}_{{\bf k}}\ ,\
\bar\theta_{{\bf k'}}\right]_+=-i\hbar \delta_{{\bf k}{\bf k'}}\ .
\end{array}
 \label{3.39}
 \end{equation}
Third is {\it the bosonization of ghost fields}, which is carried
out {\it after quantization} of (\ref{3.39}). The possibility of
the bosonization procedure is provided by Grassman algebra, which
contains Grassman units defined by relations $\bar uu=-u\bar u=1$.
Therefore, conjunctive Grassman fields can be always presented in
the following form
 \begin{equation}
\displaystyle \theta_{\bf k}=u\vartheta_{\bf k}\ , \qquad
\bar\theta_{\bf k}=\bar u\vartheta_{\bf k}^+\ , \label{3.40}
\end{equation}
where  $\vartheta_{\bf k}$ is Fourier image of complex scalar
field which is described by the usual algebra. The substitution of
(\ref{3.40}) in (\ref{3.39})  leads to the following standard Bose
commutation relations
\begin{equation}
\begin{array}{c}
 \displaystyle
\frac{a^3}{4\varkappa}\left[\dot{\vartheta}^+_{{\bf k}}\ ,\
\vartheta_{{\bf k'}}\right]_{-}=-i\hbar \delta_{{\bf k}{\bf k'}}\ ,
 \qquad
\frac{a^3}{4\varkappa}\left[\dot{\vartheta}_{{\bf k}}\ ,\
\vartheta^+_{{\bf k'}}\right]_{-}=-i\hbar \delta_{{\bf k}{\bf k'}}\ .
\end{array}
 \label{3.41}
 \end{equation}
The Hermit conjugation transforms one of them to the other.

\subsection{State Vector of the General Form}\label{stvec}

To complete the self--consistent theory of gravitons in the
isotropic Universe, one needs to present the algorithm of
introduction of the graviton--ghost ensemble into the theory.
Properties of this ensemble are defined by Heisenberg's state
vector which is expanded over the basis that has a physical
interpretation. Any possible basis is the system of eigenvectors
of an appropriate time independent Hermit operator. The existence
of such operators can be proved in a general form. Let us consider
the following operator equation which is an analog of operator
equations of gravitons and ghosts
\begin{equation}
 \displaystyle \ddot y_k+3H\dot y_k+\frac{k^2}{a^2}y_k=0\ .
  \label{3.42}
 \end{equation}
Coefficients of equation (\ref{3.42}) are continuous and
differentiated functions of time along all cosmological scales
except for the singularity. Thus, with the exception of the
singular point, the general solution of equation (\ref{3.42})
definitely exists. Below we will show that the existence of a
state vector follows only from the existence of general solution
of equation (\ref{3.42}) (see also \cite{11}).

Suppose $g_k$, $h_k$  are linear independent solutions to
(\ref{3.42}), so that their superposition with arbitrary
coefficients gives the general solution to (\ref{3.42}). With no
loss of generality, one can suppose that these solutions are
normalized in some convenient way in each concrete case. From the
theory of ordinary differential equations it is known that $g_k$,
$h_k$  functions are connected to each other by the following
relation
\begin{equation}
 \displaystyle g_k\dot h_k-h_k\dot g_k=\frac{C_k}{a^3}\ ,
 \label{3.43}
 \end{equation}
where $C_k$ is a normalization constant. The comparison of
(\ref{3.42}) with (\ref{3.30}) and (\ref{3.31}) shows that
solutions of operator equations are presented by the same
functions. For operators of graviton field we have
\begin{equation}
\displaystyle \hat\psi_{{\bf k}\sigma}=\hat A_{{\bf
k}\sigma}g_k+\hat B_{{\bf k}\sigma}h_k\ ,
 \label{3.44}
 \end{equation}
where  $\hat A_{{\bf k}\sigma},\, \hat B_{{\bf k}\sigma}$  are
operator constants of integration. Directly from these operator
constants, one needs to build the operator which gives rise to the
full set of basis vectors.

It is important to keep in mind that {\it commutation property of
operator constants $\hat A_{{\bf k}\sigma},\, \hat B_{{\bf
k}\sigma}$ and physical interpretation of basis state vectors are
determined by the choice of linear independent solutions of
equation (\ref{3.42})}. The simplest basis is that of occupation
numbers. The choice of linear independent solutions as
self--conjugated complex functions corresponds to this basis.

In accordance with (\ref{3.43}), if $g_k=f_k,\, h_k=f^*_k$ the
normalization constant is pure imaginary. Let's take $C_k=i$, so
we obtain
\begin{equation}
 \displaystyle f_k\dot f^*_k-f^*_k\dot f_k=\frac{i}{a^3}\ .
 \label{3.45}
\end{equation}
To build the graviton operator over this basis, one need to carry
out the multiplicative renormalization of operator constants
taking into account that field is real. This yield
\[
\displaystyle A_{{\bf k}\sigma}=\sqrt{4\varkappa\hbar}\hat c_{{\bf
k}\sigma}\ , \qquad B_{{\bf k}\sigma}=\sqrt{4\varkappa\hbar}\hat
c_{-{\bf k}-\sigma}^+\ .
\]
As result of these operations, we get the graviton operator and
its derivative that read
\begin{equation}
\begin{array}{c}
\displaystyle \hat\psi_{{\bf
k}\sigma}=\sqrt{4\varkappa\hbar}\left(\hat c_{{\bf
k}\sigma}f_k+\hat c^+_{-{\bf k}-\sigma}f^*_k\right)\ ,
 \qquad 
\dot{\hat\psi}^+_{{\bf
k}\sigma}=\sqrt{4\varkappa\hbar}\left(c^+_{{\bf k}\sigma}\dot
f^*_k+c_{-{\bf k}-\sigma}\dot f_k\right)\ .
\end{array}
\label{3.46}
\end{equation}
Standard commutation relations for operators of graviton creation
and annihilation are obtained by the substitution of (\ref{3.46})
into (\ref{3.37}) and taking into account (\ref{3.45}). They read
\begin{equation}
\begin{array}{c}
\displaystyle \left[\hat c_{{\bf k}\sigma}\ ,\ \hat c^+_{{\bf
k'}\sigma'}\right]_-= \delta_{{\bf k}{\bf
k'}}\delta_{\sigma\sigma'}\ ,
 \qquad
 \left[\hat c_{{\bf k}\sigma}\
,\ \hat c_{{\bf k'}\sigma'}\right]_-=0\ , \qquad \left[\hat
c^+_{{\bf k}\sigma}\ ,\ \hat c^+_{{\bf k'}\sigma'}\right]_-=0 \ .
\end{array}
 \label{3.47}
 \end{equation}
In accordance with (\ref{3.47}), the operator of occupation
numbers $\hat n_{{\bf k}\sigma}=\hat c^+_{{\bf k}\sigma}\hat
c_{{\bf k}\sigma}$ exists that gives rise to basis vectors
$|n_{{\bf k}\sigma}\rangle$ of Fock's space. Non--negative integer
numbers $n_{{\bf k}\sigma}=0,\ 1,\ 2, ...$ are eigenvalues of this
operator.

In accordance with (\ref{3.34}), the observables are additive over
modes with given ${\bf k}\sigma$. Therefore, the state vector is
of multiplicative structure that reads
\[
\displaystyle |\Psi_g\rangle=\prod_{{\bf k}\sigma}|\Psi_{{\bf k}\sigma}\rangle\ ,
\]
where $|\Psi_{{\bf k}\sigma}\rangle$ is state vector of ${\bf
k}\sigma$--subsystem of gravitons of momentum ${\bf p}=\hbar {\bf
k}$ and polarization $\sigma$. In turn, in a general case,
$|\Psi_{{\bf k}\sigma}\rangle$  is an arbitrary superposition of
vectors that corresponds to different occupation numbers but the
same ${\bf k}\sigma$ values. Suppose that $\mathcal{C}_{n_{{\bf
k}\sigma}}$ is the amplitude of probability of finding the  ${\bf
k}\sigma$--subsystem of gravitons in the state with the occupation
number $n_{{\bf  k}\sigma}$. If so, then {\it the state vector of
the general form is the product of normalized superpositions}
\begin{equation}
\displaystyle  |\Psi_g\rangle=\prod_{{\bf  k}\sigma}\sum_{n_{{\bf
k}\sigma}}\mathcal{C}_{n_{{\bf  k}\sigma}}|n_{{\bf
k}\sigma}\rangle\ ,\qquad \sum_{n_{{\bf
k}\sigma}}|\mathcal{C}_{n_{{\bf  k}\sigma}}|^2=1\ .
\label{3.48}
\end{equation}

After the bosonization in the ghost sector is done, one gets
equations of motion and commutation relations that are similar to
those for graviton. The same set of linear independent solutions
$f_k,\ f_k^*$ that was introduced for operators of graviton field
is used for operators of ghost fields. What is necessary to take
into account here is originally complex character of ghost fields,
which leads to $\vartheta_{\bf k}^+\ne \vartheta_{-\bf k}$. As a
result, operators of ghost and anti--ghosts creation and
annihilation appear in the theory. They read
\begin{equation}
\begin{array}{c}
\displaystyle \vartheta_{\bf k}=\sqrt{4\varkappa\hbar}\left(\hat
a_{\bf k}f_k+\hat b^+_{-{\bf k}}f^*_k\right)\ ,
\qquad
\dot\vartheta^+_{\bf k}=\sqrt{4\varkappa\hbar}\left(a^+_{\bf
k}\dot f^*_k+b_{-{\bf k}}\dot f_k\right)\ .
\end{array}
 \label{3.49}
 \end{equation}
The substitution of (\ref{3.49}) into (\ref{3.41}) leads to
standard commutation relations
\begin{equation}
\begin{array}{c}
\displaystyle \left[\hat a_{\bf k}\ ,\ \hat a^+_{\bf k'}\right]_-=
\delta_{{\bf k}{\bf k'}}\ ,\qquad \left[\hat a_{\bf k}\ ,\ \hat
a_{\bf k'}\right]_-= \left[\hat a^+_{\bf k}\ ,\ \hat a^+_{\bf
k'}\right]_-=0 \ ,
 \\[5mm]
\displaystyle \left[\hat b_{\bf k}\ ,\ \hat b^+_{\bf k'}\right]_-=
\delta_{{\bf k}{\bf k'}}\ , \qquad \left[\hat b_{\bf k}\ ,\ \hat
b_{\bf k'}\right]_-= \left[\hat b^+_{\bf k}\ ,\ \hat b^+_{\bf
k'}\right]_-=0 \ ,
\\[5mm]
\displaystyle \left[\hat a_{\bf k}\ ,\ \hat b_{\bf k'}\right]_-=
\left[\hat a_{\bf k}\ ,\ \hat b^+_{\bf k'}\right]_-= 0\ ,
 \qquad
\left[\hat
a^+_{\bf k}\ ,\ \hat b_{\bf k'}\right]_-=\left[\hat a^+_{\bf k}\
,\ \hat b^+_{\bf k'}\right]_-=0
\end{array}
\label{3.50}
\end{equation}
Applying the reasoning which is similar to that described above,
we conclude that in the ghost sector, the state vector of the
general form is also given by product of normalized
superpositions. It reads
\begin{equation}
\begin{array}{c}
\displaystyle  |\Psi_{gh}\rangle=\prod_{\bf  k}\sum_{n_{\bf
k}}\mathcal{A}_{n_{\bf  k}}|n_{\bf k}\rangle\prod_{\bf
k}\sum_{\bar n_{\bf k}}\mathcal{B}_{\bar n_{\bf  k}}|\bar n_{\bf
k}\rangle\ ,
\qquad
 \sum_{n_{{\bf
k}}}|\mathcal{A}_{n_{{\bf  k}}}|^2=\sum_{\bar n_{{\bf
k}}}|\mathcal{B}_{\bar n_{{\bf  k}}}|^2=1\ .
\end{array}
\label{3.51}
\end{equation}
The set of amplitudes $\mathcal{C}_{n_{{\bf  k}\sigma}}$,
$\mathcal{A}_{n_{\bf  k}}$,  $\mathcal{B}_{\bar n_{\bf  k}}$,
which parameterizes Heisenberg's state vector actually determines
the initial condition of quantum system of gravitons and ghosts.

Formulas (\ref{3.48}) and (\ref{3.51}) can be also used in case
when real functions are chosen as linear independent solutions of
equation (\ref{3.42}). The justification for this is due to the
fact that real linear independent solutions can be obtained from
complex self--conjugated ones by the following linear
transformation
\begin{equation}
\displaystyle g_k=\frac{1}{\sqrt{2}}\left(f_k+f^*_k\right)\ ,
\qquad h_k=\frac{i}{\sqrt{2}}\left(f_k-f^*_k\right)\ .
 \label{3.52}
 \end{equation}
After transition to the basis of real functions in (\ref{3.46})
and (\ref{3.49}), we get
\begin{equation}
\begin{array}{c}
\displaystyle \hat\psi_{{\bf
k}\sigma}=\sqrt{4\varkappa\hbar}\left(\hat Q_{{\bf
k}\sigma}g_k+\hat P_{{\bf k}\sigma}h_k\right)\ ,
 \qquad \hat\vartheta_{\bf
k}=\sqrt{4\varkappa\hbar}\left(\hat q_{\bf k}g_k+\hat p_{\bf
k}h_k\right)\ ,
 \end{array}
 \label{3.53}
 \end{equation}
where
\begin{equation}
\begin{array}{c}
\displaystyle \hat Q_{{\bf k}\sigma}=\hat Q^+_{-{\bf
k}-\sigma}=\frac{1}{\sqrt{2}}\left(\hat c_{{\bf
k}\sigma}+c^+_{-{\bf k}-\sigma}\right)\ ,
 \qquad
 \hat P_{{\bf
k}\sigma}=\hat P^+_{-{\bf k}-\sigma}=-\frac{i}{\sqrt{2}}\left(\hat
c_{{\bf k}\sigma}-c^+_{-{\bf k}-\sigma}\right)\ ,
 \\[5mm]
 \displaystyle
 \hat q_{\bf k}=\frac{1}{\sqrt{2}}\left(\hat a_{\bf k}+b^+_{-{\bf k}}\right)\ ,
\qquad \hat p_{\bf k}=-\frac{i}{\sqrt{2}}\left(\hat a_{\bf k}-b^+_{-{\bf k}}\right)\ .
 \end{array}
 \label{3.54}
 \end{equation}
Relations  (\ref{3.54}) allow to work with real linear independent
solutions and to use simultaneously state vectors (\ref{3.48}) and
(\ref{3.51}) for the representation of occupation numbers. Note
that in the framework of the basis of real functions, operator
constants are operators of generalized coordinates and momenta:
\begin{equation}
\displaystyle \left[\hat P^+_{{\bf k}\sigma},\ \hat Q_{{\bf
k'}\sigma'}\right]_-=-i\delta_{{\bf kk'}}\delta_{\sigma\sigma'}\
,\qquad \left[\hat p^+_{{\bf k}},\ \hat q_{{\bf
k'}\sigma'}\right]_-=-i\delta_{{\bf kk'}}\ . \label{3.55}
 \end{equation}

To complete this Section, let us discuss two problems that are
relevant to intrinsic mathematical properties of the theory. First
of all, let us mention that "bosonization"\ of ghost fields is a
necessary element of the theory because only this procedure
provides the existence of state vector in the ghost sector.
Mathematically, it is because the structure of the classic
differential equation (\ref{3.42})  and properties of its solution
(\ref{3.45}) are inconsistent with the Fermi--Dirac quantization.
In terms of original ghost fields we have
\begin{equation}
\begin{array}{c}
\displaystyle \theta_{\bf
k}=\sqrt{4\varkappa\hbar}\left(\alpha_{\bf k}f_k+\bar\beta_{-{\bf
k}}f^*_k\right)\ ,
 \qquad
 \dot{\bar\theta}_{\bf
k}=\sqrt{4\varkappa\hbar}\left(\bar\alpha_{\bf k}\dot
f^*_k+\beta_{-{\bf k}}\dot f_k\right)\ .
\end{array}
 \label{3.56}
 \end{equation}
Substitution (\ref{3.56}) into (\ref{3.39}) and taking into
account (\ref{3.45}) leads to anti--commutation relations for
operator constants that read
\[
\begin{array}{c}
\displaystyle \left[\bar\alpha_{\bf k}\ ,\ \alpha_{\bf
k'}\right]_+= -\delta_{{\bf k}{\bf k'}}\ ,\qquad \left[\beta_{\bf
k}\ ,\ \bar\beta_{\bf k'}\right]_+= \delta_{{\bf k}{\bf k'}}\ .
\end{array}
\]
The $\left[\beta_{\bf k}\ ,\ \bar\beta_{\bf k'}\right]_+=
\delta_{{\bf k}{\bf k'}}$  relation can formally be considered as
anti--commutation relation for operators giving rise the Fermi
space of ghost states. There is no such a possibility for
$\bar\alpha_{\bf k}\ ,\ \alpha_{\bf k}$ operators because their
anti--commutation is negative. If one considers these operators as
complete mathematical objects that are not subject to any
transformations, then it is impossible to build an operator over
them that gives rise to some space of states, and this is because
of non--standard anti--commutation relation. The problem is solved
by the fact of the existence of Grassman units which are necessary
elements of Grassman algebra. At the operator constants level, the
bosonization is reduced to the following transformation
\[
\displaystyle \alpha_{\bf k}=ua_{\bf k}\ , \qquad \bar\alpha_{\bf
k}=\bar ua^+_{\bf k}\ , \qquad \beta_{\bf k}=\bar ub_{\bf k}\ ,
\qquad \bar\beta_{\bf k}=ub^+_{\bf k}\ .
\]
This leads to operators with (\ref{3.50}) commutation properties.

The choice of basis is the most significant problem in the
interpretation of theory. In the theory of quantum fields of
non--stationary Universe, the choice of linear independent basis
$f_k,\ f^*_k$ is ambiguous, in principle. This differentiates it
from the theory of quantum fields in the Minkowski space. In the
latter, the separation of field into negative and positive
frequency components is Lorentz--invariant procedure \cite{15}. A
natural physical postulate in accordance to which the definition
of particle (quantum of field) in the Minkowski space must be
relativistically invariant leads mathematically to
$f_k=(2\omega_k)^{-1/2}e^{-i\omega_kt}$.  In the non--stationary
Universe with the metric (\ref{3.1}), the similar postulate can be
introduced only for conformally  invariant fields and at the level
of auxiliary Minkowski space. At the same time, the graviton field
is conformally non--invariant. This can be seen from the
following. Using the conformal transformation  $y_k=\tilde y_k/a$
and transition to the conformal time $d\eta=dt/a$, one can see
that equation (\ref{3.42}) is transformed to the equation for the
oscillator with variable frequency that reads
\begin{equation}
 \displaystyle \tilde y''_k+\left(k^2-\frac{a''}{a}\right)\tilde y_k=0\ .
  \label{3.57}
 \end{equation}
Effects of vacuum polarization and graviton creation in the
self--consistent classic gravitational field correspond to
parametric excitation of the oscillator (\ref{3.57}).

The approximate separation of field on negative and positive
frequency components is possible only in the short wavelength
limit. Regardless of the background dynamic, linearly independent
solutions of equation (\ref{3.57}) exist, and they have the
following asymptotes
\begin{equation}
 \displaystyle \tilde f_k\to \frac{1}{\sqrt{2k}}e^{-ik\eta}\ , \qquad
\tilde f^*_k\to \frac{1}{\sqrt{2k}}e^{ik\eta}\ ,\qquad k^2\gg \left|\frac{a''}{a}\right|\ .
  \label{3.58}
 \end{equation}
Effects of vacuum polarization and particle creation are
negligible for the subsystem of shortwave gravitons. In this
sector, quanta of gravitational field can be considered, with a
good accuracy, as real gravitons that are situated at their mass
shell. The conservation of the number of such real gravitons takes
also place with a good accuracy. In the shortwave limit, choosing
linear independent solutions of the (\ref{3.58}) form, occupation
numbers $n_{{\bf k}\sigma}$  are interpreted as numbers of real
gravitons with energy $\varepsilon_k=\hbar k/a$, momentum   ${\bf
p}=\hbar{\bf k}/a$ and polarization  $\sigma$. The possibility of
such an interpretation is the principle and the only argument in
favor of choice of this basis. For the subsystem of shortwave
gravitons, initial conditions are permissible not in the form of
products of superpositions but in the form of products of state
vectors with determined occupation numbers. In accordance with the
usual understanding of the status of shortwave ghosts, their state
can be chosen in the vacuum form. The gas of shortwave gravitons
is described in more detail in Section \ref{swg}.

In the $k^2\sim |a''/a|$ vicinity, there is no criterion allowing
a choice of preferable basis. It is impossible to introduce the
definition of real gravitons in this region because there is no
mass shell here. This is the reason why we will use the term
"virtual graviton of determined momentum"\ in discussions of
excitations of long wavelengths. Under the term "virtual
graviton"\ we mean a graviton whose momentum is defined but whose
energy is undefined. Each set of linear independent solutions
corresponds to the distribution of energy for the determined
momentum.  This distribution can be set up, for example by the
expansion of basis function in the Fourier integral. Thus, the
choice of basis is, at the same time, the definition of virtual
graviton. One needs to mention that different sets of probability
amplitudes $\mathcal{C}_{n_{{\bf k}\sigma}}$ correspond to
different definitions of the virtual graviton for the same initial
physical state. Note also that limitations that are defined by
asymptotes (\ref{3.58}) do not fix basis functions completely.

\subsection{One--Loop Finiteness}\label{fin}

The full system of equations of the theory consists of operator
equations for gravitons and ghosts (\ref{3.30}), (\ref{3.31}),
macroscopic Einstein equations (\ref{3.32}), (\ref{3.33}) and
formulas  (\ref{3.34}) for the energy density and pressure of
gravitons. The averaging of (\ref{3.34}) is carried out over state
vectors of general form (\ref{3.48}) and (\ref{3.51}). The
one--loop finiteness is satisfied automatically in this theory.
The finiteness is provided by the structure of ghost sector, and
it is a result of the following two facts. First, in the space
with metric (\ref{3.1}) the ghost equation (\ref{3.31}) coincides
with graviton equation (\ref{3.30}). Second, the number of
internal degrees of freedom of the complex ghost field coincides
with that of 3--tensor gravitons. We will show this by direct
calculations.

Let us introduce the graviton spectral function which is
renormalized by ghosts. It reads
\begin{equation}
\displaystyle W_{{\bf k}}=\sum_\sigma\langle \Psi_g|\hat\psi^+_{{\bf
k}\sigma}\hat\psi_{{\bf
 k}\sigma}|\Psi_g\rangle-2\langle\Psi_{gh}|\bar \theta_{{\bf k}}\theta_{{\bf
 k}}|\Psi_{gh}\rangle\ .
 \label{3.59}
 \end{equation}
Zero and first moments of this function are the most important
objects of the theory. They are
\begin{equation}
\begin{array}{c}
\displaystyle W_0=\sum_{{\bf k}}\left(\sum_\sigma\langle
\Psi_g|\hat\psi^+_{{\bf k}\sigma}\hat\psi_{{\bf
 k}\sigma}|\Psi_g\rangle-2\langle\Psi_{gh}|\bar \theta_{{\bf k}}\theta_{{\bf
 k}}|\Psi_{gh}\rangle\right)\ ,
 \\[3mm]\displaystyle
 W_1=\sum_{{\bf
k}}\frac{k^{2}}{a^{2}}\left(\sum_\sigma\langle \Psi_g|\hat\psi^+_{{\bf
k}\sigma}\hat\psi_{{\bf
 k}\sigma}|\Psi_g\rangle-2\langle\Psi_{gh}|\bar \theta_{{\bf k}}\theta_{{\bf
 k}}|\Psi_{gh}\rangle\right)\ .
\end{array}
 \label{3.60}
 \end{equation}
The energy density and pressure of gravitons that are expressed
via moments (\ref{3.60}) can be obtained by transformations
identical to (\ref{3.34}) with use of equations of motion
(\ref{3.30}) and (\ref{3.31}). They read
\begin{equation}
 \begin{array}{c}
\displaystyle \varkappa\varepsilon_g=\frac{1}{16}D+\frac14W_1\ ,
\qquad \varkappa p_g=\frac{1}{16}D+\frac{1}{12}W_1\ ,
\\[3mm]
\displaystyle D=\ddot W_0+3H\dot W_0\ .
\end{array}
 \label{3.61}
\end{equation}
In addition, the following relation between moments is derived
from equations of motion
\begin{equation}
\displaystyle  \dot D+6HD+4\dot W_1+16HW_1=0\ .
\label{3.62}
\end{equation}
This relation ensures that the graviton energy--momentum tensor is
conservative:
\[
\displaystyle \dot\varepsilon_g+3H(\varepsilon_g+p_g)=0
\]

As it was shown above, field operators can always be chosen from
the basis of complex self--conjugated functions that are the same
both for gravitons and ghosts. One needs to also mention that the
interpretation of short wave gravitons as real gravitons
determines the asymptotic of basis functions (see (\ref{3.58})).
After the commutation of operators of creation and annihilation
are done, graviton contributions to the moments of the spectral
function $W_n,\, n=0,1$ can be presented in the following form
\begin{equation}
\begin{array}{c}
\displaystyle W_{n(grav)}=
\sum_{{\bf k}}\frac{k^{2n}}{a^{2n}}
\sum_\sigma\langle \Psi_g|\hat\psi^+_{{\bf k}\sigma}\hat\psi_{{\bf k}\sigma}|\Psi_g\rangle=
 \\[5mm]
 \displaystyle
=8\varkappa\hbar\sum_{{\bf k}}\frac{k^{2n}}{a^{2n}}f_k^*f_k+
 4\varkappa\hbar\sum_{{\bf k}}\frac{k^{2n}}{a^{2n}}
\sum_\sigma \left(2\langle \Psi_g|\hat c^+_{{\bf k}\sigma}\hat c_{{\bf k}\sigma}
|\Psi_g\rangle f_k^*f_k+\langle \Psi_g|\hat c^+_{{\bf k}\sigma}\hat c^+_{-{\bf k}-
\sigma}|\Psi_g\rangle f_k^{*2}+\langle \Psi_g|\hat c_{-{\bf k}-
\sigma}\hat c_{{\bf k}\sigma}|\Psi_g\rangle f_k^{2}\right)\ .
\end{array}
\label{3.63}
\end{equation}
In the right--hand--side of (\ref{3.63}), the first term is the
functional which is independent of the structure of Heisenberg
state vector. It reads
\begin{equation}
\displaystyle W^{(0)}_{n(grav)}=8\varkappa\hbar\sum_{{\bf
k}}\frac{k^{2n}}{a^{2n}}f_k^*f_k=\frac{4\varkappa\hbar}{\pi^2a^{2n}}\int\limits_0^\infty
k^{2n+2}f_k^*f_kdk\ .
 \label{3.64}
\end{equation}

The integral (\ref{3.64}) describes the contribution of zero
oscillations whose spectrum is deformed by macroscopic
gravitational field. Asymptotic (\ref{3.58}) shows that this
integral is diverges. In such a situation, the usual way is to use
regularization and renormalization procedures. As a result of
these operations, quantum corrections to Einstein equations
appear. These corrections are the conformal anomalies and terms
that came from Lagrangian $\sim R^2\ln (R/\lambda_g^2)$ where
$\lambda_g$ is a scale parameter that comes from renormalization
(see Section \ref{ren}). The theory that we present here does not
use such operations. There is a contribution of ghost zero
oscillations in the moments of spectral function. Its sign is
opposite to (\ref{3.64}). It reads
\begin{equation}
\begin{array}{c}
\displaystyle W_{n(ghost)}=-2\sum_{{\bf
k}}\frac{k^{2n}}{a^{2n}}\langle \Psi_{gh}|\bar\theta_{\bf
k}\theta_{\bf k}|\Psi_{gh}\rangle=
 \\[5mm]
 \displaystyle
= -8\varkappa\hbar\sum_{{\bf
k}}\frac{k^{2n}}{a^{2n}}f_k^*f_k
-8\varkappa\hbar\sum_{{\bf
k}}\frac{k^{2n}}{a^{2n}}\left(\langle \Psi_{gh}|\hat a^+_{\bf
k}\hat a_{\bf k}+\hat b^+_{\bf k}\hat b_{\bf k}|\Psi_{gh}\rangle
f_k^*f_k+\langle \Psi_{gh}|\hat a^+_{\bf k}\hat b^+_{-{\bf
k}}|\Psi_{gh}\rangle f_k^{*2}+\langle \Psi_{gh}|\hat b_{-{\bf
k}}\hat a_{\bf k}|\Psi_{gh}\rangle f_k^{2}\right)\ .
\end{array}
\label{3.65}
\end{equation}
The observables (\ref{3.61}) are expressed via sums
$W_{n(grav)}+W_{n(ghost)}$. In those sums, the exact
graviton--ghost compensation takes place in the contribution from
zero oscillations.

The final expressions for the moments of spectral function are
obtained by using the explicit form of state vectors \ref{3.48})
and (\ref{3.51}). They read
\begin{equation}
\displaystyle W_n=8\varkappa\hbar\sum_{{\bf
k}}\frac{k^{2n}}{a^{2n}}\left(N_{\bf k}|f_k|^2+U^*_{\bf
k}f_k^{*2}+U_{\bf k}f_k^2\right)\ , \label{3.66}
\end{equation}
where
\begin{equation}
\displaystyle N_{\bf k}=\sum_\sigma\sum_{n_{{\bf
k}\sigma}=1}^\infty |\mathcal{C}_{n_{{\bf k}\sigma}}|^2n_{{\bf
k}\sigma}-\sum_{n_{{\bf k}}=1}^\infty |\mathcal{A}_{n_{{\bf
k}}}|^2n_{{\bf k}}-\sum_{\bar n_{{\bf k}}=1}^\infty
|\mathcal{B}_{\bar n_{{\bf k}}}|^2\bar n_{{\bf k}} \label{3.67}
 \end{equation}
and
\begin{equation}
 \begin{array}{c}
 \displaystyle  U^*_{\bf k}=
\frac12\sum_\sigma\left(\sum_{n_{{\bf
k}\sigma}=0}^\infty\mathcal{C}^*_{n_{{\bf
k}\sigma}+1}\mathcal{C}_{n_{{\bf k}\sigma}}\sqrt{n_{{\bf
k}\sigma}+1}\right)\times
 \left(\sum_{n_{{\bf
k'}\sigma'}=0}^\infty\mathcal{C}^*_{n_{{\bf
k'}\sigma'}+1}\mathcal{C}_{n_{{\bf k'}\sigma'}}\sqrt{n_{{\bf
k'}\sigma'}+1}\right)-
 \\[5mm]
 \displaystyle
-\left(\sum_{n_{\bf k}=0}^\infty\mathcal{A}^*_{n_{\bf
k}+1}\mathcal{A}_{n_{\bf k}}\sqrt{n_{\bf k}+1}\right)
\times
\left(\sum_{\bar n_{{\bf k'}}=0}^\infty\mathcal{B}^*_{\bar n_{{\bf
k'}}+1}\mathcal{B}_{\bar n_{{\bf k'}}}\sqrt{\bar n_{{\bf
k'}}+1}\right)
  \end{array}
 \label{3.68}
 \end{equation}
are spectral parameters. They are defined by initial conditions
for the chosen normalized basis of linear independent solutions of
equations (\ref{3.42}). (For sake of brevity, in (\ref{3.68}) and
below we use the following notation ${\bf k'}=-{\bf k},\
\sigma'=-\sigma$.) Note that the relation (\ref{3.66}) does not
contain divergences. Divergences in the relation (\ref{3.66}) may
appear only because of non--physical initial conditions. The
spectrum of real gravitons that slowly decreased for $k\to \infty$
is an example of such a non--physical initial conditions.

The spectral function (\ref{3.59}) depends of three arbitrary
constants as it is averaged over the state vector of general form.
It reads
\begin{equation}
\displaystyle W_{{\bf k}}=8\varkappa\hbar\left(N_{\bf
k}|f_k|^2+U^*_{\bf k}f_k^{*2}+U_{\bf k}f_k^2\right)\ .
\label{3.69}
\end{equation}
In (\ref{3.69}), the basis of normalized linear independent
solutions contains information on the dynamics of operators of
graviton-ghost field; integration constants  $N_{\bf k},\ U^*_{\bf
k}, \ U_{\bf k}$ contain information on the initial ensemble of
this field. Due to the background's homogeneity and isotropy the
moduli of the amplitudes and average occupation numbers do not
depend on the directions of wave vectors and polarizations:
\begin{equation}
\begin{array}{c}
\displaystyle \langle n_{k(g)}\rangle=\sum_{n_{{\bf
k}\sigma}=1}^\infty |\mathcal{C}_{n_{{\bf k}\sigma}}|^2n_{{\bf
k}\sigma}\ ,
\qquad
\langle n_{k(gh)}\rangle = \sum_{n_{{\bf
k}}=1}^\infty |\mathcal{A}_{n_{{\bf k}}}|^2n_{{\bf k}}\ ,\qquad
\langle \bar n_{k(gh)}\rangle= \sum_{\bar n_{{\bf k}}=1}^\infty
|\mathcal{B}_{\bar n_{{\bf k}}}|^2\bar n_{{\bf k}}\ .
\end{array}
 \label{3.70}
\end{equation}

Phase of amplitudes, in principle, may depend on the directions
and polarizations. One must bear in mind that in the pure quantum
ensembles, for which the averaging over the state vector is
defined, phases of amplitudes are determined. If the phases are
random, then the additional averaging should be conducted over
them, which corresponds to the density matrix formalism for mixed
ensembles. The question of phases of amplitudes is clearly linked
to the question of the origin of quantum ensembles. In particular,
it is natural to assume that the ensemble of long--wavelength
gravitons arises in the process of restructuring graviton vacuum.
This process is due to conformal non--invariance of the graviton
field and can be described as particle creation. In this case,
there is a correlation between the phases of states with the same
occupation numbers, but mutually opposite momenta and
polarizations: the sum of these phases is zero.

If the typical occupation numbers in the ensemble are large, then
squares of moduli of probability amplitudes are likely to be
described by Poisson distributions. For this ensemble we get
\begin{equation}
\begin{array}{c}
\displaystyle \mathcal{C}_{n_{{\bf
k}\sigma}}=\sqrt{P[n_{k(g)}]}\exp(i\varphi_{n_{{\bf k}\sigma}})\ ,
\qquad
 \mathcal{A}_{n_{{\bf
k}}}=\sqrt{P[n_{k(gh)}]}\exp(i\chi_{n_{{\bf k}}})\ ,
\qquad
\mathcal{B}_{n_{{\bf k'}}}=\sqrt{P[\bar
n_{k(gh)}]}\exp(i\chi_{\bar n_{{\bf k'}}})\ ,
\\[5mm]
\displaystyle P[n_{k}]=\frac{\langle n_k\rangle^{n_k}}{n_k!}\exp(-\langle n_k\rangle)
\ .
\end{array}
 \label{3.71}
\end{equation}
The substitution of  (\ref{3.70}), (\ref{3.71}) to (\ref{3.67}), (\ref{3.68}) leads to
\begin{equation}
\begin{array}{c}
\displaystyle
N_{\bf k}\equiv N_k=2\langle n_{k(g)}\rangle- \langle n_{k(gh)}\rangle -
\langle \bar n_{k(gh)}\rangle\ .
\end{array}
 \label{3.72}
\end{equation}
\begin{equation}
\begin{array}{c}
\displaystyle U^*_{\bf k}\equiv U^*_k=\langle
n_{k(g)}\rangle\zeta^{(g)}_ke^{i\varphi_k}-\sqrt{\langle
n_{k(gh)}\rangle\langle \bar
n_{k(gh)}\rangle}\zeta_k^{(gh)}e^{i\chi_k}\ ,
\\[5mm]
\displaystyle \zeta_k^{(g)}\leqslant 1,\qquad \zeta_k^{(gh)}\leqslant 1\ ,
\end{array}
 \label{3.73}
\end{equation}
where
\begin{equation}
\begin{array}{c}
\displaystyle
\zeta^{(g)}_ke^{i\varphi_k}=\frac12\sum_\sigma\left(\sum_{n_{{\bf
k}\sigma}}P[n_{k(g)}]\exp(i\varphi_{n_{{\bf
k}\sigma}}-i\varphi_{n_{{\bf
k}+1,\sigma}})\right)
\times\left(\sum_{n_{{\bf
k'}\sigma'}}P[n_{k(g)}]\exp(i\varphi_{n_{{\bf
k'}\sigma'}}-i\varphi_{n_{{\bf k'}+1,\sigma'}})\right)\ ,
\\[5mm]
\displaystyle \zeta^{(gh)}_ke^{i\chi_k}=\left(\sum_{n_{{\bf
k}}}P[n_{k(gh)}]\exp\left(i\chi_{n_{{\bf k}}}-i\chi_{n_{{\bf
k}+1}}\right)\right)
\times\left(\sum_{\bar n_{{\bf k'}}}P[\bar
n_{k(gh)}]\exp\left(i\chi_{\bar n_{{\bf k'}}}-i\chi_{n_{{\bf
k'}+1}}\right)\right)\
\end{array}
 \label{3.74}
\end{equation}
Limit equalities $\zeta_k^{(g)}=\zeta_k^{(gh)}=1$ are satisfied if
the phase difference between states of neighboring occupation
numbers does not depend on values of occupation numbers. It is
also easy to see that (\ref{3.72}) and (\ref{3.73}) apply, with
somewhat different definitions, to any ensemble with
$\zeta^{(g)}_ke^{i\varphi_k}$ and $\zeta^{(gh)}_ke^{i\chi_k}$
parameters.

We already mentioned above that different basis functions that
correspond to different definitions of the virtual graviton can be
used for the same initial physical state. Limitations due to the
prescriptions on the asymptotic expression (\ref{3.58})  allow to
fix only asymptotic expansions of basis functions for
$k\to\infty$. These expansions can be used, however, only for
description of shortwave modes (Section \ref{swg}). Meanwhile, all
non--trivial quantum gravity phenomena take place in spectral
region where characteristic wavelengths are of the order of the
horizon scale. The choice of basis functions to describe these
waves is not unique, and the set of amplitudes of probability
$\mathcal{C}_{n_{{\bf k}\sigma}}$ depends significantly on this
set. At the level of equations (\ref{3.72}), (\ref{3.73}), the
ambiguity in the definition of the virtual graviton reveals itself
in the ambiguity of values of parameters  $\langle
n_{k(g)}\rangle$ and $\zeta^{(g)}_ke^{i\varphi_k}$ . Similar
ambiguity exists in the ghost sector. Two conclusions follow from
that. First, it is necessary to work with the state vector of
general form, at least during the first stage of the study of the
system that contains excitations of long wavelengths.
Concretization of the amplitudes $\mathcal{C}_{n_{{\bf k}\sigma}}$
is possible only after using of additional physical considerations
that are different for each concrete case.  Second, a theory would
be extremely desirable which is invariant with respect to the
choice of linear independent solutions of equation (\ref{3.42}),
and, correspondingly, is invariant with respect to the choice of
amplitudes of probability $\mathcal{C}_{n_{{\bf k}\sigma}}$
defining the structure of Heisenberg's state vector, respectively.
In Section \ref{Bog} we will show that such a formulation of the
theory exists in the form of equation for the spectral function of
gravitons renormalized by ghosts. The mathematically equivalent
formulation of theory exists in the form of infinite BBGKY chain
or hierarchy where joint description of gravitons and ghost is
carried out in terms of moments of spectral function $W_n,\, n=0,\
1,\ 2,\ ...,\ N\to\infty$.

\section{Approximate Solutions}\label{approx}

\subsection{Gas of Short Wave Gravitons}\label{swg}

Let us consider the gas of gravitons of wavelength that is much
shorter than the distance to the cosmological horizon. We exclude
the long waves from the model. Also, the calculation of
observables is done approximately, so that non--adiabatic
evolution of quantum ensemble is not taken into account. In the
framework of these approximations, it is possible to save the pure
vacuum status of ghosts because their role is just  to provide the
one--loop finiteness of macroscopic quantities. Long wave
excitations we will consider in Section \ref{lgw}.

The calculation of observables for the gas of short wave gravitons
can be done by general formulas (\ref{3.61}), (\ref{3.66}) ---
(\ref{3.68}) after the definition of basis functions and the state
vector. For the short wave approximation, the full asymptotic
expansion of basis functions exists that satisfies the
normalization condition (\ref{3.45}) and asymptotes (\ref{3.58}).
Of course, to use the method of asymptotic expansions, basis
functions must be taken in the following form
\begin{equation}
\begin{array}{c}
\displaystyle f_k=\frac{1}{a\sqrt{2\epsilon_k}}e^{-i\phi_k}\ ,
\qquad f^*_k=\frac{1}{a\sqrt{2\epsilon_k}}e^{i\phi_k}\ , \qquad
\phi_k=\int\limits_{\eta_0}^\eta\epsilon_kd\eta\ ,
\end{array}
\label{4.1}
\end{equation}
where
\[
\displaystyle \epsilon_k=\epsilon_k(\rho,\ \rho',\ \rho'',\ ...)\ ,\qquad \rho=-\frac{a''}{a}
\]
is a real functional of scale factor and its derivatives. In the
short wave approximation, this functional is expanded into the
local asymptotic series, which satisfies to the following boundary
condition\footnote{ Note that the $\rho^{(n)}(\eta)\to 0$
asymptotic exists for cosmological solutions of usual interest.
For instance, $\rho^{(n)}(\eta_0)=0$ as $\eta_0=-\infty$ for the
inflation solution. For $\eta_0=+\infty$ it takes place for the
FRW solution for the Universe filled with ordinary matter.}
\begin{equation}
\displaystyle \epsilon_k=\epsilon_k(\rho,\ \rho',\ \rho'',\ ...)\to k\ ,\qquad
\rho,\ \rho',\ \rho'',\ ... \ \to 0\ .
\label{4.2}
\end{equation}
There are no arbitrary constants in this expansion if the (\ref{4.2}) condition is satisfied.

The following linear ordinary differential equation of the third
order with respect to $1/\epsilon_k$ functional follows from the
equation (\ref{3.57}) for $y_k=af_k,\ af^*_k$ functions
\begin{equation}
 \begin{array}{c}
  \displaystyle
  \frac12\left(\frac{1}{\epsilon_k}\right)'''
  +2\omega_k^2\left(\frac{1}{\epsilon_k}\right)'+
  \left(\omega_k^2\right)'\frac{1}{\epsilon_k}=0\ ,
  \\[5mm]
 \displaystyle \omega_k^2=k^2+\rho\ .
 \end{array}
   \label{4.3}
\end{equation}
The solution of equation (\ref{4.3}) satisfying to the asymptotic condition (\ref{4.2}) reads
 \begin{equation}
\displaystyle
  \frac{1}{\epsilon_k}=\frac{1}{\omega_k}\sum_{s=0}^{\infty}(-1)^s\hat
 J_k^s\cdot 1\ .
 \label{4.4}
\end{equation}
Powers of $\hat J_k$ operator from (\ref{4.4}) are defined as follows
 \begin{equation}
 \begin{array}{c}
  \displaystyle \hat J_k\cdot \varphi=
\frac14\int\limits_{\eta_0}^\eta\frac{d\eta}{\omega_k}\left(\frac{\varphi}{\omega_k^3}\right)'''\
  ,
    \\[5mm]
    \displaystyle
\hat J_k^0\cdot 1\equiv 1\ ,\qquad J_k\cdot
1=\frac18\left(-\frac{\rho''}{\omega_k^4}+\frac54\frac{\rho^{'2}}{\omega_k^6}\right)\
,
 \qquad
 J_k^2\cdot 1=J_k\cdot (J_k\cdot 1)\ ,
\qquad  J_k^s\cdot
1=J_k^{s-1}\cdot (J_k\cdot 1)\ .
  \end{array}
   \label{4.5}
\end{equation}
The integral is calculated explicitly for arbitrary $s$, so that
$J_k^s\cdot 1$ is a local functional of $\rho$  and its
derivatives. It follows from (\ref{4.5}) that a small parameter of
asymptotic expansion is of the order of $\sim 1/k^2$. The
(\ref{4.4})  solution is approximate because non--local effects
are not included to the local asymptotical series. Calculation of
these effects is beyond of limits of this method.

The asymptotic expansion (\ref{4.4}), (\ref{4.5}) defines the
$1/\epsilon_k$  functional, and hence, it defines basis functions
(\ref{4.1}). The substitution of (\ref{4.1}) to (\ref{3.66})
produces asymptotic expansions of moments of spectral function
that read
\begin{equation}
 \begin{array}{c}
  \displaystyle
  W_n=
\frac{4\varkappa\hbar}{a^{2+2n}}\sum_{{\bf
k}}\frac{k^{2n}}{\epsilon_k} \left\{\sum_\sigma\langle
\Psi_g|c^+_{{\bf k}\sigma}c_{{\bf k}\sigma}|\Psi_g\rangle -\langle
\Psi_{gh}|a^+_{{\bf k}}a_{{\bf k}}+b^+_{{\bf k}}b_{{\bf
k}}|\Psi_{gh}\rangle +\right.
  \\[5mm]
   \displaystyle  \left.
+\left[\frac12\sum_\sigma\langle\Psi_g|c^+_{{\bf
k}\sigma}c^+_{-{\bf k}-\sigma}|\Psi_g\rangle -
\langle\Psi_{gh}|a^+_{{\bf k}}b^+_{-{\bf
k}}|\Psi_{gh}\rangle\right] e^{2i\phi_k}+
\right.
  \\[5mm]
   \displaystyle  \left.
+\left[\frac12\sum_\sigma\langle\Psi_g|c_{-{\bf
k}-\sigma}c_{{\bf k}\sigma}|\Psi_g\rangle -
  \langle\Psi_{gh}|b_{-{\bf k}}a_{{\bf k}}|\Psi_{gh}\rangle\right] e^{-2i\phi_k}\right\}\ .
  \end{array}
\label{4.6}
\end{equation}

State vectors from (\ref{4.6}) can be concretized from the general
considerations. It was mentioned in Section \ref{stvec} that such
terms as vacuum, zero oscillations and quantum wave excitations
are well defined for the $\rho^{(n)}(\eta)\to 0$ condition. Under
the same condition, state vectors that are built on basis vectors
of the Fock space are easily interpreted. First of all, this
statement is relevant to gravitons. Eigenvalues $n_{{\bf
k}\sigma}$ and eigenvectors $|n_{{\bf k}\sigma}\rangle$ of $\hat
n_{{\bf k}\sigma}=\hat c^+_{{\bf k}\sigma}\hat c_{{\bf k}\sigma}$
operator describe real gravitons in asymptotic states. In the
short wave approximation, the concept of real gravitons is valid
for all other stages of the Universe evolution. Thus, in this
particular case, the state vector of the general form can be
reduced to the product of vectors corresponding to states with
definite graviton numbers $n_{{\bf k}\sigma}=0,\ 1,\ 2, ...$
possessing definite momentum and polarization. It reads
\begin{equation}
\displaystyle |\Psi_g\rangle=\prod\limits_{{\bf k}\sigma}|n_{{\bf
k}\sigma}\rangle\ .
\label{4.7}
\end{equation}
In asymptotic states, short wave ghosts are only used to
compensate non--physical vacuum divergences. In accordance with
such an interpretation of the ghost status, we suppose that ghosts
and anti--ghosts sit in vacuum states that read
 \begin{equation}
\displaystyle |\Psi_{gh}\rangle=\prod\limits_{{\bf k}}\prod\limits_{{\bf k'}}|0_{{\bf
k}}\rangle|\bar 0_{{\bf k'}}\rangle\ .
 \label{4.8}
 \end{equation}
Averaging over the quantum state that is defined by (\ref{4.7})
and (\ref{4.8}) vectors, we get
\[
 \begin{array}{c}
  \displaystyle
\langle \Psi_g|c^+_{{\bf k}\sigma}c_{{\bf
k}\sigma}|\Psi_g\rangle=n_{{\bf k}\sigma},
 \\[5mm]
 \displaystyle
 \langle
\Psi_{gh}|a^+_{{\bf k}}a_{{\bf k}}|\Psi_{gh}\rangle=\langle
\Psi_{gh}|b^+_{{\bf k}}b_{{\bf k}}|\Psi_{gh}\rangle=0,
 \\[5mm] \displaystyle
 \Psi_g|c^+_{{\bf k}\sigma}c^+_{-{\bf k}-\sigma}|\Psi_g\rangle=
 \Psi_g|c_{-{\bf k}-\sigma}c_{{\bf k}\sigma}|\Psi_g\rangle=
\langle\Psi_{gh}|a^+_{{\bf k}}b^+_{-{\bf k}}|\Psi_{gh}\rangle=
 \Psi_{gh}|b_{-{\bf k}}a_{{\bf k}}|\Psi_{gh}\rangle=0,
 \\[5mm]\displaystyle
  W_n=
 \frac{4\varkappa\hbar}{a^{2+2n}}\sum_{{\bf k}\sigma}\frac{k^{2n}}{\epsilon_k}
n_{{\bf k}\sigma}.
  \end{array}
\]

To calculate macroscopic observables in this approximation, it is
sufficient to keep only the first terms of expansion of moments of
spectral function that contain no higher than second derivative of
scale factor.  In this approximation, moments of spectral function
read
\begin{equation}
 \begin{array}{c}
 \displaystyle W_0=\frac{4\varkappa\hbar}{a^2}\sum_{{\bf
k}\sigma}\frac{n_{{\bf k}\sigma}}{k}\ ,\qquad D=-
\frac{8\varkappa\hbar}{a^2}\left(\dot H+H^2\right)\sum_{{\bf
k}\sigma}\frac{n_{{\bf k}\sigma}}{k}\ ,
\\[5mm]\displaystyle W_1=
\frac{4\varkappa\hbar}{a^4}\sum_{{\bf k}\sigma}kn_{{\bf
k}\sigma}+\frac{2\varkappa\hbar}{a^2}\left(\dot
H+2H^2\right)\sum_{{\bf k}\sigma}\frac{n_{{\bf k}\sigma}}{k}\ ,
 \end{array}
 \label{4.9}
\end{equation}
Taking into account (\ref{4.9}), we get energy density and
pressure of high--frequency graviton gas from (\ref{3.61}) that
read
 \begin{equation}
 \begin{array}{c}
\displaystyle
\varkappa\varepsilon_g=\frac{\varkappa\hbar}{a^4}\sum_{{\bf
k}\sigma}kn_{{\bf
k}\sigma}+\frac{\varkappa\hbar}{2a^2}H^2\sum_{{\bf
k}\sigma}\frac{n_{{\bf k}\sigma}}{k}\ ,
\\[5mm] \displaystyle
\varkappa p_g=\frac{\varkappa\hbar}{3a^4}\sum_{{\bf
k}\sigma}kn_{{\bf k}\sigma}-\frac{\varkappa\hbar}{6a^2}\left(2\dot
H+H^2\right)\sum_{{\bf k}\sigma}\frac{n_{{\bf k}\sigma}}{k}\ .
\end{array}
 \label{4.10}
\end{equation}
Relations (\ref{4.9}) and (\ref{4.10}) are valid if $a^2/\bar
k^2\sim \bar \lambda^2\ll H^{-2},\, |\dot H|^{-1}$, i.e. the
square of ratio of graviton wavelength to horizon distance is much
less than unity. In case of large occupation numbers, these
results are of the quasi--classical character and can be obtained
by the classical theory of gravitational waves \cite{20}.

As can be seen from (\ref{4.10}), the high--frequency graviton gas
differs from the ideal gas with the equation of state
$p=\varepsilon/3$ by only so--called post--hydrodynamic
corrections. In accordance with the approximation used, these
corrections are of the order of $\bar \lambda^2H^2\ll 1$ in
comparison with main terms. Thus, the following simple formula can
be used
\begin{equation}
\displaystyle \varkappa\varepsilon_g\simeq 3\varkappa p_g\simeq
\frac{C_{g1}}{a^4}\ , \qquad C_{g1}=\varkappa\hbar\sum_{{\bf
k}\sigma}kn_{{\bf k}\sigma}\ .
 \label{4.11}
\end{equation}

\subsection{Quantized Gravitons and Ghosts of Super-Long Wavelengths}\label{lgw}

In the framework of this theory, it is possible to describe the
ensemble of super--long gravitational waves ($k^2\ll |a''/a|$) by
an approximate analytical method. Such an ensemble corresponds to
the Universe whose observable part is in the chaotic bunch of
gravitational waves of wavelengths greater than the horizon
distance. The chaotic nature of the bunch is provided by non--zero
wave vectors of these waves, so that observable properties of the
Universe are formed by superposition of waves of different
polarizations and orientations in the space. Such a wave system
can produce an the isotropic spectrum and isotropic polarization
ensemble consistent with the homogeneity and isotropy of the
macroscopic space.

Such an ensemble of super--long waves can be formed only if the
size of causally--bounded region is much greater than the horizon
distance, which is possible in the framework of the hypothesis of
early inflation (or other scenarios (see, e.g. \cite{21}).
However, the problem of kinematical stability of an ensemble
exists even in the framework of the hypothesis of early inflation.
The case is due to the fact that the ensemble of long waves is
destroyed during the post--inflation epoch if the Universe is
expanded with a deceleration. When long waves come out of horizon,
they are transformed to the short waves. Below we show that the
kinematical self--stabilization of an ensemble is possible in the
framework of self--consistent theory of long waves.

Long waves under discussion correspond to virtual gravitons. To
describe them approximately, one needs to use asymptotic
expansions of basis functions over the small parameter $\sim k^2$.
As well as in the case of short waves, the basis can be chosen in
the representation of self--conjugated functions that are
parameterized by the universal real functional $\epsilon_k(a)$.
This preserves the definition (\ref{4.1}) and equation
(\ref{4.3}). However, due to of our interest in the asymptotical
expansion $1/\epsilon_k(a)$ over $\sim k^2$ parameter, it is
necessary to rewrite equation (\ref{4.3}) in the following form
 \begin{equation}
 \begin{array}{c}
 \displaystyle
\left\{a^2\left[a^2\left(\frac{1}{a^2\epsilon_k}\right)'\right]'\right\}'=
-4k^2a^2\left(\frac{1}{\epsilon_k}\right)'\
.
\end{array}
 \label{4.12}
\end{equation}
Let us introduce the geometric--dynamic time $d\tau=d\eta/a^2$ and
the following functional
\[
\displaystyle \frac{1}{2\epsilon_ka^2}\equiv\xi_k=\sum_{n=0}^\infty\xi_k^{(n)}
\]
Note that the $\tau$ time coordinate corresponds to the original
gauge $\sqrt{-g}g^{00}=1$. Equation (\ref{4.12}) and the spectral
function (\ref{3.69}) now read
\begin{equation}
 \begin{array}{c}
  \displaystyle
\frac{d^3\xi_k}{d\tau^3}=-4k^2a^2\frac{d}{d\tau}(a^2\xi_k)\ ,
\end{array}
 \label{4.13}
\end{equation}
\begin{equation}
\begin{array}{c}
\displaystyle W_{{\bf k}}=8\varkappa\hbar\xi_k\left(N_{\bf
k}+U^*_{\bf k}e^{i\Phi_k}+U_{\bf k}e^{-i\Phi_k}\right)\ ,
 \\[5mm]
 \displaystyle
\Phi_k=\int\limits_{\tau_k}^\tau\frac{d\tau}{\xi_k} \ ,
\end{array}
\label{4.14}
\end{equation}
where $\tau_k$ is a numerical parameter. Its value is unimportant
because the constant's contribution to phases of basis functions
is absorbed by phases of contributors that form vectors of the
general form (\ref{3.48}) and (\ref{3.51}). Observables
(\ref{3.61}) are expressed via moments of spectral function. The
latter read
\begin{equation}
\begin{array}{c}
\displaystyle  D=\frac{1}{a^6}\sum_{\bf k}\frac{d^2W_{\bf k}}{d\tau^2}=
\frac{8\varkappa\hbar}{a^6}\sum_{\bf
k}\left[\frac{d^2\xi_k}{d\tau^2}N_{\bf
k}+\left(\frac{d^2\xi_k}{d\tau^2}-\frac{1}{\xi_k}\right)\left(U^*_{\bf
k}e^{i\Phi_k}+U_{\bf k}e^{-i\Phi_k}\right)+\frac{i}{\xi_k}\frac{d\xi_k}{d\tau}\left(U^*_{\bf
k}e^{i\Phi_k}-U_{\bf k}e^{-i\Phi_k}\right)\right]\ ,
\\[7mm]
\displaystyle W_1=\frac{1}{a^2}\sum_{\bf k}k^2W_{\bf
k}=
 \frac{8\varkappa\hbar}{a^2}\sum_{\bf k}k^2\xi_k\left(N_{\bf
k}+U^*_{\bf k}e^{i\Phi_k}+U_{\bf k}e^{-i\Phi_k}\right)\ .
\end{array}
\label{4.15}
\end{equation}

The iteration procedure over $\sim k^2$ parameter for equation
(\ref{4.13}) is constructed accordingly to the following rules
\begin{equation}
 \begin{array}{c}
 \displaystyle
\frac{d^3\xi_k^{(0)}}{d\tau^3}=0\ , \qquad
\xi_k^{(0)}\equiv\frac{1}{a^2\epsilon_k^{(0)}}=P_{k}+R_{k}\tau+Q_{k}\tau^2\ ,
\\[5mm]
 \displaystyle
\frac{d^3\xi_k^{(n)}}{d\tau^3}=-4k^2a^2\frac{d}{d\tau}(a^2\xi^{(n-1)}_k)\
, \qquad n\geqslant 1\  .
\end{array}
 \label{4.16}
\end{equation}
In particular, we get
\begin{equation}
 \begin{array}{c}
\displaystyle \frac{d^2\xi_k^{(1)}}{d\tau^2}=-2k^2P_ka^4+...\ .
\end{array}
 \label{4.17}
 \end{equation}
The virtual graviton is defined by integration constants $P_{k},\
Q_{k},\ R_{k}$  of the main term of asymptotic expansion. Because
the  $\epsilon_k$ functional of (\ref{4.1}) is real (and therefore
the $\xi_k^{(0)}$ functional of (\ref{4.16}) is also real), we
obtain following inequality
\begin{equation}
\displaystyle 4P_kQ_k-R_k^2>0, \qquad P_k>0,\qquad Q_k>0\ .
\label{4.18}
\end{equation}
The dependence of constants $P_{k},\  Q_{k},\ R_{k}$ and phase
$\Phi_k$ on $k$ for  $k\to 0$ is defined by the finiteness
condition for $k^2W_{\bf k}$ and $d^2W_{\bf k}/d\tau^2$, and
taking into account the inequality (\ref{4.18}) we obtain
\[
\begin{array}{c}
\displaystyle P_k=\mathcal{O}(k^{-2}),\qquad
R_k=\mathcal{O}(k^{-1}),
 \qquad
 Q_k=\mathcal{O}(k^{0}),\qquad
\Phi_k=\mathcal{O}(k^{2})\ .
\end{array}
\]
The main terms of asymptotical expansions of moments (\ref{4.15}),
energy density and pressure of long wave gravitons can be obtained
from (\ref{4.16}) for $\xi_k^{(0)}$ and (\ref{4.17}) for
$\xi_k^{(1)}$. They read
\begin{equation}
\begin{array}{c}
 \displaystyle D=-\frac{16C_{g2}}{a^2}+\frac{16C_{g3}}{a^6}\ ,\qquad
 W_1=\frac{8C_{g2}}{a^2}\ ,
 \\[5mm]
\displaystyle \varkappa\varepsilon_g =
\frac{C_{g2}}{a^2}+\frac{C_{g3}}{a^6}\ , \qquad \varkappa p_g =
-\frac{C_{g2}}{3a^2}+\frac{C_{g3}}{a^6}\ .
\end{array}
\label{4.19}
\end{equation}
where
\begin{equation}
 \begin{array}{c}
 \displaystyle
C_{g2}=\varkappa\hbar\sum_{\bf k}k^2P_k\left(N_{\bf k}+U^*_{\bf k}+U_{\bf k}\right)\ ,
\qquad
C_{g3}=\varkappa\hbar\sum_{\bf k}Q_k\left(N_{\bf k}+U^*_{\bf k}+U_{\bf k}\right)\ .
\end{array}
 \label{4.20}
\end{equation}
For the first time, approximate solutions for the energy density
and pressure in the (\ref{4.19}) form were obtained for classical
long gravitational waves in \cite{42,43}. In the theory of
classical gravitational waves \cite{42,43}, the constants of
integration $C_{g2}$  and $C_{g3}$ must be positive. The crucial
formal difference between classical and quantum long gravitational
waves is in the fact that the last ones allow an arbitrary sign of
$C_{g2}$  and $C_{g3}$ (negative as well as positive). The physics
of this crucial difference will be discussed below (Section
\ref{sce}).

In a particular case of  $\delta$--type graviton spectrum, which
is localized at the region of very small conformal wave numbers,
(\ref{4.19}) can be considered as exact solutions. One needs to to
go over from summation to integration
\[
\sum_{\bf k}... \to \frac{1}{(2\pi)^3}\int d^3k...=\frac{1}{2\pi^2}\int\limits_0^\infty k^2dk...\ .
\]
After that, these solutions can be obtained by the following limits
\begin{equation}
 \begin{array}{c}
 \displaystyle  k^2P_k\to \frac{k_1}{a_1}=const(k),\qquad Q_k\to Q=const(k)\ ,
 \\[5mm] \displaystyle
 N_{\bf k}+U^*_{\bf k}+U_{\bf k}\to \frac{2\pi^2}{k^2}\mathcal{N}_0\delta(k-\kappa_0)\ ,
 \qquad
 \mathcal{N}_0=const(k)\ ,\qquad \kappa_0\to 0\ .
 \end{array}
 \label{4.21}
\end{equation}
In (\ref{4.21}) $k_1$ and $a_1$ are the constants of dimension of
conformal wave number and scale factor, respectively. They provide
the correct dimension to parameter $\displaystyle\lim_{k\to
0}k^2P_k$.

\subsection{Scenarios of Macroscopic Evolution}\label{sce}

In accordance with (\ref{4.19}), the system of long wave gravitons
behaves as a medium consisting of two subsystems whose equations
of state are $p_1=-\varepsilon_1/3$ and $p_2=\varepsilon_2$. But,
the internal structure of this substratum cannot be determined by
measurements that are conducted under the horizon of events. The
substratum effect (\ref{4.19}) on evolution of the Universe, is
seen by an observer as an energy density and pressure of the
"empty"\ (non--structured) spacetime, i.e. vacuum. The question
is: {\it does the graviton vacuum have a quasi--classic nature, or
has its quantum gravity origin been revealed in some cases?}

Let us review the situation. First, the superposition of quantum
states in state vectors of the general form (\ref{3.48}) and
(\ref{3.51}) could be essentially non--classical. Second, the
clearly non--classical ghost sector is inevitably presented in the
theory. Its properties are determined by the condition of
one--loop finiteness of macroscopic quantities (Section
\ref{fin}). The ghost sector is directly relevant to the
(\ref{4.19}) solution. Let us consider (\ref{3.72}) and
(\ref{3.73}), assuming for the sake of simplicity  $\langle
n_{k(gh)}\rangle=\langle \bar n_{k(gh)}\rangle$. Parameters of
solution (\ref{4.19}) are expressed via parameters of
graviton--ghost ensemble as follows
\begin{equation}
\begin{array}{c}
\displaystyle C_{g2}=2\varkappa\hbar\sum_{\bf
k}k^2P_k\left[\langle
n_{k(g)}\rangle(1+\zeta^{(g)}_k\cos\varphi_k)- \langle
n_{k(gh)}\rangle(1+\zeta^{(gh)}_k\cos\chi_k)\right]\ ,
\\[5mm]
\displaystyle C_{g3}=2\varkappa\hbar\sum_{\bf k}Q_k\left[\langle
n_{k(g)}\rangle(1+\zeta^{(g)}_k\cos\varphi_k)- \langle
n_{k(gh)}\rangle(1+\zeta^{(gh)}_k\cos\chi_k)\right]\ .
\end{array}
 \label{4.22}
\end{equation}
It follows from (\ref{4.22}) that  $C_{g2}>0, \ C_{g3}>0$ if the
graviton contribution dominates over ghosts in the quantum
condensate. We will name such a condensate "quasi-classical"\ .
Its energy density is positive, and it can be formed by usual
super--long gravitational waves. If the ghost contribution
dominates over gravitons in the quantum condensate, then
$C_{g2}<0, \ C_{g3}<0$. Such a condensate of negative energy
density has no classical analogy.

Summarizing the results of Sections \ref{swg} and \ref{lgw}, we
see that in cosmological applications of one--loop quantum gravity
we deal with the multi--component system consisting of short wave
graviton gas  $g1$ and two subsystems of graviton--ghost
condensate $g2,\ g3$. Taking into account (\ref{4.11}) and
(\ref{4.19}), we get the following equation for the scale factor
\begin{equation}
 \begin{array}{c}
 \displaystyle 3\frac{a'^2}{a^4}=\frac{C_{g1}}{a^4}+\frac{C_{g2}}{a^2}+\frac{C_{g3}}{a^6}\ .
 \end{array}
 \label{4.23}
\end{equation}
In the first scenario, the long wavelength condensate is of
negative energy, which means that the contribution of ghost
dominates over gravitons. The evolution of such a Universe is of
oscillating type. The solution reads
\begin{equation}
 \begin{array}{c}
\displaystyle
a^2=
\frac{C_{g1}}{2|C_{g2}|}+\frac{\sqrt{C_{g1}^2-
4C_{g2}C_{g3}}}{2|C_{g2}|}\sin\sqrt{\frac{4|C_{g2}|}{3}}\eta\
,
 \\[5mm]
\displaystyle
a^2_{1,2}=\frac{C_{g1}}{2|C_{g2}|}\mp\frac{\sqrt{C_{g1}^2-4C_{g2}C_{g3}}}{2|C_{g2}|}
 \end{array}
 \label{4.24}
\end{equation}
There is no classic analogy to the solution (\ref{4.24}). It can
be used for scenarios of evolution of the early quantum Universe.
In the region of minimal values of the scale factor $a_{min}=a_1$,
the $g3$ condensate bounces the Universe back from a singularity.
The transition from the expansion to the contraction epoch at the
region of maximal scale factor $a_{max}=a_2$ is provided by $g2$
condensate. Because of correlation of signs of  $C_{g2}<0$ and
$C_{g3}<0$, the non--singular Universe oscillates. Recent
scenarios of oscillating Universes based on condensates of
hypothetical ghost fields are under discussion in the current
literature as an alternative to the idea of inflation (see, e.g.
\cite{21})). Actually, we have shown that the same--type scenario
is constructed with the standard building blocks of quantum
gravity the well--known Faddeev--Popov's ghosts located
far from the mass shell. Thus, a very attractive idea is that one
and the same mechanism of graviton--ghost condensate formations in
the framework of one--loop quantum gravity based on the
"standard"\ Einstein equations (without hypothetical fields and
generalizations of Einstein's general relativity) could be
responsible for cyclic evolution of the early Universe (instead of inflation).

The second type of scenario applies if gravitons dominate over
ghosts in the condensate of positive energy. The solution reads
\begin{equation}
 \begin{array}{c}
\displaystyle
2\sqrt{C_{g2}(C_{g2}a^4+C_{g1}a^2+C_{g3})}+2C_{g2}a^2+C_{g1}=
\left(2\sqrt{C_{g2}C_{g3}}+C_{g1}\right)
 \exp\left(\sqrt{\frac{4|C_{g2}|}{3}}\eta\right)\ .
  \end{array}
 \label{4.25}
\end{equation}
The $g3$ condensate forms the regime of evolution in the vicinity
of singularity; meanwhile the asymptote of cosmological solution
for $\eta\to \infty$ is formed by $g2$ condensate. Short wave
gravitons $g1$ dominate during the intermediate epoch. The ratio
of graviton wavelength to horizon distance is constant during the
following asymptotical regime
\[
 \displaystyle
a\sim \exp\left(\sqrt{\frac{|C_{g2}|}{3}}\eta\right)\sim t,
\]
This means that the long wave condensate $g2$ forms the
self--consistent regime of evolution that provides its kinematic
stability.

\section{BBGKY Hierarchy (Chain)
and Exact Solutions of One--Loop Quantum Gravity
Equations}\label{Bog}

\subsection{Constructing the  Chain}\label{chain}

Approximate methods used in Sections \ref{swg} and \ref{lgw}
provide an opportunity to describe only limit cases which are
ultra shortwave gravitons and ghosts against the background of
almost stable Fock vacuum and super--long wave modes, forming
nearly stable graviton--ghost condensate. Now we are examining
self--consistent theory of gravitons and ghosts with the
wavelengths of the order of distance to the horizon:
 \begin{equation}
  \begin{array}{c}
  \displaystyle \frac{k^2}{a^2}\sim H^2,\ |\dot H|\ .
  \end{array}
  \label{5.1}
\end{equation}
When describing modes (\ref{5.1}), one should keep in mind two
factors. First, in the area of the spectrum (\ref{5.1}), there are
no reasonable approximations, which could be used to solve
equations (\ref{3.30}) and (\ref{3.31}), if the law of
cosmological expansion $a(t)$, $H(t)$ is not known in advance.
Second, the (\ref{5.1}) modes are quasi--resonant. Quantum gravity
processes of vacuum polarization, spontaneous graviton creation by
self--consistent field and graviton--ghost condensation are the
most intensive in this region of spectrum. From (\ref{5.1}) it is
also obvious that the threshold for quantum gravitational
processes involving zero rest mass gravitons and ghosts is absent.
These processes at the scale of horizon occur at any stage of
evolution of the Universe, including, in the modern Universe.

The theory that allows quantitatively describe quasi--resonant
quantum gravitational effects is constructed in the following way.
For the spectral function of gravitons and ghosts $W_{{\bf k}}$,
as defined in (\ref{3.59}), a differential equation is derived.
For this, the first equation (\ref{3.30}) is multiplied by the
$\hat\psi^+_{{\bf k}\sigma}$ (and then by the
$\dot{\hat\psi}^+_{{\bf k}\sigma}$), conjugated equation
(\ref{3.3}) is multiplied by the $\hat\psi_{{\bf k}\sigma}$ (and
then by the $\dot{\hat\psi}_{{\bf k}\sigma}$); and the equations
obtained are averaged and added. Similar action is carried out
with equations for ghosts, after which the equations for ghosts
are subtracted from the equations for gravitons. These operations
yield:
\begin{equation}
\begin{array}{c}
   \displaystyle
\ddot W_{{\bf k}}-2F_{{\bf k}}+3H\dot W_{{\bf
k}}+\frac{2k^2}{a^2}W_{{\bf k}}=0\ ,
   \end{array}
  \label{5.2}
 \end{equation}
 \begin{equation}
\begin{array}{c}
   \displaystyle
   \dot F_{{\bf k}}=-6HF_{{\bf k}}-\frac{k^2}{a^2}\dot W_{{\bf
   k}}\ ,
   \end{array}
  \label{5.3}
 \end{equation}
where
\[
\begin{array}{c}
   \displaystyle
W_{{\bf k}}=\sum_\sigma\langle
\Psi_g|\hat\psi^+_{{\bf k}\sigma}\hat\psi_{{\bf
 k}\sigma}|\Psi_g\rangle-2\langle\Psi_{gh}|{\bar\theta}_{{\bf k}}\theta_{{\bf
 k}}|\Psi_{gh}\rangle\ ,
\\[5mm]
\displaystyle
F_{{\bf k}}=\sum_\sigma\langle
\Psi_g|\dot{\hat\psi}^+_{{\bf k}\sigma}\dot{\hat\psi}_{{\bf
 k}\sigma}|\Psi_g\rangle-2\langle\Psi_{gh}|\dot{\bar\theta}_{{\bf k}}\dot\theta_{{\bf
 k}}|\Psi_{gh}\rangle\ .
 \end{array}
 \]
Further, equation (\ref{5.2}) is differentiated.  Expressions  for
$F_{{\bf k}},\ \dot F_{{\bf k}}$ via $W_{{\bf k}}$  are
substituted into the results of differentiation. For the spectral
function the third--order equation is produced
\begin{equation}
 \begin{array}{c}
    \displaystyle \stackrel{...}{W}_{{\bf k}} +9H\ddot W_{{\bf k}}
 +3\left(\dot H+6H^2\right)\dot W_{{\bf k}}+
 \frac{4k^2}{a^2}\left(
 \dot W_{{\bf k}}+2H W_{{\bf k}}\right)=0.
 \end{array}
 \label{5.4}
 \end{equation}

It is now necessary to draw attention to the fact that $W_{{\bf
k}}(t)$  is Fourier image of the two--point function, taken at
$t=t'$:
\begin{equation}
 \begin{array}{c}
    \displaystyle
    W(t,t'; {\bf{x}}-{\bf{x'}})=
    \langle
\Psi|\hat\psi_i^k(t,{\bf{x}})\hat\psi^i_k(t',{\bf{x'}})-2
\bar\theta(t,{\bf{x}})\theta(t',{\bf{x'}})|\Psi\rangle\ ,
\\[5mm]
\displaystyle W_{{\bf k}}(t)=\frac{1}{V}\int d^3y W(t,t;
{\bf{y}})e^{-i{\bf ky}}\ .
 \end{array}
 \label{5.5}
 \end{equation}
An infinite set of Fourier images is mathematically equivalent to
the infinite set of moments of the spectral function
\begin{equation}
\begin{array}{c}
\displaystyle W_n=\sum_{{\bf
k}}\frac{k^{2n}}{a^{2n}}\left(\sum_\sigma\langle
\Psi_g|\hat\psi^+_{{\bf k}\sigma}\hat\psi_{{\bf
k}\sigma}|\Psi_g\rangle-2\langle \Psi_{gh}|\bar\theta_{{\bf
k}}\theta_{{\bf k}}|\Psi_{gh} \rangle\right),
\\[5mm]
\displaystyle
 n=0,\,1,\,2,\,...,\infty.
\end{array}
 \label{5.6}
 \end{equation}
Therefore, from the equation for Fourier images (\ref{5.4}), we
can move to an infinite system of equations for the moments. For
this, equation (\ref{5.4}) is multiplied by $(k/a)^{2n}$ followed
by summation over wave numbers. The result is a
Bogoliubov--Born--Green--Kirkwood--Yvon (BBGKY) chain. Each
equation of this chain connects the neighboring moments:
\begin{equation}
\displaystyle  \dot D+6HD+4\dot W_1+16HW_1=0\ ,
\label{5.7}
\end{equation}
 \begin{equation}
\begin{array}{c}
 \displaystyle
B(1,2)\equiv \stackrel{...}{W}_1 +15H\ddot{W}_1  +3\left(22H^2
+3\dot{H} \right)\dot{W}_1+
2\left( 40H^3
+18H\dot{H}+\ddot{H} \right)W_1
+4\dot{W}_2+24HW_{2}=0,
\end{array}
\label{5.8}
\end{equation}
 \begin{equation}
\begin{array}{c}
 \displaystyle
B(n,n+1)\equiv\stackrel{...}{W}_n +3(2n+3)H\ddot{W}_n +
3\left[ \left(
4n^2+12n+6\right)H^2
+(2n+1)\dot{H} \right]\dot{W}_n+
\\[5mm]
\displaystyle
\displaystyle +2n\left[ 2\left(2n^2+9n+9\right)H^3
+6(n+2)H\dot{H}+\ddot{H} \right]W_n
+4\dot{W}_{n+1}+8(n+2)HW_{n+1}=0, \\[5mm]
\displaystyle
n=2,\,...,\,\infty\, .
\end{array}
\label{5.9}
\end{equation}
Equations (\ref{5.7}) --- (\ref{5.9}) have to be solved jointly
with the following macroscopic Einstein equations
\begin{equation}
 \begin{array}{c}
\displaystyle
\dot H=-\frac{1}{16}D-\frac16W_1\ ,
 \qquad
3H^2=\frac{1}{16}D+\frac{1}{4}W_1+\varkappa\Lambda\ .
 \end{array}
 \label{5.10}
\end{equation}

Note that an infinite chain of equations (\ref{5.7}) ---
(\ref{5.9}) contains information not only on the space--time
dynamics of field operators, but also about the quantum ensemble,
over which the averaging is done. The multitude of solutions of
the equations of the chain includes all possible self--consistent
solutions of the operator equations, averaged over all possible
quantum ensembles. Theory of gravitons presented by BBGKY chain,
conceptually and mathematically corresponds to the axiomatic
quantum field theory in the Wightman formulation (see Chapter 8 in
monograph \cite{6}). Here, as in Wightman, full information on the
quantum field is contained in an infinite sequence of averaged
correlation functions, definitions of which simply relate to the
symmetry properties of manifold, on which this field determines.

In BBGKY chain (\ref{5.7}), (\ref{5.8}) and (\ref{5.9}), unified
graviton--ghost objects appear which are moments of the spectral
function, renormalized by ghosts. The ghosts are not explicitly
labeled so that the chain is can be built formally in the model
not containing ghost fields. Mathematical incorrectness of such a
model is obvious only with a microscopic point of view because in
the quantum theory all the moments of spectral function diverge
the stronger, the more the moment number is. The system of
equations (\ref{5.7}) --- (\ref{5.9}) does not "know"\ , however,
that without the involvement of ghosts (or something other
renormalization procedure) it applies to the mathematically
non--existent quantities. The three following mathematical facts
are of  principal importance.

(i) {\it In one--loop quantum gravity, the BBGKY chain can be
formally introduced at an axiomatic level;}

(ii) {\it The internal properties of equations (\ref{5.7}) ---
(\ref{5.10}) provide the existence of finite solutions to this
system;}

(iii) {\it In finite solutions, there are solutions which do not
meet the "classic"\ condition of positiveness of moments  (see
Sections \ref{ggh_cond} and \ref{Sitt}).}

It follows from these facts that there should be an opportunity
and the need to implement a renormalization procedure to the
theory. This procedure should be able to redefine the moments of
the spectral function to finite values, but that leaves them
sign--undefined. As it can be seen from the theory which is
presented in Sections \ref{qg} and \ref{scgt}, in the one--loop
quantum gravity such a procedure is contained within the theory
under condition that the ghost sector automatically provides the
one--loop finiteness.

We found three exact self--consistent solutions of the system of
equations consisting of   the BBGKY chain (\ref{5.7}) ---
(\ref{5.9}) and macroscopic given below in Sections \ref{ggh_cond}
and \ref{Sitt}. The existence of exact solutions can be obtained
through direct substitution into the original system of equations.
The microscopic nature of these solutions, i.e. dynamics of
operators and structure of state vector is described in Sections
\ref{exact}, \ref{inst}.

\subsection{Graviton--Ghost Condensates of Constant Conformal Wavelength}\label{ggh_cond}

In Section \ref{lgw} the exact solution was found for the
graviton--ghost condensate, consisting of spatially uniform modes
(see (\ref{4.19}) --- (\ref{4.21})). This solution satisfies to
the first two BBGKY equations (\ref{5.7}), (\ref{5.8}) for an
arbitrary law of evolution $H(t)$ and under condition that $W_n=0$
for $n\geqslant 2$. (Recall that in this solution $D$ and  $W_1$
must be understood as the result of limit transition $k^2\to 0$;
and equality to zero of higher moments follows from the spatial
uniformity of modes.) Now we describe the exact self--consistent
solutions for the system, in which in addition to spatially
uniform modes, quasi--resonant modes with a wavelength equal to
the distance to the horizon of events are taken into account. In
terms of moments of the spectral function, the structure of
solutions under discussion is
\begin{equation}
 \begin{array}{c}
\displaystyle
D=D(g2)+D(g3)+D(g4)\ ,
\\[5mm]
\displaystyle W_1=W_1(g2)+W_1(g4)\ , \qquad W_1(g2)=\frac{8C_{g2}}{a^2}\
, \qquad W_{n}=W_{n}(g4),\; n\geqslant 2\ ,
 \\[5mm]
\displaystyle D(g3)=\frac{16C_{g3}}{a^6}\ ,\qquad
D(g2)=-\frac{16C_{g2}}{a^2}\ , 
\qquad
D(g4)=-\frac{48C_{g4(1)}}{a^2}\ln\frac{a_0}{e^{1/4}a}\ , \\[5mm]
\displaystyle
W_n(g4)=\frac{24C_{g4(n)}}{a^{2n}}\ln\frac{a_0}{a},\qquad
n=1,\,...,\,\infty\ .
 \end{array}
 \label{5.11}
\end{equation}
Here $C_{g3},\ C_{g2},\ C_{g4(n)},\ a_0$  are numerical
parameters. Restrictions on their values follow from the condition
of the existence of the exact self--consistent solution.

The solution is found by using of the consistency of functions
(\ref{5.11}) with the relations arising from the macroscopic
Einstein's equations (we are discussing model with $\Lambda=0$):
\begin{equation}
 \begin{array}{c}
\displaystyle
H^2=\frac{C_{g3}}{3a^6}+\frac{C_{g2}}{3a^2}+\frac{C_{g4}}{a^2}\ln\frac{e^{1/4}a_0}{a}\
,
\\[5mm]
\displaystyle
\dot H= -\frac{C_{g3}}{a^6}-\frac{C_{g2}}{3a^2}-\frac{C_{g4}}{a^2}\ln\frac{e^{3/4}a_0}{a}\ ,
\\[5mm]
\displaystyle \ddot
H=2H\left(\frac{3C_{g3}}{a^6}+\frac{C_{g2}}{3a^2}+
\frac{C_{g4}}{a^2}\ln\frac{e^{5/4}a_0}{a}\right)\
.
\end{array}
 \label{5.12}
\end{equation}
In (\ref{5.12}) as well as further, we use notation
$C_{g4(1)}\equiv C_{g4}$. Functions $D$ and $W_1$ from
(\ref{5.11}) transform the equation (\ref{5.7}) to an identity.
The substitution of  $W_1$ and $W_2$ into (\ref{5.8}), taking into
account (\ref{5.12}), leads to the following expression
\begin{equation}
 \begin{array}{c}
 \displaystyle
B(1,2)= H\frac{48}{a^4}\left[
4(C_{g4(2)}-C_{g4}^2)\ln\frac{a_0}{a}-\frac43C_{g2}C_{g4}+C_{g4}^2-2C_{g4(2)}\right]=0\
.
\end{array}
\label{5.13}
\end{equation}
The infinite chain (\ref{5.9}), in contrast to the equation
(\ref{5.8}), contains moments of spectral functions of
quasi--resonant modes. Nevertheless, it does result, only
including (\ref{5.13}) as a particular case
\begin{equation}
 \begin{array}{c}
 \displaystyle
B(n,n+1)= H\frac{48}{a^{2n+2}}\left[
4(C_{g4(n+1)}-C_{g4}C_{g4(n)})\ln\frac{a_0}{a}-\frac43C_{g2}C_{g4(n)}+C_{g4}C_{g4(n)}-2C_{g4(n+1)}\right]=0\
,
\\[5mm]
n=2,\,...,\,\infty.
\end{array}
\label{5.14}
\end{equation}
The following relations between parameters follow from
(\ref{5.13}) and (\ref{5.14})
\begin{equation}
 \begin{array}{c}
 \displaystyle
 C_{g4(n)}=C_{g4}^n\ , \qquad C_{g2}=-\frac34C_{g4}\ .
\end{array}
\label{5.15}
\end{equation}
Thus, moments of the spectral function of quasi--resonant modes
satisfy to the following recurrent relation
\begin{equation}
 \begin{array}{c}
 \displaystyle
 W_{n+1}(g4)=\frac{C_{g4}}{a^2}W_n(g4)=\left(\frac{C_{g4}}{a^2}\right)^{n}W_1(g4)\ .
\end{array}
\label{5.16}
\end{equation}
Comparison of (\ref{5.16}) with (\ref{5.6}) shows that in the
exact solution under discussion all quasi--resonant modes have the
same wavelength $\lambda=a/\sqrt{|C_{g4}|}\equiv a/k_0$. In other
words, in the space of conformal wave numbers the spectrum of
quasi--resonant wave modes is localized in the vicinity of the
fixed value $|{\bf k}|=k_0$.

Depending on the sign of $C_{g4}$, we get two exact solutions to
the macroscopic observables of graviton--ghost media in the form
of functionals of scale factor.

(i) {\it Oscillating Universe.}

Suppose that $C_{g4}>0$. In accordance with (\ref{5.15}), in this
case all  $C_{g4(n)}>0$. The positive sign of all moments
$W_n(g4)>0$ suggests that gravitons dominate over ghosts in the
ensemble of quasi--resonant modes.  We also see that the parameter
of spatially uniform mode $g2$ is negative, i.e. $C_{g2}<0$.  As
was shown in Section \ref{sce}, signs of parameters of $g2$ and
$g3$  modes are the same, so $C_{g3}<0$. From this it follows that
ghosts are dominant in case of spatially uniform modes. The energy
density and pressure of graviton--ghost substratum read
\begin{equation}
\begin{array}{c}
\displaystyle
\varkappa\varepsilon_g=-\frac{|C_{g3}|}{a^6}+\frac{3C_{g4}}{a^2}\ln\frac{a_0}{a}\
,\qquad \varkappa p_g=-\frac{|C_{g3}|}{a^6}-\frac{C_{g4}}{a^2}\ln\frac{a_0}{ea}\ .
\end{array}
\label{5.17}
\end{equation}
The parameter $C_{g2}$ is not explicitly showed up in (\ref{5.17})
because it is expressed via $C_{g4}$ in accordance with
(\ref{5.15}). There is an oscillating solution to the Einstein
equation $3H^2=\varkappa\varepsilon_g$ if solutions for the
turning points $a_m=a_{min},\, a_{max}$ exist, i.e.
\begin{equation}
\displaystyle b=\frac{3C_{g4}a_0^4}{4|C_{g3}|}>e,\qquad
\left(\frac{a_0}{a_m}\right)^4=b\ln\left(\frac{a_0}{a_m}\right)^4\
. \label{5.18}
\end{equation}
In the vicinity of turning points energy density is formed by
contributions of ghosts and gravitons, which are comparable in
their absolute values, but have opposite signs. Far from turning
points, graviton quasi--resonant modes dominate. Simplifying the
situation, we can say that in the oscillating Universe spatially
uniform modes have essentially quantum nature, and quasi--resonant
modes allow semi--classical interpretation.

In the absence of a spatially homogeneous subsystem  $g3$, the
infinite sequence of oscillations degenerates into one
semi--oscillation. Indeed, with  $C_{g3}=0$ the scale factor, as a
function of cosmological time, reads
\begin{equation}
\displaystyle a(\eta)=a_0\exp\left(-\frac{C_{g4}\eta^2}{4}\right)\ ,\qquad C_{g4}>0\ .
\label{5.19}
\end{equation}
In accordance with (\ref{5.19}), the Universe originates from a
singularity, reaches the state of maximal scale factor
$a_{max}=a_0$ and then collapses again to singularity.

(ii)    {\it Birth in Singularity and Accelerating Expansion.}

Accordingly to (\ref{5.16}), moments of the spectral function of
quasi--resonant modes form an alternating sequence if $C_{g4}<0$.
It reads
\begin{equation}
\displaystyle W_n(g4)=-(-1)^{n}\frac{24|C_{g4}|^n}{a^{2n}}\ln\frac{a}{a_0}\ ,
\qquad n=1,\,...,\,\infty \ .
\label{5.20}
\end{equation}
It is clear that the result (\ref{5.20}) can not be obtained for
the quasi--classical ensemble of gravitational waves. The
microscopic nature of this solution is discussed in Section
\ref{exact}. It is appropriate here to emphasize one more time
that the theory, which is formulated in the most common way in the
BBGKY form, captures the existence of such a solution.

It is not difficult to notice that the solution which we are now
discussing is in a sense, an alternative to the previous solution.
With $C_{g4}<0$, parameters of spatially homogeneous modes are
positive $C_{g2}>0,\ C_{g3}>0$. Thus, spatially uniform modes
admit semi--classical interpretation, but quasi--resonant modes
have essentially quantum nature. The energy density and pressure
of graviton--ghost substratum are
 \begin{equation}
 \begin{array}{c}
\displaystyle
\varkappa\varepsilon_g=\frac{C_{g3}}{a^6}+\frac{3|C_{g4}|}{a^2}\ln\frac{a}{a_0}\
,\qquad \varkappa
  p_g=\frac{C_{g3}}{a^6}-\frac{|C_{g4}|}{a^2}\ln\frac{ea}{a_0}\ .
  \end{array}
\label{5.21}
\end{equation}
Specific properties of solutions to Einstein's equations
$3H^2=\varkappa\varepsilon_g$ depend on initial conditions and
relations between the parameters of graviton--ghost substratum.
First of all, let us mention a scenario that corresponds to a
singular origin with the strong excitation of spatially uniform
modes
\begin{equation}
\displaystyle C_{g3}\ne 0,\qquad H>0, \qquad \frac{3|C_{g4}|a_0^4}{4C_{g3}}<e\ .
\label{5.22}
\end{equation}
In the case  (\ref{5.22}), the Universe is born in the singularity
and fairly quickly reaches the area of large scale factor values,
where it expands with the acceleration:
\begin{equation}
\begin{array}{c}
\displaystyle a\simeq |C_{g4}|^{1/2}t\ln^{1/2}\frac{t}{t_0}\ ,
\qquad \frac{\ddot{a}}{a}\simeq \frac{|C_{g4}|}{2a^2},\qquad a\gg a_0,\;
\left(\frac{C_{3g}}{3|C_{4g}|}\right)^{1/4}\ .
\end{array}
\label{5.23}
\end{equation}
Branch of the same solution, with $H<0$  describes the collapsing
Universe with a singular end--state.

Two other scenarios correspond to the weak excitation of graviton spatially uniform modes
\begin{equation}
\displaystyle C_{g3}\ne 0, \qquad \frac{3|C_{g4}|a_0^4}{4C_{g3}}>e\ .
\label{5.24}
\end{equation}
In the case of  (\ref{5.24}), the region of legitimate values of
the scale factor is divided into two sub--regions $0\leqslant
a\leqslant  a_1$ and $a_2\leqslant a<\infty$  separated by a
barrier of finite width $a_2>a_1$. In the sub--region of small
values of the scale factor, the Universe is born in a singularity,
reaches the state with a maximum value of $a=a_1$, and then
returns to the singularity. In the  limit  $C_{3g}\to 0$ the
possibility of such an evolution disappears because of $a_1\to 0$.
In sub--region of the large scale factor, the evolution of the
Universe starts at the infinite past from the state of zero
curvature. At the stage of compression, the Universe reaches the
state with a minimum value of $a=a_2$, and then turns into an
accelerated mode of expansion. With $C_{g3}=0$, this branch of
cosmological solution is described by the following function of
cosmological time
\begin{equation}
\displaystyle a(\eta)=a_0\exp\left(\frac{|C_{g4}|\eta^2}{4}\right)\ ,\qquad C_{g4}<0\ .
\label{5.25}
\end{equation}
Note that degenerate solutions (\ref{5.19}) and (\ref{5.25})
differ only in the sign under of exponent.

\subsection{Self--Polarized Graviton--Ghost Condensate in De Sitter Space}\label{Sitt}

It is easy to find that the system of equations (\ref{5.7}) ---
(\ref{5.10}) has a simple stationary solution  $H=const$,
$D=const$,  $W_n=const$. This solution describes the highly
symmetrical graviton--ghost substratum that fills the De Sitter
space. It reads
 \begin{equation}
 \begin{array}{c}
 \displaystyle  H^2=\frac{1}{36}W_1+\frac13\varkappa\Lambda\ ,\qquad
 a=a_0e^{Ht}\ ,
 \\[5mm]
\displaystyle \varepsilon_{g}=-p_{g}=\frac{1}{12}W_{1}\ .
 \end{array}
\label{5.26}
\end{equation}
This solution exists both for the $\Lambda=0$ case and for
$\Lambda\ne 0$. The first moment of the spectral function
satisfies the inequality $W_1>-12\varkappa\Lambda$ is the only
independent parameter of the solution. The remaining moments are
expressed through  by recurrence relations:
 \begin{equation}
 \begin{array}{c}
  \displaystyle D=-\frac{8}{3}W_1\ ,\qquad
 W_{n+1}=-\frac{n(2n+3)(n+3)}{2(n+2)}H^2W_n\ ,
 \qquad n\geqslant 1\ .
\end{array}
\label{5.27}
 \end{equation}

From  (\ref{5.26}) and (\ref{5.27}) it clearly follows that the
solution has essentially {\it vacuum and quantum} nature. The
first can be seen from the equation of state $p_g=-\varepsilon_g$.
The second can be seen from the fact that the signs of the moments
$W_{n+1}/W_n<0$ alternate. Another sign of the quantum nature of
the effect is contained in the properties of graviton spectrum.
The first of recurrence relations allows estimating of wavelengths
of gravitons and ghosts that play a dominant part in the formation
of observables
\begin{equation}
\displaystyle \lambda\sim \frac{a}{\overline{k}}\sim \sqrt{
 \frac{W_1}{|W_2|}}=\frac{1}{H}\sqrt{\frac{3}{10}}=const\ .
 \label{5.28}
\end{equation}
As can be seen from (\ref{5.28}), during the exponential expansion
of the Universe typical values of $\overline{k}$ rapidly shift to
the region of exponentially large conformal wave numbers. The
physical wavelength and macroscopic observables are unchanged in
time. Such a situation occurs if the following two conditions
apply.

(i) In the  ${\bf k}-$ space of conformal wave numbers spectra of
graviton vacuum fluctuations are flat;

(ii) In the integration over the flat spectrum, divergent
components of integrals excluded for reason to be discussed in
Section \ref{S}. Observables are formed by finite residuals of
these integrals.

In Section \ref{S} we will show that these conditions are actually
satisfied on the exact solution of operator equations of motion,
with special choice of Heisenberg's state vector of
graviton--ghost vacuum. Microscopic calculation also allows
expressing the first moment of spectral function through the
curvature of De Sitter space
\begin{equation}
 \displaystyle W_1=\frac{9\varkappa\hbar N_g}{2\pi^2}H^4\ ,
 \label{5.29}
 \end{equation}
where  $N_g$ is a functional of parameters of state vector, which
is of the order of the number of virtual gravitons and ghosts that
are situated under the horizon of events. Their wavelengths are of
the order of the distance to the horizon. It must be stressed that
the number of gravitons and ghosts $N_g$ is a macroscopic value.

The order of magnitude of $N_g$ is determined by graviton and
ghost numbers in the condensate. Let us emphasize that {\it
numbers of gravitons and ghosts and hence, $N_g$ parameters are
macroscopic qualities.} Further down in this section it is assumed
that the gravitons dominate in the condensate and that the
parameter $N_g>0$.

Note that the result (\ref{5.29})  can be easily predicted from
the general considerations, including considerations of dimension.
Indeed, the general formula (\ref{3.66}) shows that the moment
$W_1$ is of dimension  $[W_1]=[l]^{-2}$ ($[l]$ is of dimension of
length). It also contains the square of the Planck length as a
coefficient. Because $W_1$ is a functional of the metric, desired
dimension can be obtained only using metric's derivatives. It
follows from this that $W_1=C\cdot \varkappa\hbar H^4$  where $C$
dimensionless constant that contains parameters of vacuum
condensate. Given (\ref{5.29}), the solution in its final form is
as follows:
 \begin{equation}
 \begin{array}{c}
 \displaystyle
 D=-\frac{12\varkappa\hbar N_g}{\pi^2}H^4\ ,
\qquad 
 W_n=\frac{(-1)^{n+1}}{2^{2n}}
 (2n-1)!(2n+1)(n+2)\times
  \frac{2\varkappa\hbar N_g}{\pi^2}H^{2n+2} ,\quad n\geqslant 1\
  .
   \end{array}
 \label{5.30}
 \end{equation}
\begin{equation}
\displaystyle \varepsilon_g=-p_g=\frac{3\hbar
 N_g}{8\pi^2}H^4\ ,
 \label{5.31}
 \end{equation}
The macroscopic Einstein's equation is transformed into the
equation for the inflation exponent
\begin{equation}
\displaystyle 3H^2=\frac{3\varkappa\hbar
 N_g}{8\pi^2}H^4+\varkappa\Lambda\ .
 \label{5.32}
 \end{equation}

Because $N_g$ is a macroscopic parameter, the solution under
discussion can be directly relevant to the asymptotic future of
the Universe. In this case, the number of gravitons and ghosts
under the horizon of events and $\Lambda$--term in the equation
(\ref{5.32}) should be considered as parameters, whose values were
formed during the earlier stages of cosmological evolution.
According to Zel'dovich \cite{22},  $\Lambda$--term is the total
energy density of equilibrium vacuum subsystems of
non--gravitational origin. The problem of the $\Lambda$--term
formation is so complex that little has changed since the
excellent review of Weinberg \cite{23}. We are limited only to
showing the order of magnitude of $\Lambda\sim 3 \cdot
10^{-47}\hbar^{-3}$ GeV$^4$ allowed by observational data.

Some possibilities of co--existence of graviton condensate and
$\Lambda$--term will be discussed for $\Lambda\geqslant 0,\
N_g>0$. (For other possibilities see Section \ref{S}.) The
curvature of the De Sitter space for the asymptotical state of the
Universe is calculated by means of the solution to the equation
(\ref{5.32}). It reads
 \begin{equation}
\displaystyle H_\infty^2=\frac{4\pi^2}{\varkappa\hbar}
  \left(\frac{1}{N_g}\pm\sqrt{\frac{1}{N_g^2}-\frac{\varkappa^2\hbar\Lambda}{6\pi^2N_g}}\right)\ ,\qquad R=-12H_\infty^2\ .
 \label{5.33}
 \end{equation}
The energy density of vacuum in this state contains contributions
of subsystems formed by all physical interactions including the
gravitational one
\begin{equation}
\displaystyle \varepsilon_{vac}^{(\infty)}=\frac{3\hbar
 N_g}{8\pi^2}H_{\infty}^4+\Lambda\ .
 \label{5.34}
 \end{equation}
The relative input of graviton--ghost condensate into asymptotic
energy density of the vacuum depends on parameters of the
Universe. If the following inequality
 \begin{equation}
\displaystyle \frac{\varkappa^2\hbar\Lambda N_g}{6\pi^2}\ll 1\ ,
\label{5.35}
 \end{equation}
applies because of a small number of gravitons and ghosts, then
the quantum--gravitational term is small and one must use the
following solution
 \begin{equation}
\displaystyle H_{\infty}^2\simeq
\frac13\varkappa\Lambda\left(1+\frac{\varkappa^2\hbar\Lambda}
{24\pi^2}N_g\right)\ .
 \label{5.36}
 \end{equation}
If the inequality (\ref{5.35}) is satisfied because of a small
$\Lambda$--term then the asymptotic state is mostly formed by the
graviton--ghost condensate
\begin{equation}
\displaystyle H_{\infty}^2\simeq \frac{8\pi^2}{\varkappa\hbar
N_g}-\frac{\varkappa\Lambda}{3}\ .
 \label{5.37}
 \end{equation}
It can be seen from (\ref{5.33}) for $\Lambda>0$,  the number of
gravitons and ghosts that can appear in the Universe is limited by
maximum value
 \begin{equation}
\displaystyle N_{g(max)}=\frac{6\pi^2}{\varkappa^2\hbar\Lambda}\sim 10^{122}\
.
  \label{5.38}
 \end{equation}
In this limiting case (\ref{5.38}), the equipartition of the
vacuum energy takes place between graviton--ghost and
non--gravitational vacuum subsystems
\begin{equation}
\displaystyle H_\infty^2=\frac{4\pi^2}{\varkappa\hbar
 N_{g(max)}}=\frac23\varkappa\Lambda\ ,\qquad\qquad \varepsilon_g^{(\infty)}=\Lambda=\frac12
 \varepsilon_{vac}^{(\infty)}\ .
 \label{5.39}
 \end{equation}

\subsection{The Problem of Quantum--Gravity Phase Transitions}\label{pt}

Three exact solutions of the equations of quantum gravity (with no
matter fields and in the absence of $\Lambda$--term) are, in our
view, impressive illustrations of physical content of the theory.
(Of course, we can not exclude the existence of other exact
solutions). The sets of basic formulas (that characterize each of solutions) have the form:

1) {\bf Oscillating Universe},
\begin{equation}
 \begin{array}{c}
\displaystyle
t=\int\limits_{a_{min}}^a\frac{da}{H_\sI(a)}\ , \qquad H^2_\sI(a_{min})=0\ ,
\qquad
 H^2_\sI(a)=\frac{C^{(\sI)}_{g3}}{3a^6}+\frac{C^{(\sI)}_{g2}}{3a^2}+\frac{C^{(\sI)}_{g4}}{a^2}\ln\frac{e^{1/4}a^{(\sI)}_0}{a}\
,
\\[5mm]
\displaystyle
C^{(\sI)}_{g4}>0\ ,\qquad C^{(\sI)}_{g2}=-\frac34C^{(\sI)}_{g4}<0\ ,\qquad C^{(\sI)}_{g3}\eqslantless 0\ ,
\\[5mm]
\displaystyle
3\frac{{\dot a}^2}{a^2}=\varkappa\varepsilon_g=-\frac{|C^{(\sI)}_{g3}|}{a^6}+\frac{3C^{(\sI)}_{g4}}{a^2}\ln\frac{a^{(\sI)}_0}{a}\
, \qquad -6\frac{{\ddot a}}{a}=\varkappa(\varepsilon_g+3p_g)=-\frac{4|C^{(\sI)}_{g3}|}{a^6}+\frac{3C^{(\sI)}_{g4}}{a^2}\ .
 \end{array}
 \label{5.40}
\end{equation}

2) {\bf Birth in Singularity and Accelerating Expansion},
\begin{equation}
 \begin{array}{c}
\displaystyle
t=\int\limits_{0}^a\frac{da}{H_\sII(a)}\ ,
\qquad H^2_\sII(a)=\frac{C^{(\sII)}_{g3}}{3a^6}+\frac{C^{(\sII)}_{g2}}{3a^2}+\frac{C^{(\sII)}_{g4}}{a^2}\ln\frac{e^{1/4}a^{(\sII)}_0}{a}\
,
\\[5mm]
\displaystyle
C^{(\sII)}_{g4}<0\ ,\qquad C^{(\sII)}_{g2}=-\frac34C^{(\sII)}_{g4}>0\ ,\qquad C^{(\sII)}_{g3}\geqslant 0\ ,
\\[5mm]
\displaystyle
3\frac{{\dot a}^2}{a^2}=\varkappa\varepsilon_g=\frac{C^{(\sII)}_{g3}}{a^6}+\frac{3|C^{(\sII)}_{g4}|}{a^2}\ln\frac{a}{a^{(\sII)}_0}\
,  -6\frac{{\ddot a}}{a}=\varkappa(\varepsilon_g+3p_g)=\frac{4C^{(\sII)}_{g3}}{a^6}-\frac{3|C^{(\sII)}_{g4}|}{a^2}\ .
 \end{array}
 \label{5.41}
\end{equation}

3) {\bf De Sitter Universe},
\begin{equation}
 \begin{array}{c}
\displaystyle
a=a^{(\sIII)}_{0}e^{H_{\sIII}t}\ , \qquad  H^2_{\sIII}=\frac{8\pi^2}{\varkappa\hbar N_g}\ ,
\\[5mm]
\displaystyle
3\frac{{\dot a}^2}{a^2}=\varkappa\varepsilon_g=\frac{24\pi^2}{\varkappa\hbar N_g}\
, \qquad -6\frac{{\ddot a}}{a}=\varkappa(\varepsilon_g+3p_g)=-\frac{48\pi^2}{\varkappa\hbar N_g}\ .
 \end{array}
 \label{5.42}
\end{equation}

IIf arbitrary shifts in time axis are excluded, then (\ref{5.40}) and
(\ref{5.41}) are 3--parameter solutions ($a_0,\ C_{g4},\ C_{g3}$). Meanwhile
(\ref{5.42}) does contain one free parameter $N_g$. Also one can see that three exact solutions
correspond to the spaces of different symmetries. The solution
(\ref{5.42}) describes 4--space of constant curvature, with
the highest possible symmetry. Solution (\ref{5.41}) (in the
version of appropriate unlimited expansion) describes 4--space,
the geometry of which tends asymptotically to the geometry of the
Milln space. Finally, the solution (\ref{5.40}) (in the version
corresponding to oscillations) describes 3--geometry, which is
translation--invariant along the axis of time. Different symmetries of different solutions are the rationale for the introduction of phases of graviton-ghost vacuum. It is supposed to be continuous phase transitions between phases with different symmetries.

Representations of phase transitions are, of course, only
heuristic nature. In the one--loop quantum gravity,
multi--particle correlations in the system of gravitons and ghosts
are not taken into account. For this reason, in this theory it is
impossible to define the order parameter that plays the role of
the master parameter when choosing a phase state. Phase
transitions that were discussed above, were actually initiated by
disparity between the choice of the asymptotic state and set of
the initial conditions. Of course, such operations are meaningful
only within the suggestion that the effect of {\it
non--equilibrium} phase transition will be contained in future
theory.

Staying on the heuristic level, we can use the exact solutions
(\ref{5.40}),  (\ref{5.41}),  (\ref{5.42}) to demonstrate
in principle the possibility of the existence of equilibrium phase
transitions. Let us consider the exact solutions as the various
branches of a general solution. A rough phase transition model is
the passage from one branch to another while maintaining
continuity of scale factor and its first and second derivatives. As can be
seen from (\ref{5.40}),  (\ref{5.41}),  (\ref{5.42}),
these conditions provide the equalities of volumes, energies and pressures of
graviton--ghost systems on both sides of the transition point. It is easy to see that these conditions correspond to the phase transitions of the second kind. The
microscopic theory makes it possible to see  that at the
point of transition the internal structure of graviton--ghost
substratum is changed (see Sections \ref{exact}, \ref{inst}).

Consider consistently simplified models of all of the phase
transitions. Graviton--ghost vacuum is of the lowest symmetry in phase (\ref{5.40}). This phase is invariant under condition that the shift on the time axis is of the oscillation period only. Phase (\ref{5.41}) is of higher symmetry. It is Milln space asymptotically in time at $t\to \infty$.
Phase (\ref{5.42}) (graviton-ghost vacuum in the De Sitter space) has the highest symmetry.

Suppose that the symmetry of the graviton--ghost vacuum increases in the process of the universe evolution, that is, the phase transitions occur in the sequence $I\to II\to III$. According to (\ref{5.40}),  (\ref{5.41}), the point of transition from the initial state of the oscillating universe $I$ to the state of unlimited expansion $II$ is determined by the following relations
\begin{equation}
 \begin{array}{c}
\displaystyle
\varepsilon_g^{(\sI)}(a_{c1})=\varepsilon_g^{(\sII)}(a_{c1})\quad \to \quad
 C^{(\sI)}_{g4}\ln\frac{a^{(\sI)}_0}{a_{c1}}+|C^{(\sII)}_{g4}|\ln\frac{a^{(\sII)}_0}{a_{c1}}=
\frac{1}{3a_{c1}^4}\left(C^{(\sII)}_{g3}+|C^{(\sI)}_{g3}|\right)\ ,
\\[5mm]
\displaystyle
\varepsilon_g^{(\sI)}(a_{c1})+3p_g^{(\sI)}(a_{c1})=\varepsilon_g^{(\sII)}(a_{c1})+3p_g^{(\sII)}(a_{c1})\quad \to \quad
C^{(\sI)}_{g4}+|C^{(\sII)}_{g4}|=
\frac{4}{3a_{c1}^4}\left(C^{(\sII)}_{g3}+|C^{(\sI)}_{g3}|\right)\ ,
 \end{array}
 \label{5.43}
\end{equation}
where $a_{c1}$ is the value of the scale factor at the fitting point,
common to the two phases. From (\ref{5.43}) one can get the formula for the fitting point and relationship between the parameters of different phases:
\begin{equation}
 \begin{array}{c}
\displaystyle \left[a_0^{(\sI)}\right]^{C_{g4}^{(\sI)}/(C^{(\sI)}_{g4}+|C^{(\sII)}_{g4}|)}
\cdot\left[a_0^{(\sII)}\right]^{|C_{g4}^{(\sII)}|/(C^{(\sI)}_{g4}+|C^{(\sII)}_{g4}|)}=
\left[\frac{4(|C^{(\sI)}_{g3}|+C^{(\sII)}_{g3})}{3(C^{(\sI)}_{g4}+|C^{(\sII)}_{g4}|)}\right]^{1/4}=a_{c1}e^{1/4}\ .
 \end{array}
\label{5.44}
\end{equation}
As we know, in phase $I$ gravitons dominate in quasi--resonant modes, and ghosts dominate
in spatially uniform modes. Following the transition, in phase
$II$ quasi--resonant modes are dominated by ghosts, but spatially
uniform modes are dominated by gravitons. Formulas (\ref{5.44}) provide constraints on the range of allowed values of the transition point $a_{c1}$ and the parameters $C^{(\sII)}_{g3},\ |C^{(\sII)}_{g4}|,\ a_0^{(\sII)}$ of the phase $II$ for given values of the parameters $|C^{(\sI)}_{g3}|,\ C^{(\sI)}_{g4},\ a_0^{(\sI)}$ of phase $I$. {\it According to (\ref{5.44}), whatever the parameters of phase $I$ are there is a some set of parameters of a phase $II$. Thus, a continuous phase transition $I\to II$ from the state of oscillating Universe to the state of the Universe in a phase of the unlimited expansion and the asymptotic acceleration is inevitable.}

Further, let us consider the phase transition $II\to III$. The conditions of sewing together of solutions
(\ref{5.41}) and (\ref{5.42}) read
\begin{equation}
 \begin{array}{c}
\displaystyle
\varepsilon_g^{(\sII)}(a_{c2})=\varepsilon_g^{(\sIII)}\quad \to \quad
\frac{C^{(\sII)}_{g3}}{a_{c2}^6}+\frac{3|C^{(\sII)}_{g4}|}{a_{c2}^2}\ln\frac{a_{c2}}{a^{(\sII)}_0}=\frac{24\pi^2}{\varkappa\hbar N_g}\ ,
\\[5mm]
\displaystyle
\varepsilon_g^{(\sII)}(a_{c2})+3p_g^{(\sII)}(a_{c2})=\varepsilon_g^{(\sIII)}+3p_g^{(\sIII)}\quad \to \quad
-\frac{2C^{(\sII)}_{g3}}{a_{c2}^6}+\frac{3|C^{(\sII)}_{g4}|}{2a_{c2}^2}=\frac{24\pi^2}{\varkappa\hbar N_g}\
 \end{array}
 \label{5.45}
\end{equation}
In this case, we have the following formulas for the transition point and the relationship between the parameters of the phases:
\begin{equation}
 \begin{array}{c}
\displaystyle
\ln\frac{e^2[a^{(\sII)}_0]^4}{a_{c2}^4}=\frac{4C^{(\sII)}_{g3}}{a_{c2}^4|C^{(II)}_{g4}|}\ ,
\qquad
\frac{|C^{(\sII)}_{g4}|}{a_{c2}^2}\ln\frac{e^{1/4}a_{c2}}{a^{(\sII)}_0}=\frac{12\pi^2}{\varkappa\hbar N_g}\ .
 \end{array}
 \label{5.46}
\end{equation}
According to (\ref{5.46}), a continuous phase transition $II\to III$ is possible if the parameters of the phase $II$ satisfy the inequality
\begin{equation}
 \begin{array}{c}
\displaystyle
\frac{e[a^{(\sII)}_0]^4|C^{(II)}_{g4}|}{4C^{(\sII)}_{g3}}\geqslant 1\ .
 \end{array}
 \label{5.47}
\end{equation}
If the phase transition took place, then in phase $III$ the number of gravitons under the
horizon of events is unambiguously defined by parameters of phase $II$.
The phase transition looks
like a "freezing" of the distance to the horizon and of the value
of the physical wavelength of quasi--resonant modes.

Finally, we note that a continuous phase transition $I\to III$ from the oscillating Universe to De Sitter space is possible if the following conditions are met
\begin{equation}
 \begin{array}{c}
\displaystyle
\ln\frac{e^2[a^{(\sI)}_0]^4}{a_{c3}^4}=\frac{4|C^{(\sI)}_{g3}|}{a_{c3}^4C^{(I)}_{g4}}\ ,\qquad
\frac{e[a^{(\sI)}_0]^4C^{(I)}_{g4}}{4|C^{(\sI)}_{g3}|}\geqslant 1\ ,
\\[5mm]
\displaystyle
\frac{C^{(\sI)}_{g4}}{a_{c3}^2}\ln\frac{e^{1/4}a_{c3}}{a^{(\sI)}_0}=\frac{12\pi^2}{\varkappa\hbar N_g}\ .
 \end{array}
 \label{5.48}
\end{equation}

\subsection{Gravitons in the Presence of Matter. Nonlinear Representation of the BBGKY Chain}\label{gm}

The full system of equations of self--consistent theory of
gravitons in the isotropic Universe consists of the BBGKY chain
(\ref{5.7}) --- (\ref{5.9}) and macroscopic Einstein equations. In
equations (\ref{5.7}) --- (\ref{5.9}), the Hubble function $H$ and
its derivatives $\dot H,\ \ddot H$ are coefficients multiplied by
the moments of the spectral function. In such a form the chain
conserves its form even if besides of gravitons, other physical
fields are also sources of the macroscopic gravitational field. We
are interesting in the evolution of the flat isotropic Universe at
a stage when the contributions of gravitons and non--relativistic
particles, baryons and neutralinos, are quantitatively
significant. (The latter are presumably carriers of the mass of
Dark Matter.) We assume also that non--gravitational physical
interactions created the equilibrium vacuum subsystems with full
energy (an effective $\Lambda$--term) of the order of $\Lambda\sim
3 \cdot 10^{-47}\hbar^{-3}$ GeV$^4$.   The macroscopic Einstein
equations containing all sources mentioned above read
 \begin{equation}
 \begin{array}{c}
\displaystyle
R_0^0-\frac12R=\varkappa\varepsilon_{tot}\quad\to\quad
H^2=\frac{1}{48}D+
\frac{1}{12}W_1+\frac{\varkappa}{3}\left(\Lambda+\frac{M}{a^3}\right)\
,
 \end{array}
 \label{8.1}
\end{equation}
\begin{equation}
 \begin{array}{c}
\displaystyle
R_0^0-\frac14R=\frac{3\varkappa}{4}\left(\varepsilon_{tot}+p_{tot}\right)\quad\to\quad
 \dot H=-\frac{1}{16}D-\frac16W_1-\frac{\varkappa M}{2a^3}\ .
 \end{array}
 \label{8.2}
\end{equation}
Equation (\ref{8.2}) should be differentiated with respect to
time, and then $\dot D$ from (\ref{5.7}) should be substituted
into the result of differentiation. These operations produce one
more equation
\begin{equation}
 \begin{array}{c}
\displaystyle
 \ddot
H=H\left(\frac{3}{8}D+W_1+\frac{3\varkappa M}{2a^3}\right)+\frac{1}{12}\dot W_1\ .
 \end{array}
 \label{8.3}
\end{equation}

The BBGKY chain (\ref{5.7}) --- (\ref{5.9}) takes into account the
interaction of gravitons with the self--consistent classical
gravitational field which is represented by the Hubble function
and its derivatives. According to Einstein equations (\ref{8.1})
--- (\ref{8.3}), a self--consistent gravitational field is created
by gravitons and other components of cosmological medium, i.e. by
the matter and non--gravitational vacuum subsystems. Therefore,
the self--consistent gravitational field is a way of describing of
significantly non--linear properties of the system that are the
result of gravitational interaction of elements of the system.
After excluding higher derivatives of the metric from the BBGKY
chain (\ref{5.8}) and (\ref{5.9}), the true non--linear character
of the theory emerges. Substitution of (\ref{8.1}) --- (\ref{8.3})
into (\ref{5.8}) and (\ref{5.9}) gives the non--linear
representation of BBGKY chain:
\begin{equation}
\begin{array}{c}
\displaystyle  \dot D+6HD+4\dot W_1+16HW_1=0\ , \\[3mm]
 \displaystyle
\stackrel{...}{W}_n +3(2n+3)H\ddot{W}_n+
\left[\frac{1}{16}(4n^2+6n+3)D+(n+1)^2W_1+(8n^2+18n+9)\frac{\varkappa
M}{2a^3}+2(2n^2+6n+3)\varkappa\Lambda\right]\dot{W}_n
+ \\[5mm]
\displaystyle +\frac{n}{3}\left\{\frac12\dot
W_1+H\left[\frac{n^2}{2}D+(2n^2+3n+3)W_1+(8n^2+18n+9)\frac{\varkappa
M}{a^3}+4(2n^2+9n+9)\varkappa\Lambda\right]\right\}W_n+
\\[5mm]
\displaystyle +
4\dot{W}_{n+1}+8(n+2)HW_{n+1}=0, \\[5mm]
\displaystyle
\displaystyle n=1,\,...,\,\infty\, .
\end{array}
\label{8.4}
\end{equation}

In the general case, the system of equations (\ref{8.2}) and
(\ref{8.4}) (to which the definition $\dot a/a=H$ is added) should
be solved numerically with initial conditions determined by the
scale factor, moments of the spectral function and their
derivatives
 \begin{equation}
\displaystyle a(0);\quad D(0); \quad W_n(0),\quad \dot
W_n(0),\quad \ddot W_n(0),\\[5mm]
\displaystyle n=1,\,...,\,\infty\ .
 \label{8.5}
\end{equation}
The initial condition for the Hubble function should be calculated
via the equation of the constraint (\ref{8.1})
\begin{equation}
\displaystyle H(0)=+\sqrt{\frac{1}{48}D(0)+\frac{1}{12}W_1(0)+
\frac13\varkappa\Lambda+\frac{\varkappa M}{3a^3(0)}}\ .
 \label{8.6}
\end{equation}
Any solution of equations (\ref{8.2}) and (\ref{8.4}), which
corresponds to initial conditions (\ref{8.5}), (\ref{8.6}),
satisfies the identity which is local in time
\begin{equation}
\displaystyle H^2(t)=\frac{1}{48}D(t)+\frac{1}{12}W_1(t)+
\frac13\varkappa\Lambda+\frac{\varkappa M}{3a^3(t)}\ ,
 \label{8.7}
\end{equation}

\section{Exact Solutions: Dynamics of Operators and Structure of State Vectors}\label{exact}

In this section, we get the exact solutions for field operators
and expressions for the state vectors that correspond to exact
analytical solutions of BBGKY chain (\ref{5.40}) and
(\ref{5.42}). Microscopic studies of exact solutions allow
greater detail to identify their physical content. Solutions
(\ref{5.40}) and (\ref{5.42}) are formed as a result of
certain spectrally dependent correlations between graviton and
ghost contributions to the observables. These are full
graviton--ghost compensation of contributions of zero oscillations
(one--loop finiteness); full compensation of contributions in all
parts of the spectrum, except the region of quasi--resonant (QR)
and spatially homogeneous (SH) modes; incomplete compensation of
contributions of QR and SH modes with non--zero occupation
numbers; correlations between excitation levels and
graviton--ghost contents of  QR and SH modes, and, finally, some
correlations of phases in quantum superpositions of graviton and
ghost state vectors.

The physical nature of solution (\ref{5.41}) turned out to be
unexpected and nontrivial. In Section \ref{inst} it will be shown
that mathematically this solution describes instanton condensate,
which physically corresponds to the system of correlated
fluctuations arising during tunneling of graviton--ghost medium
between states with fixed difference of graviton and ghost
numbers. We explain also that self--polarized graviton--ghost
condensate in the De Sitter space also allows instanton
interpretation.

\subsection{Condensate of Constant Conformal Wavelength}\label{K}

Let us consider the solution (\ref{5.40}) for $C_{3g}=0, \,
C_{4g}=k_0^2$:
 \begin{equation}
 \begin{array}{c}
\displaystyle
H^2=\frac{k_0^2}{a^2}\ln\frac{a_0}{a}\ , \qquad a=a_0\exp\left(-\frac{k_0^2\eta^2}{4}\right)\ .
 \end{array}
 \label{6.1}
\end{equation}
The graviton wave equation with the (\ref{6.1}) background reads
 \begin{equation}
 \begin{array}{c}
\displaystyle \hat\psi''_{{\bf k}\sigma}-k^2_0\eta\hat\psi'_{{\bf k}\sigma}+
k^2\hat\psi_{{\bf k}\sigma}=0\ .
 \end{array}
 \label{6.2}
\end{equation}
The equation for the ghosts looks similar. Fundamental solutions
of equation (\ref{6.2}) are degenerate hypergeometric functions.
It is unnecessary to consider those solutions for all possible
values of the parameter $k^2$. First of all, it is obvious that
the macroscopic observables can be formed only by simplest
hypergeometric functions. Values   $k^2$  that are $k^2=0$
(spatially uniform modes) and  $k^2=k_0^2$ (quasi--resonant modes)
stand out. For all other modes there is a precise graviton--ghost
compensation. The reason why it is a mathematically possible
follows from the general formulas  (\ref{3.69}), (\ref{3.72}),
(\ref{3.73}) \footnote{Formally, all modes except with  $k^2=0$
and $k^2=k_0^2$, look like "frozen"\ degrees of freedom, which are
excluded from consideration by the model postulate. By virtue of
the principle of uncertainty, postulates of this type are outside
the formalism of quantum field theory. We want to emphasize that
in the finite one--loop quantum gravity there is no need to
"freeze"\ degrees of freedom not participating in the formation of
particular exact solutions. Instead of mathematically incorrect
operation of "freezing"\ , the formalism of the theory offers
mathematically consistent operations of graviton--ghost
compensations.}.

Let us start with quasi--resonant modes. Exact solutions of the
equation (\ref{6.1}) and similar equation for ghosts for
$k^2=k_0^2$ read
\begin{equation}
\begin{array}{c}
\displaystyle
\hat\psi_{{\bf k}\sigma}=
\frac{\sqrt{4\varkappa\hbar k_0}}{a_0}\left[-\eta\left(\hat Q_{{\bf k}\sigma}+
k_0\hat P_{{\bf k}\sigma}\int\limits_0^\eta e^{k_0^2\eta^2/2}d\eta\right)+\frac{\hat P_{{\bf k}\sigma}}{k_0}e^{k_0^2\eta^2/2}\right]=
-\sqrt{\frac{16\varkappa\hbar}{k_0a_0^2}}\left[\hat Q_{{\bf k}\sigma}+
\hat P_{{\bf k}\sigma}F(a)\right]\ln^{1/2}\frac{a_0}{a} \ ,
\end{array}
 \label{6.3}
\end{equation}
\begin{equation}
\begin{array}{c}
\displaystyle
\hat\vartheta_{{\bf k}}=\frac{\sqrt{4\varkappa\hbar k_0}}{a_0}\left[-\eta\left(\hat q_{{\bf k}}+
k_0\hat p_{{\bf k}}\int\limits_0^\eta e^{k_0^2\eta^2/2}d\eta\right)+\frac{\hat p_{{\bf k}}}{k_0}e^{k_0^2\eta^2/2}\right]=
-\sqrt{\frac{16\varkappa\hbar}{k_0a_0^2}}\left[\hat q_{{\bf k}}+
\hat p_{{\bf k}}F(a)\right]\ln^{1/2}\frac{a_0}{a} \ ,
\end{array}
 \label{6.4}
\end{equation}
where $\hat Q_{{\bf k}\sigma},\, \hat P_{{\bf k}\sigma}$ and $\hat
q_{{\bf k}},\, \hat p_{{\bf k}}$ are operators whose properties
are defined in (\ref{3.53}), (\ref{3.54}), (\ref{3.50});
\[
 \displaystyle F(a)=a_0^2\int\limits_{a_0}^a\frac{da}{a^3\displaystyle\ln^{1/2}\frac{a_0}{a}}-
 \frac{a_0^2}{2a^2\displaystyle\ln^{1/2}\frac{a_0}{a}}\ .
 \]

Note that one of fundamental solutions to equation (\ref{6.2}) is
the Hermite polynomial $H_1(\eta)$, which corresponds to positive
eigenvalue $k^2/k_0^2=1$. In the reproduction of solutions
(\ref{5.41}.I) at the microscopic level, this fact is crucial. We
will show that the choice of a state vector, satisfying the
condition of coherence leads to the fact that only this solution
takes part in the formation of the observables. The second
solution, containing a function $F(a)$, is a mathematical
structure that does not correspond to the exact solution to the
BBGKY chain.

Averaging of bilinear forms of operators (\ref{6.3}) and
(\ref{6.4}) over the state vector of the general form leads to the
following spectral function
\begin{equation}
\begin{array}{c}
\displaystyle
W_{{\bf k}}=\sum_\sigma\langle\Psi_g|\hat\psi_{{\bf k}\sigma}^+\hat\psi_{{\bf k}\sigma}
|\Psi_g\rangle-2\langle\Psi_{gh}|\hat\vartheta_{{\bf k}}^+\hat\vartheta_{{\bf k}}
|\Psi_{gh}\rangle =
\frac{16\varkappa\hbar}{ k_0a_0^2}\left[A_{\bf k}+B_{\bf k}F^2(a)+
C_{\bf k}F(a)\right]\ln\frac{a_0}{a}
\ .
\end{array}
 \label{6.5}
\end{equation}
The constants appearing in (\ref{6.5}) are expressed through
averaged quadratic forms of operators of generalized coordinates
and momentums:
\begin{equation}
\begin{array}{c}
\displaystyle A_{\bf k}=
\sum_\sigma\langle\Psi_g|\hat Q^+_{{\bf k}\sigma}Q_{{\bf k}\sigma}|\Psi_g\rangle-
2\langle\Psi_{gh}|\hat q_{{\bf k}}^+\hat q_{{\bf k}}|\Psi_{gh}\rangle\ ,
\qquad B_{\bf k}=
\sum_\sigma\langle\Psi_g|\hat P^+_{{\bf k}\sigma}P_{{\bf k}\sigma}|\Psi_g\rangle-
2\langle\Psi_{gh}|\hat p_{{\bf k}}^+\hat p_{{\bf k}}|\Psi_{gh}\rangle\ ,
\\[5mm]
\displaystyle  C_{\bf k}=
\sum_\sigma\langle\Psi_g|\left(\hat Q^+_{{\bf k}\sigma}P_{{\bf k}\sigma}+
\hat P^+_{{\bf k}\sigma}Q_{{\bf k}\sigma}\right)|\Psi_g\rangle-
2\langle\Psi_{gh}|\left(\hat q_{{\bf k}}^+\hat p_{{\bf k}}+
p_{{\bf k}}^+\hat q_{{\bf k}}\right)|\Psi_{gh}\rangle\ .
\end{array}
 \label{6.6}
\end{equation}
Following the transition to the ladder operators in formulas
(\ref{3.54}) and calculations, carried out similar to (\ref{3.66})
--- (\ref{3.74}), we get
\begin{equation}
\begin{array}{c}
\displaystyle
A_{\bf k}=2\langle n_{k(g)}\rangle(1+\zeta^{(g)}_k\cos\varphi_k)-
2\langle n_{k(gh)}\rangle(1+\zeta_k^{(gh)}\cos\chi_k)\ ,
\\[5mm]
\displaystyle B_{\bf k}=2\langle n_{k(g)}\rangle(1-\zeta^{(g)}_k\cos\varphi_k)-
2\langle n_{k(gh)}\rangle(1-\zeta_k^{(gh)}\cos\chi_k)\ ,
\\[5mm]
\displaystyle C_{\bf k}=0\ .
\end{array}
 \label{6.7}
\end{equation}
For sake of simplicity, in (\ref{6.7}) average numbers of ghosts
and anti--ghosts are assumed to be the same:  $\langle
n_{k(gh)}\rangle=  \langle \bar n_{k(gh)}\rangle$.

Let us go back to the expression (\ref{6.5}). Obviously, the
spectral function (\ref{6.5}) creates moments (\ref{5.16}) only if
$B_{\bf k}=C_{\bf k}=0$. The condition $C_{\bf k}=0$ is satisfied
automatically as a consequence of isotropy of macroscopic state,
i.e. because of independence of average occupation numbers of the
direction of vector ${\bf k}$. $B_{\bf k}=0$ imposes the
conditions on amplitudes and phases of quantum superpositions of
state vectors with different occupation numbers. It is necessary
to draw attention to the fundamental fact: {\it the solution under
discussion does not exist, if phases of superpositions are
random.} Indeed, averaging the expression (\ref{6.7}) over phases,
we see that condition $B_{\bf k}=0$ is satisfied only if $\langle
n_{k(g)}\rangle=\langle n_{k(gh)}\rangle$. The last equality
automatically leads to $A_{\bf k}=0$, i.e. which eliminates the
nontrivial solution.

Thus, the condition of the existence of the solution under
discussion is the coherence of the quantum state. It is easy to
notice (see (\ref{3.74})), that equality  $B_{\bf k}=0$, as a
condition of coherence, is satisfied for zero phase difference of
states with the neighboring occupation numbers of gravitons and
ghosts:
\begin{equation}
\begin{array}{c}
\displaystyle \zeta^{(g)}_k\cos\varphi_k=\zeta_k^{(gh)}\cos\chi_k=1 \quad \to \quad
\zeta^{(g)}_k=\zeta_k^{(gh)}=1,\qquad \cos\varphi_k=\cos\chi_k=1\ .
\end{array}
\label{6.8}
\end{equation}
Taking into account (\ref{6.8}), we get the following final
expression (\ref{6.9}) for the spectral function of
quasi--resonant gravitons and ghosts
\begin{equation}
\begin{array}{c}
\displaystyle
W_{{\bf k}}\equiv W_k=
\frac{64\varkappa\hbar}{ k_0a_0^2}\left(\langle n_{k(g)}\rangle-
\langle n_{k(gh)}\rangle\right)\ln\frac{a_0}{a}
\ .
\end{array}
 \label{6.9}
\end{equation}
In calculating moments, summation over wave numbers is replaced by
integration. Account is taken of that the spectrum as the
delta--form with respect to the modulus of $k=|{\bf k}|$.  Also a
new parameter  $N_g$ is introduced where $N_g$ is the difference
of numbers of gravitons and ghosts in the unit volume of $V=\int
d^3x=1$  in the 3--space, which is conformally similar to the
3--space of expanding Universe. Index "$g$"\ in designation of
$N_g$  parameter indicates the dominance of gravitons in
quasi--resonant modes. In accordance with this definition, the
following replacement is performed
\begin{equation}
\displaystyle \langle n_{k(g)}\rangle-\langle n_{k(gh)}\rangle \to
\frac{2\pi^2}{k^2}N_g\delta(k-k_0)\ ,
 \label{6.10}
\end{equation}
Results of calculating of moments are equated to the relevant
expressions of (\ref{5.11}) and (\ref{5.16}), which were obtained
by exact solution of the BBGKY chain:
\begin{equation}
\begin{array}{c}
\displaystyle W_n(g4)=\frac{1}{2\pi^2a^{2n}}\int\limits_0^\infty W_kk^{2n+2}dk=
\frac{64\varkappa\hbar N_g k_0^{2n-1}}{a^2_0a^{2n}}\ln\frac{a_0}{a}=
\frac{24k_0^{2n}}{a^{2n}}\ln\frac{a_0}{a}\ ,
\\[5mm]
\displaystyle D(g4)=\frac{1}{a^2}\left(W_0''+2\frac{a'}{a}W_0'\right)=
-\frac{128\varkappa\hbar N_g k_0}{a^2_0a^{2}}\ln\frac{a_0}{e^{1/4}a}=
-\frac{48k_0^{2}}{a^{2}}\ln\frac{a_0}{e^{1/4}a}\ .
\end{array}
\label{6.11}
\end{equation}
In accordance with (\ref{6.11}), there is a relation between
parameters $k_0,\ a_0$   and $N_g$ that  appear in the microscopic
solution
\begin{equation}
 \displaystyle N_g=\frac{3k_0a_0^2}{8\varkappa\hbar}\ .
 \label{6.12}
\end{equation}
Recall that in the solution under discussion, the Universe was
born in singularity, expands to a state with a maximum scale
factor  $a_{max}=a_0$, and then is again compressed to the
singularity. In this scenario, value $a_0$ can be defined as the
size of the Universe, accessible for observation in the end stage
of expansion. As can be seen from (\ref{6.12}), if  $a_0$ is a
macroscopic value, the difference in numbers gravitons and ghosts
$N_g\gg 1$ is also a macroscopic value.

Contributions of SH modes to the expressions for the moments are
shown in (\ref{5.11}), and the relation between the parameters
$C_{2g}$ and $C_{4g}$ is shown in (\ref{5.15}). As a part of the
microscopic approach, the construction of exact solutions for
these modes is performed by the method of transaction to the
limit, described at the end of Section \ref{lgw}. The parameter of
spatially homogeneous condensate is introduced similarly to
(\ref{6.10}):
\begin{equation}
\begin{array}{c}
\displaystyle
\langle n_{0(gh)}\rangle(1+\zeta^{(gh)}_0\cos\varphi_0)-
\langle n_{0(g)}\rangle(1+\zeta^{(g)}_0\cos\chi_0)\quad
\to \quad
\frac{2\pi^2}{k^2}N_{gh}\delta(k-\kappa_0)\ ,\qquad \kappa_0\to 0\ .
\end{array}
 \label{6.13}
\end{equation}
The index "$gh$"\ in $N_{gh}>0$ indicates the dominance of ghosts
over the gravitons in the spatially homogeneous  condensate. The
moments are:
\begin{equation}
\displaystyle W_1(g2)=-\frac{16\varkappa\hbar k_1 N_{gh}}{a_1^2a^2}\ ,
\qquad D(g2)=\frac{32\varkappa\hbar k_1N_{gh}}{a_1^2a^2}\ .
 \label{6.14}
\end{equation}
Definitions of parameters $k_1$ and  $a_1$ are given in
(\ref{4.21}). The energy density and pressure of the system of QR
and SH modes are given by (\ref{6.11}) and (\ref{6.14}):
\begin{equation}
 \begin{array}{c}
 \displaystyle
\varkappa\varepsilon_g=\frac{8\varkappa\hbar
k_0N_g}{a_0^2a^2}\ln\frac{a_0}{a}+
\frac{2\varkappa\hbar}{a^2}\left(\frac{k_0N_g}{a^2_0}-
\frac{k_1N_{gh}}{a^2_1}\right)=
\frac{8\varkappa\hbar
k_0N_g}{a_0^2a^2}\ln\frac{a_0}{a}\ ,
\\[5mm]
\displaystyle \varkappa p_g=-\frac{8\varkappa\hbar
k_0N_g}{3a_0^2a^2}\ln\frac{a_0}{ea}-
\frac{2\varkappa\hbar}{3a^2}\left(\frac{k_0N_g}{a^2_0}-
\frac{k_1N_{gh}}{a^2_1}\right)=
-\frac{8\varkappa\hbar
k_0N_g}{3a_0^2a^2}\ln\frac{a_0}{ea}\ .
\end{array}
 \label{6.15}
\end{equation}
In formulas (\ref{6.15}), the terms in brackets are eliminated by
the condition (\ref{5.15}), which is rewritten in terms of
microscopic parameters
\begin{equation}
\displaystyle \frac{k_0N_g}{a^2_0}=\frac{k_1N_{gh}}{a^2_1}\ .
\label{6.16}
\end{equation}
The solution (\ref{6.15}), (\ref{6.16}) describes a quantum
coherent condensate of quasi--resonant modes with graviton
dominance, parameters of which are consistent with that of
spatially homogeneous  condensate with the ghost dominance.

\subsection{Condensate of Constant Physical Wavelength}\label{S}

The De Sitter solution for plane isotropic Universe reads
\begin{equation}
\displaystyle a=a_0e^{Ht}=-\frac{1}{H\eta}\ , \qquad H=const\ .
\label{6.17}
\end{equation}
For the background (\ref{6.17}), the gravitons and ghost equations
and their solutions read
\begin{equation}
 \begin{array}{c}
 \displaystyle
\hat \psi''_{{\bf k}\sigma}-\frac{1}{\eta}\hat \psi'_{{\bf k}\sigma}+
k^2\hat \psi_{{\bf k}\sigma}
=0\ ,
\qquad
\hat \psi_{{\bf k}\sigma}=\frac{1}{a}\sqrt{\frac{2\varkappa\hbar}{k}}
 \left[c_{{\bf k}\sigma}f(x)+c^+_{{\bf -k}-\sigma}f^*(x)\right]\ ,
\end{array}
 \label{6.18}
\end{equation}
\begin{equation}
 \begin{array}{c}
 \displaystyle
\hat \vartheta''_{{\bf k}}-\frac{1}{\eta}\hat \vartheta'_{{\bf k}}+k^2\hat \vartheta_{{\bf k}}=0\ ,
\qquad
\hat \vartheta_{{\bf k}}=\frac{1}{a}\sqrt{\frac{2\varkappa\hbar}{k}}
 \left[a_{{\bf k}}f(x)+b^+_{{\bf -k}}f^*(x)\right]\ ,
\end{array}
 \label{6.19}
\end{equation}
where
\[
\displaystyle f(x)=\left(1-\frac{i}{x}\right)e^{-ix}\ , \qquad x=k\eta\ .
\]
Ladder operators in (\ref{6.18}), (\ref{6.19}), have the standard
property of (\ref{3.47}), (\ref{3.50}), which allow their use of
in constructing build basic vectors for the Fock space from which
the general state vectors are constructed.

The self--consistent dynamics of gravitons and ghosts in the De
Sitter space are not trivial in the sense that the averaged
bilinear forms of operators (\ref{6.18}), (\ref{6.19}) which are
explicitly and essentially depending on time, must lead to
time--independent macroscopic observables. It must be emphasized,
that the existence of such, at first glance unlikely solution, is
guaranteed by the existence of the solution for the BBGKY chain.
The key to the solution lies in the structure of the state vectors
of gravitons and ghosts.

Substitution of operator functions (\ref{6.18}), (\ref{6.19}) into
the general expression for the moments (\ref{5.6}) yields:
\begin{equation}
  \begin{array}{c}
 \displaystyle
 W_n=\frac{2\varkappa\hbar}{\pi^2}H^{2n+2}\int\limits_0^{\infty}dxx^{2n+1}
 \biggl\{U_{{\bf k}(wave)}|f(x)|^2 +U_{{\bf k}(cr)}[f^*(x)]^2+U_{{\bf k}(ann)}[f(x)]^2\biggr\}
 \ ,
\end{array}
 \label{6.20}
 \end{equation}
where
 \begin{equation}
 \begin{array}{c}
 \displaystyle
  N_{{\bf k}}\equiv U_{{\bf k}(wave)}=
  \sum_\sigma\langle \Psi_g|c^+_{{\bf k}\sigma}c_{{\bf
 k}\sigma}|\Psi_g\rangle-
  \langle \Psi_{gh}|a^+_{{\bf k}}a_{{\bf
 k}}|\Psi_{gh}\rangle-\langle \Psi_{gh}|b^+_{{\bf k}}b_{{\bf
 k}}|\Psi_{gh}\rangle\ ;
\end{array}
\label{6.21}
 \end{equation}
\begin{equation}
 \begin{array}{c}
\displaystyle U^*_{{\bf k}}\equiv U_{{\bf k}(cr)}=
\frac12\sum_\sigma\langle \Psi_g| c^+_{{\bf
k}\sigma}c^+_{{\bf
 -k}-\sigma}|\Psi_g\rangle-\langle\Psi_{gh}| a^+_{{\bf
k}}b^+_{{\bf
 -k}}|\Psi_{gh} \rangle\ ;
\\[5mm]
\displaystyle U_{{\bf k}}\equiv U_{{\bf k}(ann)}=
\frac12\sum_\sigma\langle \Psi_g| c_{{\bf
-k}-\sigma}c_{{\bf
 k}\sigma}|\Psi_g\rangle-\langle \Psi_{gh}| b_{{\bf
-k}}a_{{\bf
 k}}|\Psi_{gh}\rangle\equiv U^*_{{\bf k}(cr)}\ .
\end{array}
\label{6.22}
 \end{equation}
Here $U_{{\bf k}(wave)}$  is the spectral parameter of quantum
waves, which become real gravitons if $k\eta\gg 1$; $U_{{\bf
k}(cr)}$, $U_{{\bf k}(ann)}$ are the spectral parameters of
quantum fluctuations that emerge in the processes of graviton (and
ghost) creation from the vacuum and graviton (and ghost)
annihilation to the vacuum.

Obviously, at the first stage of calculations we assume that the
averaging in (\ref{6.21}), (\ref{6.22}) is conducted over the
state vectors of the general form (\ref{3.48}), (\ref{3.51}). This
allows us to go to formulas (\ref{3.67}), (\ref{3.68}) or
(\ref{3.72}) --- (\ref{3.74}). Then it is necessary to take into
account that the moments  $W_n$ must not depend on time, and that
they also should be free of divergences. When analyzing the
conditions for these demands, the specific form of the expression
(\ref{6.20}) plays an important part. The measure of integration
and the dependence of field operators on the wave number and time
can be represented in the terms of the variable $x=k\eta$. A
separate (additional) dependence on the wave number can be
connected with the structure of spectral parameters. After
substitution of the variable $k=x/\eta$   in the equation
(\ref{6.21}), it is seen that the first term in (\ref{6.20}) is
time-independent only if $U_{{\bf k}(wave)}$ is independent on the
wave number. This means that the graviton and ghost spectra must
be flat. However, with the flat spectrum there is danger of
divergences: if $U_{{\bf k}(wave)}=const\ ({\bf k})\ne 0$, then
the first integral in (\ref{6.20}) does not exist, because
$|f(x)|^2\to 1$ with $x\to \infty$.

The divergences can be avoided only with exact compensation of
contributions from gravitons and ghosts to the spectral parameter
$U_{{\bf k}(wave)}$. Let us point out, that in that case we are
not talking about zero oscillations but about the contributions
from the states with non--zero occupation numbers. The
compensation condition leading to $U_{{\bf k}(wave)}=0$ is:
\begin{equation}
\displaystyle  |\mathcal{C}_{n_{{\bf k}\sigma}}|=|\mathcal{A}_{n_{{\bf
k}}}|=|\mathcal{B}_{n_{-{\bf k}}}|\ .
\label{6.23}
\end{equation}
The result (\ref{6.23}) has a simple physical interpretation. The
quantum waves of gravitons and ghosts with the equation of state
which differs from $p=-\varepsilon$ can not be carriers of energy
in the De Sitter space with the self--consistent geometry. The
total energy of quantized waves is equal to zero due to exactly
the same number of gravitons and ghosts in all regions of the
spectrum:
\begin{equation}
\displaystyle  \langle n_{{\bf k}\sigma_1}\rangle +\langle n_{{\bf k}\sigma_2}\rangle=
\langle n_{{\bf k}}\rangle+\langle\bar n_{{\bf k}}\rangle\ .
\label{6.24}
\end{equation}
With equal polarizations of gravitons and the equality of numbers
of ghosts and anti--ghosts, it follows from (\ref{6.24})  that
$\langle n_{{\bf k}(g)}\rangle=\langle n_{{\bf k}(gh)}\rangle$.
Exact equality of the average number of gravitons and ghost is a
characteristic feature of the De Sitter space with the
self--consistent geometry. Let us mention that for the solution
discussed in the previous section \ref{K}, that equality is absent
in principle. It means that different solutions have different
microscopic structures of the graviton--ghost condensate.

Based on the reasoning analogous to the one described above,
spectrum parameters $U_{{\bf k}(cr)}$, $U_{{\bf k}(ann)}$  also
must not depend on the wave vector ${\bf k}$. However, the
corresponding integrals in the second and third terms of
(\ref{6.20}) are not divergent. The absence of divergences is due
to the fact that with $x\to \infty$  the integration is taken over
the fast oscillating functions $\sim e^{\pm 2ix}$. To calculate
these integrals, they should be additionally defined as follows:
\begin{equation}
 \begin{array}{c}
 \displaystyle
 \lim_{\zeta\to \ 0}\int\limits_0^\infty dxx^{2n\pm 1}e^{-(\zeta-2i)x}=
 \mp (-1)^n\frac{(2n\pm 1)!}{2^{2n+1\pm 1}}\ ,
 \qquad
 2i\lim_{\zeta\to \ 0}\int\limits_0^\infty dxx^{2n}e^{-(\zeta-2i)x}=
  (-1)^{n+1}\frac{(2n)!}{2^{2n}}\ .
\end{array}
\label{6.25}
 \end{equation}
At every instant of time, the procedure of re-definitions of
integrals (\ref{6.25}) selects the contributions from virtual
gravitons and ghosts with a characteristic wavelength (\ref{5.28})
and eliminate the contributions of all other graviton--ghost
modes. This redefining procedure provides the existence of
recursive relations (\ref{5.27}) in the exact solution of the
BBGKY chain.

Thus, in (\ref{6.20}) we have a flat spectrum of gravitons and
ghosts, $U_{{\bf k}(wave)}\equiv 0$,  $U_{{\bf k}(cr)}=U^*_{{\bf
k}(ann)}=U=const(k)$. The expression for the spectral parameter
takes the form:
\begin{equation}
 \begin{array}{c}
 \displaystyle  U=\left(\sum_n\mathcal{C}^*_{n+1}\mathcal{C}_n\sqrt{n+1}\right)^2-
 \left(\sum_n\mathcal{A}^*_{n+1}\mathcal{A}_n\sqrt{n+1}\right)
 \left(\sum_n\mathcal{B}^*_{n+1}\mathcal{B}_n\sqrt{n+1}\right)\ ,
 \\[5mm]
 \displaystyle  |\mathcal{C}_n|=|\mathcal{A}_n|=|\mathcal{B}_n|\equiv \sqrt{\mathcal{P}_n} \ ,
 \end{array}
 \label{6.26}
 \end{equation}
where  $\mathcal{P}_n$ is a normalized statistical distribution.
The average value of the number of gravitons and ghosts, having
the wavelength in the vicinity of characteristic values
(\ref{5.28}), are calculated by the formula
\begin{equation}
\displaystyle\langle n_{g}\rangle=\langle n_{gh}\rangle=\langle n\rangle=
\sum_{n=0}^\infty n\mathcal{P}(n)\ .
\label{6.27}
 \end{equation}
Using the Poisson distribution in (\ref{6.26}), (\ref{6.27}), the
values of integrals (\ref{6.25}) and the formulas (\ref{3.73}),
(\ref{3.74}), we get the moments
\begin{equation}
 \begin{array}{c}
 \displaystyle
 D=-\frac{12\varkappa\hbar N_g}{\pi^2}H^4\ ,
 \qquad 
 W_n=\frac{(-1)^{n+1}}{2^{2n}}
 (2n-1)!(2n+1)(n+2)\times
  \frac{2\varkappa\hbar N_g}{\pi^2}H^{2n+2} ,\quad  n\geqslant 1\ ,
\end{array}
 \label{6.28}
 \end{equation}
where
\begin{equation}
 \begin{array}{c}
\displaystyle N_g= \langle n\rangle(\zeta_{g}\cos\varphi- \zeta_{gh}\cos\chi)\ ,
\\[5mm]
\displaystyle \varphi=\varphi_{n_{{\bf k}\sigma}}-\varphi_{n_{{\bf k}+1,\sigma}},\qquad \chi=\chi_{n_{{\bf
k}}}-\chi_{n_{{\bf k}+1}}\ .
\end{array}
\label{6.29}
\end{equation}
Zero moment $W_0$, which has an infrared logarithmic singularity,
is not contained in the expressions for the macroscopic
observables, and for that reason, is not calculated. In the
equation for  $W_0$, the functions are differentiated in the
integrand and the derivatives are combined in accordance with the
definition $D=\ddot W_0+3H\dot W_0$. At the last step the
integrals that are calculated, already posses no singularities.

Averaging of the parameter (\ref{6.29}) over the phases yields
$N_g=0$.  Therefore {\it the solution under discussion does not
exist if the superposition of the phases are random.} The
coherence  of the quantum ensemble, i.e. the correlation of phases
in the quantum superposition of the basic vectors, corresponding
to the different occupation numbers, points to the fact that the
medium is in the graviton--ghost condensate state. The gravitons
are dominant in the condensate if $N_g>0$, and the ghosts are
dominant if $N_g<0$.

The duality of the condensate and the indeterminate sign of the
$\Lambda$--term create different evolutional scenarios. Of course,
all these scenarios are present in the expression (\ref{5.33}),
which is obtained as a solution of the macroscopic Einstein
equation (\ref{5.32}). In addition to the scenarios described in
the Section \ref{Sitt}, we will show the possibility of strong
renormalization of energy of non--gravitational vacuum subsystems
by the energy of the graviton--ghost
condensate\footnote{Mechanisms that are able to drive the
cosmological constant to zero have been discussed for decades (see
\cite{23,79} for a review). Any particular scenarios were
considered in \cite{PBV2, Dolgov1983, Ford1987, R2, ST}.
}.

We have in mind a situation, in which the modulus of
$\Lambda$--term exceeds the density of vacuum energy in the
asymptotic state of the Universe by many orders of magnitude:
\begin{equation}
\displaystyle \frac{|\Lambda|}{\varepsilon_{vac}^{(\infty)}}\equiv
\frac{\varkappa |\Lambda|}{3H^2_\infty}={\mathcal{N}}\gg 1\ ,
\label{6.30}
 \end{equation}
where ${\mathcal{N}}$ is a huge macroscopic number. From
(\ref{5.33}) it follows that the effect of  strong renormalization
takes place if
\begin{equation}
\begin{array}{c}
\displaystyle \frac{\Lambda}{N_g}<0\ , \qquad
|N_g|\gg \frac{6\pi^2}{\varkappa^2\hbar |\Lambda|}\ ,
\qquad  \varepsilon_{vac}^{(\infty)}\simeq 2\pi\sqrt{\frac{6|\Lambda|}{\varkappa^2\hbar |N_g|}}\ .
\end{array}
\label{6.31}
\end{equation}
Let us mention that the strong renormalization of the positive
$\Lambda$--term is provided by a condensate in which the ghosts
are dominant, and for the negative  $\Lambda$--term --- by a
condensate for which the gravitons are dominant.

For clarity and for the evaluations let us introduce the Plank
scale $M_{Pl}=(8\pi\hbar/\varkappa)^{1/2}=1.22\cdot 10^{19}$ GeV,
the scale of  $\Lambda$--term
$M_\Lambda=(\hbar^3|\Lambda|)^{1/4}$, and the scale of the density
of Dark Energy in the asymptotical state of the Universe,
$M_{{\scriptscriptstyle D}{\scriptscriptstyle
E}}=(\hbar^3\varepsilon_{vac}^{(\infty)})^{1/4}$. We discuss the
case when $M_{{\scriptscriptstyle D}{\scriptscriptstyle E}}\ll
M_\Lambda$.

If non--gravitational contributions to $\Lambda$--term are
self--compensating, then a realistic estimate of the
$M_\Lambda$--scale can be based on the Zeldovich remark \cite{22}.
According to \cite{22}, non--gravitational $\Lambda$--term is
formed by gravitational exchange interaction of quantum
fluctuations on the energy scale of hadrons. In terms of
contemporary understanding of hadron's vacuum, the focus should be
on non--perturbative fluctuations of quark and gluon fields,
forming a quark--gluon condensate (see \cite{38, P}). In
this case, $\Lambda$--term is expressed only through the minimum
and maximum scales of particle physics which are the QCD scale
$M_{{\scriptscriptstyle Q}{\scriptscriptstyle
C}{\scriptscriptstyle D}}\simeq 215\ \text{MeV}$ and Planck scale
$M_{Pl}=1.22\cdot 10^{19}\ \text{GeV}$:
\begin{equation}
\displaystyle \hbar^3|\Lambda|=M_\Lambda^4=
\frac{M_{{\scriptscriptstyle Q}{\scriptscriptstyle C}{\scriptscriptstyle D}}^6}{M^2_{Pl}}\simeq
10^{-42}\ \text{GeV}^4\ .
\label{6.32}
\end{equation}
In terms of these scales, it is turns out that a large number of
${\mathcal{N}}=M_\Lambda^4/M_{{\scriptscriptstyle
D}{\scriptscriptstyle E}}^4\sim 10^5$, which is defined in
(\ref{6.30}), can be obtained by the huge number of $|N_g|^{1/2}$,
for the same number of orders of magnitude greater than the ratio
$(M_{Pl}/M_\Lambda)^2$. Indeed, choosing ${\mathcal{N}}$, we find
the value of $|N_g|$, which determines the ratio of vacuum energy
density to the true cosmological constant in the asymptotic state:
\begin{equation}
\displaystyle {\mathcal{N}}=\frac{M_\Lambda^2}{M^2_{Pl}}\sqrt{\frac{2|N_g|}{3}}\ .
\label{6.33}
\end{equation}
The vacuum energy density of asymptotical state is calculated as
follows
\begin{equation}
\begin{array}{c}
\displaystyle \varepsilon_{vac}^{(\infty)}\simeq
\hbar^{-3}M_{Pl}^2M_\Lambda^2\sqrt{\frac{3}{2|N_g|}}\ .
   \end{array}
 \label{6.34}
 \end{equation}

Thus, {\it the macroscopic effect of quantum gravity --- the
condensation of gravitons and ghosts into the state with a certain
wavelength of the order of the horizon scale --- plays a
significant role in the formation of the asymptotic values of
energy density of cosmological vacuum.} The current theory
explains {\it how} the strong renormalization of the vacuum energy
occurs, but, unfortunately, it does not explain {\it why} this
happens and {\it why} the quantitative characteristics of the
phenomenon are those that are observed in the modern Universe. Of
the general considerations one can suggest that the coherent
graviton--ghost condensate occurs in the quantum--gravitational
phase transition (see Section \ref{pt}), and the answers to
questions should be sought in the light of the circumstances.

\section{Gravitons and Ghosts as Instantons}\label{inst}

\subsection{Self--Consistent Theory of Gravitons in Imaginary Time}\label{git}

\subsubsection{Invariance of Equations of the Theory
 with Respect to Wick Rotation of Time Axis}\label{vick}

As has been repeatedly pointed out, the complete system of
equations of the theory consists of the BBGKY chain (\ref{5.7})
--- (\ref{5.9}) and macroscopic Einstein's equations (\ref{5.10}).
On the basis of common mathematical considerations, it can be
expected that  solutions to these equations covers every possible
self--consistent states of quantum subsystem of gravitons and
ghosts and the classical subsystem of macroscopic geometry as
well. In examining the model that operates with the pure gravity
(no matter fields and $\Lambda$--term), one can identify the
following unique property of the theory. {\it Equations of the
theory (\ref{5.7}) --- (\ref{5.10}) are invariant with respect to
the Wick time axis rotation, conducted jointly with the
multiplicative transformation of moments of the spectral
function}:
\begin{equation}
\begin{array}{c}
\displaystyle t\to i\tau,\qquad H\to -i\mathcal{H},
\\[5mm]
\displaystyle  D\to
-\mathcal{D},\qquad W_n\to (-1)^n\mathcal{W}_n\ .
   \end{array}
 \label{7.1}
 \end{equation}
Rules of transformation of time derivatives are obtained from
(\ref{7.1})
\begin{equation}
\begin{array}{c}
\displaystyle \dot H \to -{\mathcal{H}}^{^\centerdot}\ ,
\qquad \ddot H \to i{\mathcal{H}}^{{^\centerdot}{^\centerdot}}\ ,
\qquad \dot D\to i{\mathcal{D}}^{^\centerdot}\ ,
\\[5mm]
\displaystyle \dot W_n\to -i(-1)^n{\mathcal{W}}_n^{^\centerdot}\ , \qquad
\ddot W_n \to -(-1)^n{\mathcal{W}}_n^{{^\centerdot}{^\centerdot}}\ ,
\qquad
\stackrel{...}{W}_n\to i(-1)^n{\mathcal{W}}_n^{{^\centerdot}{^\centerdot}{^\centerdot}}\ .
\end{array}
\label{7.2}
 \end{equation}
In (\ref{7.2}) and further on we use the notation
${\mathcal{F}}^{^\centerdot}=d\mathcal{F}/d\tau$. The statement
about the invariance of the theory can proved by direct
calculations. As a matter of fact, transformations of quantities
that appear in (\ref{5.7}) --- (\ref{5.10}) by the use of the
rules (\ref{7.1}) and (\ref{7.2}) lead to the BBGKY chain with
imaginary time
 \begin{equation}
\displaystyle  {\mathcal{D}}^{^\centerdot}+6\mathcal{H}\mathcal{D}+
4{\mathcal{W}}_1^{^\centerdot}+16\mathcal{H}\mathcal{W}_1=0\ ,
\label{7.3}
\end{equation}
\begin{equation}
\begin{array}{c}
 \displaystyle
{\mathcal{W}}_n^{{^\centerdot}{^\centerdot}{^\centerdot}} +
3(2n+3)\mathcal{H}{\mathcal{W}}_n^{{^\centerdot}{^\centerdot}}+
3\left[ \left(
4n^2+12n+6\right)\mathcal{H}^2
+(2n+1){\mathcal{H}}^{^\centerdot} \right]{\mathcal{W}}_n^{^\centerdot}+ \vspace{5mm} \\
\displaystyle +2n\left[ 2\left(2n^2+9n+9\right)\mathcal{H}^3
+6(n+2)\mathcal{H}{\mathcal{H}}^{^\centerdot}+
{\mathcal{H}}^{{^\centerdot}{^\centerdot}} \right]{\mathcal{W}}_n
+
4{\mathcal{W}}^{^\centerdot}_{n+1}+8(n+2)\mathcal{H}{\mathcal{W}}_{n+1}=0,
\qquad \displaystyle
n=1,\,...,\,\infty\, ,
\end{array}
\label{7.4}
\end{equation}
and to macroscopic Einstein's equations with imaginary time
\begin{equation}
 \begin{array}{c}
\displaystyle
{\mathcal{H}}^{^\centerdot}=-\frac{1}{16}\mathcal{D}-\frac16\mathcal{W}_1\ ,
 \qquad
3\mathcal{H}^2=\frac{1}{16}\mathcal{D}+\frac{1}{4}\mathcal{W}_1\ .
 \end{array}
 \label{7.5}
\end{equation}
It is easy to see that for  $\Lambda=0$ equations (\ref{5.7}) ---
(\ref{5.10}) identically coincide with (\ref{7.3}) --- (\ref{7.5})
after some trivial renaming.

The invariance of the theory with respect to the Wick rotation of
the time axis leads to the nontrivial consequence.  {\it Having
only self--consistent solution of the BBGKY chain and macroscopic
Einstein's equations, we can not say whether this solution is in
real or imaginary time.} Nevertheless, having a concrete solution
of BBGKY chain, we can view the status of time during further
study. To do so, it is necessary to explore the opportunity to
obtain the same solution at the level of operator functions and
state vectors. If this opportunity exists, the  appropriate
self--consistent solution of BBGKY chain and macroscopic
Einstein's equations is recognized as existing in real time. In
the previous Section \ref{exact}, we showed that two exact
solutions (\ref{5.41}.I) and (\ref{5.41}.III) really exist at the
level of operators and vectors, and thus have a physical
interpretation of standard notions of quantum theory.

The problem is: {\it What a physical reality reflects the
existence of solutions to the equations (\ref{5.7}) ---
(\ref{5.10}) (or that the same thing, (\ref{7.3}) ---
(\ref{7.5})), not reproducible in real time at the level of
operators and vectors?} The existence of the problem is explicitly
demonstrated by the example of exact solutions (\ref{5.41}.II).
Assume that this solution for  $C_{3g}=0, \, C_{4g}=-k_0^2<0, \,
C_{g2}=3k_0^2/4$ exists in real time:
 \begin{equation}
 \begin{array}{c}
\displaystyle
H^2=\frac{k_0^2}{a^2}\ln\frac{a}{a_0}\ , \qquad a=a_0\exp\left(\frac{k_0^2\eta^2}{4}\right)\ .
 \end{array}
 \label{7.6}
\end{equation}
The wave equation for gravitons with the (\ref{7.6}) background
reads
 \begin{equation}
 \begin{array}{c}
\displaystyle \hat\psi''_{{\bf k}\sigma}+k^2_0\eta\hat\psi'_{{\bf k}\sigma}+
k^2\hat\psi_{{\bf k}\sigma}=0\ .
 \end{array}
 \label{7.7}
\end{equation}
The equation for the ghosts looks similar. Equation (\ref{7.7})
differs from (\ref{6.2}) just in the sign of coefficient before
the first derivative. However, this difference is crucial: if
$k^2/k_0^2>0$ it is impossible to allocate the finite Hermit
$H_1(\eta)$  polynomial from degenerate hypergeometric functions
that correspond to solutions of the equation (\ref{7.7}). We have
been left with the infinite series only. These series and
integrals over spectrum of products of these series can not be
made consistent with the simple mathematical structure of the
exact solutions (\ref{5.41}.II). For this reason the solution
(\ref{5.41}.II), as the functional of scale factor is not relevant
to solving operator equations in real time.

\subsubsection{Imaginary Time Formalism}\label{formit}

As is known, the imaginary time formalism is used in
non--relativistic Quantum Mechanics (QM) (examples see, e.g., in
book \cite{24}), in the instanton theory of Quantum Chromodynamics
(QCD) \cite{25, 26, 27, 28, 28a, Sh} and in the axiomatic quantum field
theory (AQFT) (See Chapter 9 in the monograph \cite{6}). The
instanton physics in Quantum Cosmology was discussed in \cite{R3,
R4}.

In QM and QCD the imaginary time formalism is a tool for the study
of tunnelling, uniting classic independent states that are
degenerate in energy, in a single quantum state. In AQFT, the
Schwinger functions are defined in the four--dimensional Euclidian
space --- Euclid analogues of Wightman functions defined over the
Minkowski space. It is believed that using properties of
Euclid--Schwinger functions after their analytical continuation to
the Minkowski space, one can reconstruct the properties of
Wightman functions, and thereby restore the physical meaning of
the appropriate model of quantum field theory.

All prerequisites for the use of the formalism of imaginary time
in the QM and QCD on the one hand, and in AQFT, on the other hand,
are united in the self--consistent theory of gravitons.
Immediately, however, the specifics of the graviton theory under
discussion should be noted. Macroscopic space--time in
self--consistent theory of gravitons, unlike the space--time in
the QM, QCD  and AQFT, is a classical dynamic subsystem, which
actually evolved in real time. If in QCD and AQFT Wick's turn is
used to examine the significant properties of quantum system
expressed in {\it the probabilities}  of quantum processes, then
in relation to {\it the deterministic } evolution of classical
macroscopic subsystem this turn makes no sense. Therefore, after
solving equations of the theory in imaginary time, we are obliged
to apply (to the solution obtained) the operation of analytic
continuation of the space for the positive signature to the space
of negative signature. It is clear from the outset that the
operation is not reduced to the opposite Wick turn, but is an
independent postulate of the theory.

Before discussing the physical content of the theory, let us
define its formal mathematical scheme. The theory is formulated in
the space with metric
\begin{equation}
\begin{array}{c}
\displaystyle ds^2=-d\tau^2-a^2(\tau)(dx^2+dy^2+dz^2)\ .
\end{array}
 \label{7.8}
\end{equation}
Note that in our theory, that is suppose to do with cosmological
applications (as opposed to QCD and AQFT), one of the coordinates
is singled out simply because the scale factor depends on it. This
means that in the classical sector of the theory time $\tau$,
despite the fact that it is imaginary, is singled out in
comparison with the 3--spatial coordinates. In the quantum sector
the $\tau$ coordinate also has a special status. {\it Operators}
of graviton and ghost fields with nontrivial commutation
properties are defined over the space (\ref{7.8}). Symmetry
properties of space (\ref{7.8}) allow us to define the Fourier
images of the operators by coordinates $x,\ y,\ z$, and to
formulate the canonical commutation relations in terms of
derivatives of operators with respect to the imaginary time
$\tau$:
 \begin{equation}
\begin{array}{c}
 \displaystyle  \frac{a^3}{4\varkappa}\left[\frac{d\hat\psi^+_{{\bf k}\sigma}}{d\tau}\ ,\
\hat\psi_{{\bf k'}\sigma'}\right]_{-}=-i\hbar \delta_{{\bf k}{\bf
k'}}\delta_{\sigma\sigma'}\ .
\end{array}
\label{7.9}
 \end{equation}
 \begin{equation}
\begin{array}{c}
 \displaystyle
\frac{a^3}{4\varkappa}\left[\frac{d{\hat\vartheta}^+_{{\bf k}}}{d\tau}\ ,\
\hat\vartheta_{{\bf k'}}\right]_{-}=-i\hbar \delta_{{\bf k}{\bf k'}}\ ,
\qquad
\frac{a^3}{4\varkappa}\left[\frac{d{\hat\vartheta}_{{\bf k}}}{d\tau}\ ,\
\hat\vartheta^+_{{\bf k'}}\right]_{-}=-i\hbar \delta_{{\bf k}{\bf k'}}\ .
\end{array}
 \label{7.10}
 \end{equation}
Note that(\ref{7.9}), (\ref{7.10}) are introduced by the newly
independent postulate of the theory, and not derived from standard
commutation relations (\ref{3.37}), (\ref{3.41}) by conversion of
$t\to i\tau$. (Such a conversion would lead to the disappearance
of the imaginary unit from the right hand sides of the commutation
relations.) Thus, the imaginary time formalism can not be regarded
simply as another way to describe the graviton and ghost fields,
i.e. as a mathematically equivalent way for real time description.
In this formalism the new specific class of quantum phenomena is
studied.

The system of self--consistent equations is produced by variations
of action, as defined in 4--space with a positive signature:
\begin{equation}
\begin{array}{c}
 \displaystyle S=\frac{1}{\varkappa}\int d\tau\left\{3\left[\frac{a^2}{N}\frac{d^2a}{d\tau^2}-
 \frac{a^2}{N^2}\frac{dN}{d\tau}\frac{da}{d\tau}
 +\frac{a}{N}\left(\frac{da}{d\tau}\right)^2\right]+\right.
 \\[5mm]
 \displaystyle \left.
 +\frac{1}{8}\sum_{{{\bf k}\sigma}}
 \left(\frac{a^3}{N}\frac{d{\hat \psi}_{{\bf k}\sigma}^+}{d\tau}
 \frac{d{\hat \psi}_{{\bf k}\sigma}}{d\tau}+Nak^2\hat \psi_{{\bf k}\sigma}^+
 \hat \psi_{{\bf k}\sigma}\right)-\frac{1}{4}\sum_{\bf k}
 \left(\frac{a^3}{N}\frac{d{\hat\vartheta}_{\bf k}^+}{d\tau}
 \frac{d{\hat\vartheta}_{\bf k}}{d\tau}+
 Nak^2\hat\vartheta_{\bf k}\hat\vartheta_{\bf k}\right)\right\}\ .
 \end{array}
\label{7.11}
 \end{equation}
Note that the full derivative with respect to the imaginary time
is not excluded from Lagrangian. In (\ref{7.11}) the integrand
contains the density of  invariant $\sqrt{\hat g}\hat R$. The
Lagrange multiplier $N$ after the completion of the variation
procedure is assumed to be equal to unity. The system of equations
corresponding to the action (\ref{7.11}) can also be obtained from
the system of equations in real time by conversion of $t\to
i\tau$. Quantum equations of motion for field operators in the
imaginary time read
 \begin{equation}
 \displaystyle
 \frac{d^2\hat\psi_{{\bf
k}\sigma}}{d\tau^2}+3\mathcal{H}\frac{d\hat\psi_{{\bf
k}\sigma}}{d\tau}-\frac{k^2}{a^2}\hat\psi_{{\bf k}\sigma}=0\ ,
 \label{7.12}
 \end{equation}
\begin{equation}
 \displaystyle
 \frac{d^2\hat\vartheta_{{\bf k}}}{d\tau^2}+3\mathcal{H}\frac{d\hat\vartheta_{{\bf
k}}}{d\tau}-\frac{k^2}{a^2}\hat\vartheta_{{\bf k}}=0\ ,
 \label{7.13}
\end{equation}
where ${\mathcal{H}}=a^{^\centerdot}/a$.

Equations (\ref{7.12}), (\ref{7.13}) differ from (\ref{3.30}),
(\ref{3.31}) by only replacement of $k^2\to -k^2$. At the level of
analytic properties of solutions of the equations this difference,
of course, is crucial. However, formal transformations, not
dependent on the properties of analytic solutions to equations
(\ref{3.30}), (\ref{3.31}) and (\ref{7.12}), (\ref{7.13}), look
quite similar. Therefore, all operations to construct the equation
for the spectral function in imaginary time (analogue to equation
(\ref{5.4})) and the subsequent construction of BBGKY chain
coincide with that described in Section \ref{chain} with the
replacement of $k^2\to -k^2$. Replacing  $k^2\to -k^2$ changes the
definition of moments only parametrically: instead of (\ref{5.6})
we get
 \begin{equation}
\begin{array}{c}
\displaystyle \mathcal{W}_n=\sum_{{\bf
k}}\left(\frac{-k^2}{a^2}\right)^n\left(\sum_\sigma\langle
\Psi_g|\hat\psi^+_{{\bf k}\sigma}\hat\psi_{{\bf
k}\sigma}|\Psi_g\rangle-2\langle \Psi_{gh}|\hat\vartheta^+_{{\bf
k}}\hat\vartheta_{{\bf k}}|\Psi_{gh} \rangle\right),\qquad
 n=0,\,1,\,2,\,...,\infty\ ,
 \\[5mm]
 \displaystyle \mathcal{D}=
 \frac{d^2\mathcal{W}_0}{d\tau^2}+3\mathcal{H}\frac{d\mathcal{W}_0}{d\tau}\ .
\end{array}
 \label{7.14}
 \end{equation}
Further actions lead obviously to the BBGKY chain (\ref{7.3}),
(\ref{7.4}) and to the macroscopic Einstein equations (\ref{7.5}).

To solve equations (\ref{7.12}) and (\ref{7.13}), we will be using
only the real linear--independent basis
\begin{equation}
\begin{array}{c}
 \displaystyle \hat\psi_{{\bf k}\sigma}=
 \sqrt{4\varkappa\hbar}\left(\hat Q_{{\bf k}\sigma}g_k+\hat P_{{\bf k}\sigma}h_k\right)\ ,
 \qquad \hat\vartheta_{\bf k}=\sqrt{4\varkappa\hbar}\left(\hat q_{\bf k}g_k+
 \hat p_{\bf k}h_k\right)\ ,
 \\[3mm]
 \displaystyle  g_k h_k^{^\centerdot}-h_k g_k^{^\centerdot}=\frac{1}{a^3}\ .
 \end{array}
 \label{7.15}
 \end{equation}
As will be seen below, one of the basic solutions satisfies the
known definition of instanton: an instanton is a solution to the
classical equation, which is localized in the imaginary time and
corresponds to the finite action in the 4--space with a positive
signature. We will call the operator functions (\ref{7.15}) {\it
the quantum instanton fields of gravitons and ghosts.} Operator
constants of integration  $\hat Q_{{\bf k}\sigma},\ \hat P_{{\bf
k}\sigma}$  and  $q_{\bf k},\ p_{\bf k}$ satisfy commutation
relations (\ref{3.55}). Ladder operators are imposed by the the
equations (\ref{3.54}) and then used in the procedure for
constructing the state vectors over the basis of occupation
numbers. State vectors of the general form in graviton and ghost
sectors are already familiar structure (\ref{3.48}) and
(\ref{3.51}). Only the interpretation of occupation numbers is
changed: now it is number of instantons $n_{{\bf k}\sigma}$,
$n_{\bf k}$, $\bar n_{\bf k}$  of graviton, ghost and anti--ghost
types, respectively.

Direct calculation of the moments of the spectral function leads
to the expression:
\begin{equation}
\begin{array}{c}
\displaystyle
\mathcal{W}_n=4\varkappa\hbar (-1)^n\sum_{\bf k}\left(\frac{k^2}{a^2}\right)^n\left(A_kg_k^2+
B_kh_k^2\right)\ ,
\end{array}
 \label{7.16}
\end{equation}
where
\begin{equation}
\begin{array}{c}
\displaystyle A_k=
\sum_\sigma\langle\Psi_g|\hat Q^+_{{\bf k}\sigma}Q_{{\bf k}\sigma}|\Psi_g\rangle-
2\langle\Psi_{gh}|\hat q_{{\bf k}}^+\hat q_{{\bf k}}|\Psi_{gh}\rangle=
2\langle n_{k(g)}\rangle(1+\zeta^{(g)}_k\cos\varphi_k)-
2\langle n_{k(gh)}\rangle(1+\zeta_k^{(gh)}\cos\chi_k)\ ,
\end{array}
 \label{7.17}
\end{equation}
\begin{equation}
\begin{array}{c}
\displaystyle
B_k=\sum_\sigma\langle\Psi_g|\hat P^+_{{\bf k}\sigma}P_{{\bf k}\sigma}|\Psi_g\rangle-
2\langle\Psi_{gh}|\hat p_{{\bf k}}^+\hat p_{{\bf k}}|\Psi_{gh}\rangle=
2\langle n_{k(g)}\rangle(1-\zeta^{(g)}_k\cos\varphi_k)-
2\langle n_{k(gh)}\rangle(1-\zeta_k^{(gh)}\cos\chi_k)\ .
\end{array}
 \label{7.18}
\end{equation}
The term containing products of basis functions $g_kh_k$ is
eliminated from (\ref{7.16}) by the condition of homogeneity of
3--space. In (\ref{7.17}) and (\ref{7.18}) average values of
numbers of instantons of ghost and anti--ghost types are assumed
to be equal: $\langle n_{k(gh)}\rangle=  \langle \bar
n_{k(gh)}\rangle$. One needs to pay attention to the multiplier
$(-1)^n$ in (\ref{7.16}): {\it the alternating sign  of moments is
a common symptom of instanton nature of the spectral function.}

Instanton equations of motion (\ref{7.12}), (\ref{7.13}) are of
the hyperbolic type. This fact determines the form of asymptotics
of basis function for $k|\xi|\gg 1$ where $\xi=\int d\tau/a$ is
conformal imaginary time. One of basis functions is localized in
the imaginary time and the other is increasing without limit with
the increasing of modulus of the imaginary time
\begin{equation}
\displaystyle g_k\sim \frac{e^{-k\xi}}{a\sqrt{2k}}\ ,
\qquad h_k\sim \frac{e^{k\xi}}{a\sqrt{2k}}\ , \qquad k|\xi|\gg 1\ .
 \label{7.19}
\end{equation}
In this situation, it is necessary to differentiate between stable
and unstable instanton configurations. We call a configuration
stable, if moments of the spectral function are formed by
localized basis functions only. Without limiting generality, we
assigned $h_k$ to the class of increasing functions. It is easy to
see that the condition of stability $B_k=0$ that eliminates
contributions of $h_k$ from (\ref{7.16}) {\it is reduced to the
condition of quantum coherence of instanton condensate}:
\begin{equation}
\begin{array}{c}
\displaystyle \zeta^{(g)}_k\cos\varphi_k=\zeta_k^{(gh)}\cos\chi_k=1 \quad \to \quad
\zeta^{(g)}_k=\zeta_k^{(gh)}=1,\qquad \cos\varphi_k=\cos\chi_k=1\ .
\end{array}
\label{7.20}
\end{equation}
Expressions for the moments are simplified and read
\begin{equation}
\begin{array}{c}
\displaystyle
\mathcal{W}_n=4\varkappa\hbar (-1)^n\sum_{\bf k}A_{\bf k}\left(\frac{k^2}{a^2}\right)^ng_k^2\ .
\end{array}
 \label{7.21}
\end{equation}
Exact solutions, with the stable instanton configurations, are
described in the following Sections \ref{ins1} and \ref{ins2}. In
principle, for a limited imaginary time interval, there might be
unstable configurations, but in the present work such
configurations are not discussed. (The example of the unstable
instanton configuration see in \cite{R5}.)

Note that moments (\ref{7.21}) can be obtained within the
classical theory, limited, as generally accepted, to the solutions
localized in imaginary time. In doing so, $A_{k}$ acts as a
constant of integration of classical equation.

The above approach is the quantum theory of instantons in
imaginary time. Here are present all the elements of quantum
theory: operator nature of instanton field; quantization on the
canonical commutation relations; basic vectors in the
representation of instanton occupation numbers; state vectors of
physical states in the form of superposition of basic vectors.
With the quantum approach, a significant feature of instantons is
displayed, which clearly is not visible in the classical theory.
It is the nature of instanton stable configurations as coherent
quantum condensates.

Construction of the formalism of the theory is completed by
developing a procedure to transfer the results of the study of
instantons to real time. It is clear that this procedure is
required to match the theory with the experimental data, i.e. to
explain the past and predict the future of the Universe. As
already noted, the procedure of transition to real time is not an
inverse Wick rotation. This is particularly evident in the quantum
theory: in (\ref{7.9}), (\ref{7.10}) the reverse Wick turn leads
to the commutation relations for non--Hermitian operators, which
can not be used to describe the graviton field.

The procedure for the transition to real time has the status of an
independent theory postulates. We will formulate this postulate as
follows.

(i) Results of solutions of quantum equations of motion
(\ref{7.12}), (\ref{7.13}), together with the macroscopic
Einstein's equations (\ref{7.5}) {\it after calculating of the
moments} (that is, after averaging over the instanton state
vector) should be represented in the functional form
\begin{equation}
\begin{array}{c}
\displaystyle \mathcal{D}=\mathcal{D}(a, \mathcal{H}, \mathcal{H}^{^\centerdot},...)\ ,
\qquad \mathcal{W}_n=\mathcal{W}_n(a, \mathcal{H}, \mathcal{H}^{^\centerdot},...)\ .
\end{array}
 \label{7.22}
\end{equation}

(ii) It is postulated that {\it functional} dependence of the
moments of the spectral function on functions describing the
macroscopic geometry must be identical in the real and imaginary
time. Thus, {\it at the level of the moments of the spectral
function}, the transition to the real time is reduced to a change
of notation
\begin{equation}
\begin{array}{c}
\displaystyle  \mathcal{D}(a, \mathcal{H}, \mathcal{H}^{^\centerdot},...)
\qquad \to \qquad D(a,H,\dot H,...)\ ,
\\[5mm]
\displaystyle \mathcal{W}_n(a, \mathcal{H}, \mathcal{H}^{^\centerdot},...)
\qquad \to \qquad W_n(a,H,\dot H,...)\ .
\end{array}
 \label{7.23}
\end{equation}

(iii) Moments $D(a,H,\dot H,...)$ and $W_1(a,H,\dot H,...)$
obtained by operations (\ref{7.23}), are substituted to right hand
side of macroscopic Einstein equations that are considered now as
equations in real time. Formally this means that the transition to
the real time in the left hand side of equations (\ref{7.5}) is
reduced to changing of the following notations
\begin{equation}
\begin{array}{c}
\displaystyle \mathcal{H}^{^\centerdot} \qquad \to \qquad \dot H,\qquad\qquad
{\mathcal{H}}^2 \qquad \to \qquad  H^2\ .
\end{array}
 \label{7.24}
\end{equation}

Thus, the acceptance of postulates (\ref{7.22}) --- (\ref{7.24})
is equivalent to the suggestion that in real time the
self--consistent evolution of classic geometry and quantum
instanton system is described by the following equations
\begin{equation}
 \begin{array}{c}
\displaystyle
\dot H=-\frac{1}{16}D(a,H,\dot H,...)-\frac16W_1(a,H,\dot H,...)\ ,
 \\[5mm]
\displaystyle
3H^2=\frac{1}{16}D(a,H,\dot H,...)+\frac{1}{4}W_1(a,H,\dot H,...)\ ,
 \end{array}
 \label{7.25}
\end{equation}
under the condition that the form of functionals in right hand
sides of  (\ref{7.25}) is established by microscopic calculations
in imaginary time. It is obvious also that in the framework of
these postulates {\it any solution of equations consisting of
BBGKY chain and macroscopic Einstein equations (obtained without
use of microscopic theory) can be considered as the solution in
real time.}

\subsubsection{Physics of Imaginary Time}\label{physit}

Mathematical and physical motivation to look for the formalism of
imaginary time comes from the fact that there are degenerate
states separated by the classical impenetrable barrier. In
non--relativistic quantum mechanics the barriers are considered,
that have been formed by classical force fields and for that
reason they have the obvious interpretation. It is well known,
that the calculation of quantum tunnelling across the classical
impenetrable barrier can be carried out in the following order:
(1) the solution of classical equation of motion inside the
barrier area is obtained with imaginary time; (2) from the
solution obtained for the tunnelling particle, one calculates the
action $S$ for the imaginary time; (3) the tunnelling probability,
coinciding with the result of the solution for Schrodinger
equation in the quasi--classical approximation, is equal
$w=e^{-S}$. Obviously, the  sequence described bears a formal
character and cannot be interpreted operationally. Nevertheless, a
strong argument toward the use of the formalism of imaginary time
in the quantum mechanics is the agreement between the calculations
and experimental data for the tunnelling micro--particles.

A new class of phenomena arises in the cases when tunnelling
processes form a macroscopic quantum state. The Josephson effect
is a characteristic example: fluctuations of the electromagnetic
field arise  when a superconductive condensate is tunnelling
across the classically impenetrable non--conducting barrier. Here,
{\it the tunnelling can be formally described as a process
developing in imaginary time, but the fluctuations arise and exist
in the real space--time.} Experimental data show that regardless
of the description, the tunnelling process forms a physical
subsystem in the real space--time, with perfectly real
energy--momentum.

In Quantum Chromodynamics (QCD) physically similar phenomena are
studied by similar methods \cite{Sh}. The vacuum degeneration is an internal
property of QCD: different classical vacuums of gluon field are
not topologically equivalent. In the framework of the classic
dynamics any transitions between different vacuums are impossible.
In that sense the topological non--equivalence plays role of the
classical impenetrable barrier. There is an heuristic hypothesis
in quantum theory --- that the probability of tunnelling
transition between different vacuums can be calculated as
$w=e^{-S}$, where $S$ is the action of the classical instanton.
The instanton is defined as a solution of gluon--dynamic equations
localized in the Euclidian space--time connecting configurations
with different topologies. As in the case of Josephson Effect, it
is assumed that the tunnelling processes between topologically
non--equivalent vacuums are accompanied by  generation of
non--perturbative fluctuations of gluon and quark fields in real
space--time. Let us notice that {\it in QCD the instanton
solutions, analytically continued into real space--time, are used
to evaluate the amplitude of fluctuations.} The fluctuations in
real space--time are considered as a quark--gluon condensate
(QGC). The existence of QGC with different topological structure
in "off-adrons"\ and "in-adrons"\ vacuums, is confirmed by
comparison of theoretical predictions with experimental data. One
of remarkable facts is that the carrier of approximately the half
of nucleon mass is in fact the energy of the reconstructed QGC.

Now let us go back to the self--consistence theory of gravitons.
In that theory, due to its one--loop finiteness, all observables
are formed by the difference between graviton and ghost
contributions. That fact is obvious both from the general
expressions for the observables (see (\ref{3.72}), (\ref{3.73})),
and from the exact and approximate solutions (described in the
previous sections) as well. The same final differences of
contributions may correspond to the totally different graviton and
ghost contributions themselves. All quantum states are degenerated
with respect to mutually consistent transformations of gravitons
and ghosts occupation numbers, but providing unchanged values of
observable quantities. Thus {\it the multitude of state vectors of
the general form, averaging over which leads to the same values of
spectral function, is a direct consequence of the internal
mathematical structure of the self--consistent theory of
gravitons, satisfying the one--loop finiteness condition.}

In that situation, it is very natural to introduce a hypothesis
about the tunnelling of the graviton--ghost system between quantum
states corresponding to the same values of macroscopic
observables. By the analogy with the effects described above, one
may suggest that 1) the tunnelling processes unite degenerate
quantum states into a single quantum state; 2) tunnelling is
accompanied by creation of specific quantum fluctuations of
graviton and ghost fields in real space--time. With regard to the
mathematical method used to describe these phenomena, today we may
use only those methods that have been tested in adjacent brunches
of quantum theory. It is easy to see that this program has been
realized in Sections \ref{vick}, \ref{formit}. We solve the
equations of the theory for imaginary time, but the amplitude of
the arising fluctuations we evaluate by the analytical
continuation (\ref{7.22}) --- (\ref{7.24}), analoguos to the ones
used in QCD. The specific of our theory lie in the fact that at
the final step of calculations we use the classical Einstein
equations (\ref{7.25}) describing the evolution of the macroscopic
space in real time. The possibility of using these equations is
determined by the action (\ref{7.11}), which, when calculated by
means of the instanton solutions and averaged over the state
vector of instantons, is identically equal zero. As a matter of
fact, after using instanton equations (\ref{7.12}), (\ref{7.13})
and averaging, the action (\ref{7.11}) is reduced to the form:
\begin{equation}
\begin{array}{c}
\displaystyle \langle\Psi|S|\Psi\rangle=
\frac{1}{\varkappa}\int d\tau a^3\left[3\left({\mathcal{H}}^{^\centerdot}+{\mathcal{H}}^2\right)
+\frac{1}{16}\mathcal{D}\right]\ .
 \end{array}
\label{7.26}
\end{equation}
The integrand in (\ref{7.26}) is equal zero in the Einstein
equations with imaginary time (\ref{7.5}). The fact that
$w=\exp(-\langle\Psi|S|\Psi\rangle)=1$  means that the macroscopic
evolution of the Universe is determined. That feature allows the
use of equations (\ref{7.25}), after the moments are analytically
continued into the real time.

\subsection{ Instanton Condensate in the De Sitter Space}\label{ins1}

Among exact solutions of the one--loop quantum gravity, a special
status is given to De Sitter space if the space curvature of this
space is self--consistent with the quantum state of gravitons and
ghosts. In Section \ref{S} it was shown that in the
self--consistent solution, gravitons and ghosts can be interpreted
as quantum wave fields in real space--time. Nevertheless, it
should be mentioned, that the alternating sign of the moments
(\ref{6.28}) points to a possibility of instanton interpretation
of that solution. Methods described in Sections \ref{vick},
\ref{formit}, when applied to De Sitter space, show that such
interpretation is really possible.

We will work with the imaginary conformal time $\xi=\int d\tau/a$.
The cosmological solution is:
\begin{equation}
\displaystyle a=a_0e^{\mathcal{H}\tau}=-\frac{1}{\mathcal{H}\xi}\ ,
\qquad -\infty<\xi\leqslant 0\ .
\label{7.27}
\end{equation}
At the level of the BBGKY chain, due to the fact that the theory
is invariant with respect to the Wick rotation, the calculations
performed to get the solutions coincide  with the those described
in Section \ref{Sitt}. At the microscopic level we use the exact
solutions (\ref{7.12}), (\ref{7.13}) with the background
(\ref{7.27}):
\begin{equation}
 \begin{array}{c}
 \displaystyle
\hat \psi_{{\bf k}\sigma}=\frac{1}{a}\sqrt{\frac{2\varkappa\hbar}{k}}
 \left[Q_{{\bf k}\sigma}g(x)+P_{{\bf k}\sigma}h(x)\right]\ ,
\qquad
\hat \vartheta_{{\bf k}}=\frac{1}{a}\sqrt{\frac{2\varkappa\hbar}{k}}
 \left[q_{{\bf k}}g(x)+p_{{\bf k}}h(x)\right]\ ,
\end{array}
 \label{7.28}
\end{equation}
where 
\[
\displaystyle g(x)=\left(1-\frac{1}{x}\right)e^{x}\ ,\qquad h(x)=
\left(1+\frac{1}{x}\right)e^{-x}\ ,\qquad x=k\xi<0\ .
\]
The expressions for the moments of the spectral function are reduced to the form:
\begin{equation}
\begin{array}{c}
\displaystyle
\mathcal{W}_n=(-1)^{n}\frac{\varkappa\hbar}{\pi^2}{\mathcal{H}}^{2n+2}\int\limits^0_{-\infty}
dxx^{2n+2}\left(A_kg^2+B_kh^2\right)\ .
\end{array}
 \label{7.29}
\end{equation}
Equations for  $A_k,\ B_k$ are given in (\ref{7.17}),
(\ref{7.18}). From (\ref{7.29})  it is obvious that the
self--consistent values $\mathcal{W}_n=const$  can be obtained
only for a flat spectrum of instantons. However, with the flat
specter and $B_k\ne 0$, the second term in (\ref{7.29}) creates a
meaningless infinity. Therefore $B_k=0$, and that, in turn, leads
to the condition (\ref{7.20}), i.e. to quantum coherence of the
instanton condensate.  The quantitative characteristics of the
condensate are formed by instantons only, localized in imaginary
time.

It is easy to calculate of the converging integrals in
(\ref{7.29}):
\begin{equation}
\begin{array}{c}
\displaystyle \int\limits_{-\infty}^0dxx^{2n+2}\left(1-\frac{1}{x}\right)^2e^{2x}=
\frac{1}{2^{2n+1}}(2n-1)!!(2n+1)(n+2)\ .
\end{array}
 \label{7.30}
\end{equation}
After  analytical continuation into the real space--time,
following the rules (\ref{7.22}) --- (\ref{7.24}), we obtain the
final result:
\begin{equation}
 \begin{array}{c}
 \displaystyle
 D=-\frac{12\varkappa\hbar N_{inst}}{\pi^2}H^4\ ,
\qquad
 W_n=\frac{(-1)^{n+1}}{2^{2n}}
 (2n-1)!(2n+1)(n+2)\times
  \frac{2\varkappa\hbar N_{inst}}{\pi^2}H^{2n+2} ,\qquad n\geqslant 1\ ,
\end{array}
 \label{7.31}
 \end{equation}
where
\begin{equation}
\displaystyle N_{inst}= \langle n_{gh}\rangle-\langle n_{g}\rangle\  .
\label{7.32}
\end{equation}

The comparison of the two models of graviton--ghost condensate in
the De Sitter space reveals some interesting features. In both
cases we deal with the effect of quantum coherence. Expressions
(\ref{7.31}) differ from (\ref{6.28}) only in the formal
substitution $N_g\to N_{inst}$. However the conditions leading to
the quantum coherence are different in these models. According to
(\ref{6.29}), in the condensate of virtual gravitons and ghosts,
the average value of graviton and ghost occupational numbers are
the same, and the non--zero effect appears due to the fact that
the phase correlation in the quantum superposition in the
graviton's and ghost's sectors are formed differently. As it
follows from (\ref{7.20}), (\ref{7.32}), in the instanton
condensate the phases in the graviton and ghost sectors correlate
similarly, but the non--zero effect appears due to the difference
of average occupation numbers for graviton's and ghost's
instantons. The absence of the macroscopic structure of the
condensates does not allow the detection of the differences by
macroscopic measurements. In both cases the graviton--ghost vacuum
possess equal energy--momentum characteristics.

The question about the actual nature of the De Sitter space is
lies in the formal mathematical domain. In these circumstances one
should pay attention to the following facts. While describing the
condensate of virtual gravitons and ghosts, we were forced to
introduce an additional definition of the mathematically
non--existent integrals (\ref{6.25}), i.e. to introduce into the
theory some operations that were not present from the beginning.
It is the additional operations that have provided a very specific
property of the solution --- the alternating signs in the sequence
of the moments of the spectral function. By contrast, the theory
of the instanton condensate has a completely different formal
mathematics. The theory is motivated by the concrete property of
the graviton--ghost system which is degeneration of quantum
states, and the construction of the theory is constructed by the
introduction of mathematically non--contradictory postulates. The
moments of the spectral function's with alternating signs is an
internal property of the graviton--ghost instanton theory.  When
we considered the instanton condensate in the De Sitter space, no
additional mathematical redefinitions were necessary (compare the
formulas (\ref{6.25}) and (\ref{7.30})). We have the impression
that the instanton version of the De Sitter space is more
mathematically comprehensive. Therefore, one may suggest that {\it
the key role in the formation of the De Sitter space (the
asymptotic state of the Universe) belongs to the instanton
condensate, appearing in the tunnelling processes between
degenerated states of the graviton--ghost vacuum.}\footnote{A cosmological scenario based on this solution was proposed in \cite{A}. In this scenario, birth of the flat inflationary Universe can be thought of as a quantum tunneling from "nothing". As the Universe ages and is emptied, the same mechanism of tunneling that gave rise to the empty Universe at the beginning, gives now birth to dark energy. The emptying Universe should possibly complete its evolution by tunneling back to "nothing".}

\subsection{Instanton Condensate of Constant Conformal Wavelength}\label{ins2}

The exact solution (\ref{5.41}.II) has a pure instanton nature.
Now we will obtain that solution with the value $C_{3g}=0$. One
can rewrite the formulas (\ref{7.6}), (\ref{7.7}) for the
imaginary time:
 \begin{equation}
 \begin{array}{c}
\displaystyle
{\mathcal{H}}^2=\frac{k_0^2}{a^2}\ln\frac{a}{a_0}\ ,
\qquad a=a_0\exp\left(\frac{k_0^2\xi^2}{4}\right)\ .
 \end{array}
 \label{7.33}
\end{equation}
 \begin{equation}
 \begin{array}{c}
\displaystyle \frac{d^2\hat\psi_{{\bf k}\sigma}}{d\xi^2}+k^2_0\xi
\frac{d\hat\psi_{{\bf k}\sigma}}{d\xi}-k^2\hat\psi_{{\bf k}\sigma}=0,\qquad
\frac{d^2\hat\vartheta_{{\bf k}}}{d\xi^2}+
k^2_0\xi\frac{d\hat\vartheta_{{\bf k}}}{d\xi}-k^2\hat\vartheta_{{\bf k}}=0\ .
 \end{array}
 \label{7.34}
\end{equation}
As we already know, the spatially homogeneous modes participate in
the formation of the solution for the equation (\ref{5.41}.II). As
follows from (\ref{7.34}), when $k^2\to 0$, the description of the
spatially homogeneous modes in imaginary time does not differ from
their description in real time. The contribution from modes $g2$
is present in (\ref{7.33}), with the relations $C_{g4}=-k_0^2<0,
\, C_{g2}=3k_0^2/4$ taken into account. These relations are
necessary to provide the existence of the self--consistent
solution. In what follows we are considering the quasi--resonant
modes only.

For   $k^2=k^2_0$, the signs of the last terms in the equations
(\ref{7.34}) provide the existence of instanton solutions we are
looking for:
\begin{equation}
\begin{array}{c}
\displaystyle
\hat\psi_{{\bf k}\sigma}=
\frac{\sqrt{4\varkappa\hbar k_0}}{a_0}\left[\xi\left(\hat Q_{{\bf k}\sigma}+
k_0\hat P_{{\bf k}\sigma}\int\limits_0^\xi e^{-k_0^2\xi^2/2}d\xi\right)+
\frac{\hat P_{{\bf k}\sigma}}{k_0}e^{-k_0^2\xi^2/2}\right]=
\sqrt{\frac{16\varkappa\hbar}{k_0a_0^2}}\left[\hat Q_{{\bf k}\sigma}+
\hat P_{{\bf k}\sigma}F(a)\right]\ln^{1/2}\frac{a}{a_0} \ ,
\end{array}
 \label{7.35}
\end{equation}
\begin{equation}
\begin{array}{c}
\displaystyle
\hat\vartheta_{{\bf k}}=\frac{\sqrt{4\varkappa\hbar k_0}}{a_0}\left[\xi\left(\hat q_{{\bf k}}+
k_0\hat p_{{\bf k}}\int\limits_0^\xi e^{-k_0^2\xi^2/2}d\xi\right)+
\frac{\hat p_{{\bf k}}}{k_0}e^{-k_0^2\eta^2/2}\right]=
\sqrt{\frac{16\varkappa\hbar}{k_0a_0^2}}\left[\hat q_{{\bf k}}+
\hat p_{{\bf k}}F(a)\right]\ln^{1/2}\frac{a}{a_0} \ ,
\end{array}
 \label{7.36}
\end{equation}
where
\[
 \displaystyle F(a)=a_0^2\int\limits_{a_0}^a\frac{da}{a^3\displaystyle\ln^{1/2}\frac{a}{a_0}}+
 \frac{a_0^2}{2a^2\displaystyle\ln^{1/2}\frac{a}{a_0}}\ .
 \]
Calculations which follow contain the same mathematical operations
we have already described several times in the previous sections.
After we remove contributors to the spectral function which
contains $F(a)$, we obtain the condition for the coherence of the
condensate. Some details of the calculations is related to the
alternating signs  of the moments, i.e. with the multiplier
$(-1)^n$, characteristic for the instanton theory. Particularly,
in the expression for $\mathcal{W}_1(g4)$, there is a general sign
"minus"\ . But, according to the Einstein equations in imaginary
time  $\mathcal{W}_1(g4)>0$. The positive sign of the first moment
is provided by the dominant contribution of ghost instantons over
the contribution of graviton instantons. With that taken into
account, we obtain the final equations for the moments of
quasi--resonant modes, obtained after the analytic continuation
into the real space--time:
\begin{equation}
\begin{array}{c}
\displaystyle W_n(g4)= (-1)^{n+1}\frac{64\varkappa\hbar
N_{inst}^{(gh)} k_0^{2n-1}}{a^2_0a^{2n}}\ln\frac{a}{a_0}=
(-1)^{n+1}\frac{24k_0^{2n}}{a^{2n}}\ln\frac{a}{a_0}\ ,
\\[5mm]
\displaystyle D(g4)=
-\frac{128\varkappa\hbar N_{inst}^{(gh)} k_0}{a^2_0a^{2}}\ln\frac{e^{1/4}a}{a_0}=
-\frac{48k_0^{2}}{a^{2}}\ln\frac{e^{1/4}a}{a_0}\ .
\end{array}
\label{7.37}
\end{equation}
Here the following definition has been used:
\[
\displaystyle \langle n_{k(gh)}\rangle-\langle n_{k(g)}\rangle
\to \frac{2\pi^2}{k^2}N_{inst}^{(gh)} \delta(k-k_0)\ , \quad N_{inst}^{(gh)}=
\frac{3k_0a_0^2}{8\varkappa\hbar}\ .
\]
The graviton instantons are dominant for the spatially homogeneous modes:
\begin{equation}
\displaystyle W_1(g2)=\frac{16\varkappa\hbar k_1
N_{inst}^{(g)}}{a_1^2a^2}\ , \qquad D(g2)=-\frac{32\varkappa\hbar
k_1N_{inst}^{(g)}}{a_1^2a^2}\ .
 \label{7.38}
\end{equation}
The parameter of the spatially homogeneous condensate is defined as follows:
\[
\begin{array}{c}
\displaystyle \langle
n_{0(g)}\rangle(1+\zeta^{(g)}_0\cos\varphi_0)- \langle
n_{0(gh)}\rangle(1+\zeta^{(gh)}_0\cos\chi_0)\quad \to\quad
\frac{2\pi^2}{k^2}N_{inst}^{(g)}\delta(k-q_0)\ ,\qquad q_0\to 0\ .
\end{array}
 \]
From expressions (\ref{7.37}), (\ref{7.38}), one gets energy
density and pressure for the system of quasi--resonant and
spatially homogeneous instantons:
\begin{equation}
 \begin{array}{c}
 \displaystyle
\varkappa\varepsilon_g=\frac{8\varkappa\hbar
k_0N_{inst}^{(gh)}}{a_0^2a^2}\ln\frac{a}{a_0}+
\frac{2\varkappa\hbar}{a^2}\left(\frac{k_0N_{inst}^{(gh)}}{a^2_0}-
\frac{k_1N_{inst}^{(g)}}{a^2_1}\right)=
\frac{8\varkappa\hbar
k_0N_{inst}^{(gh)}}{a_0^2a^2}\ln\frac{a}{a_0}\ ,
\\[5mm]
\displaystyle \varkappa p_g=-\frac{8\varkappa\hbar
k_0N_{inst}^{(gh)}}{3a_0^2a^2}\ln\frac{ea}{a_0}-
\frac{2\varkappa\hbar}{3a^2}\left(\frac{k_0N_{inst}^{(gh)}}{a^2_0}-
\frac{k_1N_{inst}^{(g)}}{a^2_1}\right)=
-\frac{8\varkappa\hbar
k_0N_{inst}^{(gh)}}{3a_0^2a^2}\ln\frac{ea}{a_0}\ .
\end{array}
 \label{7.39}
\end{equation}
In formulas (\ref{7.39}), the terms in brackets are eliminated by
the condition (\ref{5.15}), which is rewritten in terms of
macroscopic parameters
\begin{equation}
\displaystyle
\frac{k_0N_{inst}^{(gh)}}{a^2_0}=\frac{k_1N_{inst}^{(g)}}{a^2_1}\
. \label{7.40}
\end{equation}
Solutions (\ref{7.39}), (\ref{7.40}) describe a quantum coherent
condensate of quasi--resonant instantons with the ghost dominance.
The parameters of the condensate are in accordance with parameters
of a spatially homogeneous condensate with graviton dominance.

\section{Discussion}\label{con}

From the formal mathematical point of view, the above theory is
identical to transformations of equations, determined by the
original gauged path integral (\ref{2.1}), leading to exact
solutions for the model of self--consistent theory of gravitons in
the isotropic Universe. To assess the validity of the theory, it
is useful to discuss again but briefly the three issues of the
theory that are missing in the original path integral.

(i) {\it The hypothesis of the existence of classic spacetime with
deterministic, but self--consistent geometry is introduced into
the theory.} It is not necessary to discuss in detail this
hypothesis because it simply reflects the obvious experimental
fact (region of Planck curvature and energy density is not a
subject of study in the theory under discussion). Note, however,
that the introduction of this hypothesis into the formalism of the
theory leads to a rigorous mathematical consequence: the strict
definition of the operation of separation of classical and quantum
variables uniquely captures the exponential parameterization of
the metric.

(ii) {\it The transfer to the one--loop approximation is conducted
in the self--consistent classical and quantum system of
equations.} Formally, this approximation is of a technical nature
because the equations of the theory are simplified only in order
to obtain specific approximate solutions. After classical and
quantum variables are identified, the procedure of transition to
the one--loop approximation is of a standard and known character
\cite{7}. In reality, of course, the situation in the theory is
much more complex and paradoxical. On the one hand, the quantum
theory of gravity is a non--renormalized theory (see, e.g.
\cite{36}). Specific quantitative studies of effects off one--loop
approximation are simply impossible. On the other hand, the
quantum theory of gravity without fields of matter is finite in
the one--loop approximation \cite{3}. The latter means that the
results obtained in the framework of one--loop quantum gravity
pose limits to its applicability that is mathematically clear and
physically significant. The existence of a range of validity for
the one--loop quantum gravity without fields of matter is a
consequence of two facts. First, there are supergravity theories
with fields of matter which are finite beyond the limits of
one--loop approximation. Second, the quantum graviton field is the
only physical field with a unique combination of such properties
as conformal non--invariance and zero rest mass. For this field
only there is no threshold for the vacuum polarization and
particle creation in the isotropic Universe. Therefore, in the
stages of evolution of the Universe, where $H^2,\ |\dot H|\ll m^2$
($m$ is mass of any of the elementary particles), quantum
gravitational effects can occur only in the subsystem of
gravitons. It is also clear that in any future theory that unifies
gravity with other physical interactions, equations of theory of
gravitons in one--loop approximation will not be different from
those we discuss in this work. Therefore the self--consistent
theory of gravitons has the right to lay claim be a reliable
description of the most significant quantum gravity phenomena in
the isotropic Universe.

(iii) {\it The need to use the Hamilton gauge, which provides a transition from the path integral to the Heisenberg representation \cite{VM}, and then to the self-consistent theory of gravitons in the macroscopic spacetime.} It is important that in the Hamilton gauge the dynamic properties of the ghost fields automatically provide one--loop finiteness of the theory off mass shell of
gravitons and ghosts.
The condition of
one--loop finiteness off the mass shell largely determines the
mathematical and physical content of the theory. Given that the
main results of this work are exact solutions and exact
transformations, the evaluation of he proposed approach is reduced
to a discussion of this point of the theory. Let us enumerate once
more logical and mathematical reasons, forcing us to include the
condition of one--loop finiteness off the mass shell into the
structure of the theory.

a)  Future theory that will unify quantum gravity with the theory
of other physical interactions may not belong to renormalizability
theories. If such a theory exists, it may only be a finite theory.
One--loop finiteness of quantum gravity with no fields of matter
that is fixed on the mass shell \cite{3} can be seen as the
prototype of properties of the future theory.

b) Because of their conformal non--invariance and zero rest mass,
gravitons and ghosts fundamentally can not be located exactly on
the mass shell in the real Universe. Therefore, the problem of
one--loop finiteness off the mass shell is contained in the
internal structure of the theory.

c) In formal schemes, which do not meet the one--loop finiteness,
divergences arise in terms of macroscopic physical quantities. To
eliminate these divergences, one needs to modify the Lagrangian of
the gravity theory, entering quadratic invariants. This, in turn,
leads to abandonment of the original definition of the graviton
field that generates these divergences. The logical inconsistency
of such a formal scheme is obvious. (The mathematical proof of
this claim is contained at Section \ref{nonren}.)

d) In the self--consistent theory of gravitons, one--loop
finiteness off the mass shell can be achieved only through mutual
compensation of divergent graviton and ghost contributions in
macroscopic quantities. The existence of gauges, automatically
providing such a compensation, is an intrinsic property of the
theory.

From our perspective, the properties of the theory identified in
points a), b), c) and d), clearly dictate the need to use only the
formulation of self--consistent theory of gravitons, in which the
condition of one--loop finiteness off the mass shell (the
condition of internal consistency of the theory) is performed
automatically. We also want to emphasize that, as it seems to us,
the scheme of the theory given below has no alternative both
logically and mathematically.

{\it Gauged path integral $\Longrightarrow$ choosing the Hamilton gauge, which provides one--loop finiteness of the theory off mass shell of
gravitons and ghosts
 $\Longrightarrow$  factorization of
classic and quantum variables, which ensures the existence of a
self--consistent system of equations $\Longrightarrow$  transition
to the one--loop approximation, taking into account the
fundamental impossibility of removing the contributions of ghost
fields to observables} --- appears to us logically and mathematically as the
only choice.

As part of the theories preserving macroscopic spacetime being
clearly one of its components, we see two topics for further
discussions. The first of these is the replication of the results
of this work by mathematically equivalent formalisms of one--loop
quantum gravity. Here we can note that, for example, in the
formalism of the extended phase space with BRST symmetry, our
results are reproduced, even though the mathematical formalism is
more cumbersome. The second topic is the reproduction of our
results in more general theories than the one--loop quantum
gravity without fields of matter. Here is meant a step beyond the
limits of one--loop approximation as well as a description of
quantum processes involving gravitons, while taking into account
the existence of other quantum fields of spin $J\leqslant 3/2$. In
the framework of discussion on this topic, we can make only one
assertion: in the one--loop $N=1$  supergravity containing
graviton field and one gravitino field, the results of our work
are fully retained. This is achieved by two internal properties of
$N=1$ supergravity: (i) The sector of gravitons and graviton
ghosts in this theory is exactly the same as in the one--loop
quantum gravity without fields of matter; (ii) The physical
degrees of freedom of gravitino with chiral $h=\pm 3/2$ in the
isotropic Universe are dynamically separated from the
non--physical degrees of freedom and are conformally invariant;
(iii) The gauge of gravitino field can be chosen in such a way
that the gravitino ghosts automatically provide one--loop
finiteness of $N=1$ supergravity. As for multi--loop calculations
in the $N=1$ supergravity and more advanced theoretical models, we
have not explored the issue.

Of course, a rather serious problem of the physical nature of
ghosts remains. The present work makes use in practice only of
formal properties of quantum gravity of Faddeev--Popov--De Witt,
which point to the impossibility in principle of removing
contributions of ghosts to observable quantities off the mass
shell. A deeper analysis undoubtedly will address the foundations
of quantum theory. In particular, one should point out the fact
that the formalism of the path integral of Faddeev--Popov--De Witt
is mathematically equivalent to the assumption that observable
quantities can be expressed through derivatives of
operator--valued functions defined on the classical spacetime of a
given topology. On the other hand, finiteness of physical
quantities is ensured in the axiomatic quantum field theory by
invoking limited field operators smoothed over certain small areas
of spacetime. Extrapolation of this idea to quantum theory of
gravity immediately brings up the question on the role of
spacetime foam \cite{37} (fluctuations of topology on the
microscopic level) in the formation of smoothed operators, and
consequently, observable quantities. To make this problem more
concrete, a question can be posed on collective processes in a
system of topological fluctuations that form the foam. It is not
excluded that the non--removable Faddeev--Popov ghosts  in
ensuring the one--loop finiteness of quantum gravity are at the
same time a phenomenological description of processes of this
kind.

Study of equations of self--consistent theory of gravitons,
automatically satisfying the condition of one--loop finiteness,
leads to the discovery  of a new class of physical phenomena which
are {\it macroscopic effects of quantum gravity.} Like the other
two macroscopic quantum phenomena of superconductivity and
superfluidity, macroscopic effects of quantum gravity occur  on
the macroscopic scale of the system as a whole, in this case, on
the horizon scale of the Universe. Interpretation of these effects
is made in terms of {\it gravitons--ghost condensates arising from
the interference of quantum coherent states.} Each of coherent
states is a state of gravitons (or ghosts) with a certain
wavelength of the order of the distance to the horizon and a
certain occupation number. The vector of the physical state is a
coherent superposition of vectors with different occupation
numbers.

A key part in the formalism of self--consistent theory of
gravitons is played by the BBGKY chain for the spectral function
of gravitons, renormalized by ghosts. It is important that
equations of the chain may be introduced at an axiomatic level
without specifying explicitly field operators and state vectors.
It is only necessary to assume the preservation of the structure
of the chain equations in the process of elimination of
divergences of the moments of the spectral function. Three exact
solutions of one--loop quantum gravity are found in the framework
of BBGKY formalism. The invariance of the theory with respect to
the Wick rotation is also shown. This means that the solutions of
the chain equations, in principle, cover two types of condensates:
condensates of virtual gravitons and ghosts and condensates of
instanton fluctuations.

All exact solutions, originally found in the BBGKY formalism, are
reproduced at the level of exact solutions for field operators and
state vectors. It was found that exact solutions correspond to
various condensates with different graviton--ghost microstructure.
Each exact solution we found is compared to a phase state of
graviton--ghost medium; quantum--gravity phase transitions are
introduced.

We suspect that the manifold of exact solutions of one--loop
quantum gravity is not exhausted by three solutions described in
this paper. Search for new exact solutions and development of
algorithms for that search, respectively, is a promising research
topic within the proposed theory. Of great  interest will also be
approximate solutions, particularly those that describe
non--equilibrium and unstable graviton--ghost and instanton
configurations.

\section{Conclusion}

1. The equations of quantum gravity in the Heisenberg representation and the equations of semi--quantum/semi--classical self--consistent theory of gravitons in the macroscopic Riemann space, respectively, can exist only in the exponential parameterization and Hamilton gauge of the density of the contravariant metric.

2. Equations of semi--quantum/semi--classical theory necessarily contain the ghost sector in the form of a complex scalar field providing one--loop finiteness to the theory.

3. In case of isotropic Universe, in one--loop approximation the theory can be presented as a set of equations including Einstein's equations for the macroscopic metric with the energy--momentum tensor for gravitons and ghosts and BBGKY chain for the moments of the spectral function of gravitons renormalized by ghosts. Three exact solutions to the set of these equations are obtained which describe the various states of the graviton--ghost substratum.

 4. Each exact solution to the BBGKY chain put in correspondence to the exact solutions of operator equations and observables averaged over the Heisenberg state vector. It was found that various exact solutions describe various graviton, ghost and instanton condensates on the horizon scale of the Universe

5. It is shown that continuous phase transitions are possible between different the states of graviton--ghost condensate.

\section{APPENDIX. Renormalizations and Anomalies}\label{an}

In Sections \ref{ren}, \ref{nonren} we discuss a self--consistent
theory of gravitons in isotropic Universe with the ghost sector
not taken into account. As has been repeatedly stated, we believe
that such a model is not mathematically sound. Gauges, completely
removing the degeneracy, are absent in the theory of gravity.
Thus, in the self--consistent theory of gravitons the ghost sector
is inevitable present. Now, however, let us assume for the moment
that the self--consistent theory of gravitons without ghosts is
worth at least as a model of mathematical physics. The purpose of
this Section is to get the properties of this model and to show
that it is mathematically and physically internally inconsistent.

\subsection{Gravitons with no Ghosts. Vacuum Einstein Equations
with Quantum Logarithmic Corrections}\label{ren}

It clear from the outset that in the non--ghost model the
calculation of observables will be accompanied by the emergence of
divergences. It is therefore necessary to formulate the theory in
such a way that the regularization and renormalization operations
are to be contained in its mathematical structure from the very
beginning. We talk here about changes in the mathematical
formulation of the theory. The relevant operations should be
introduced into the theory with care: first, in the amended
theory, coexistence of classical and quantum equations should be
ensured automatically; second, the enhanced theory should not
contain objects initially missing from the theory of gravity.

The dimensional regularization satisfies both above--mentioned
conditions. Important, however, is the following fact: the use of
dimensional regularization suggests that the self--consistent
theory of gravitons in the isotropic Universe is originally
formulated in a spacetime of dimension $D=1+d$, where $1$ is the
dimension of time; $d=3-2\varepsilon$ is the dimension of space.
The special status of the time is due to the two factors: (i) all
the events in the Universe, regardless of its actual dimension,
are ordered along the one--dimensional temporal axis; (ii) the
canonical quantization of the graviton field in terms of the
commutation relations for generalized coordinates and generalized
momenta also presuppose the existence of the one--dimensional
time. As for the space dimension, the limit transition to the true
dimension $d=3$ is implemented after the regularization and
renormalization.

Thus, we are working in a space with a metric
 \begin{equation}
\begin{array}{c}
 \displaystyle
 ds^2=a^2(\eta)(d\eta^2-\gamma_{\alpha\beta}dx^\alpha dx^\beta)\ ,
 \qquad \gamma^{\alpha\beta}\gamma_{\alpha\beta}=d\ ,
 \\[5mm]
\displaystyle \sqrt{|g_{(d)}|}=a^{d+1}\ ,\qquad R_{(d)}=
-\frac{d}{a^2}\left(
2\frac{a''}{a}+(d-3)\frac{a^{'2}}{a^2}\right)\ .
\end{array}
 \label{12.1}
\end{equation}
To avoid mathematical contradictions that could arise at the limit
$d\to 3$, Einstein equations in $D$-dimensional spacetime should
be written down in exactly the form in which they were obtained
from the variational principle:
 \begin{equation}
 \begin{array}{c}
\displaystyle
 \frac{1}{\varkappa_d}\sqrt{|g_{(d)}|}\left(R_{0(d)}^0-\frac12R_{(d)}\right)\equiv
  \frac1{2\varkappa_d}d(d-1)a^{d-3}{a'}^2=
 \frac{1}{8\varkappa_d}a^{d-1}\sum_{{{\bf k}\sigma}}
 \langle\Psi_g|\hat \psi_{{\bf k}\sigma}^{+'}\hat \psi'_{{\bf k}\sigma}+
 k^2\hat \psi_{{\bf k}\sigma}^+\hat \psi_{{\bf k}\sigma}|\Psi_g\rangle\ ,
 \\[5mm]
 \displaystyle
 -\frac{d-1}{2\varkappa_d}\sqrt{|g_{(d)}|}R_{(d)}\equiv
\frac1{2\varkappa_d}d(d-1)\left[2a^{d-2}{a''}+(d-3)a^{d-3}{a'}^2\right]=
 -\frac{d-1}{8\varkappa_d}a^{d-1}\sum_{{{\bf k}\sigma}}\langle\Psi_g|
 \hat \psi_{{\bf k}\sigma}^{+'}\hat \psi'_{{\bf k}\sigma}-
 k^2\hat \psi_{{\bf k}\sigma}^+\hat \psi_{{\bf k}\sigma}|\Psi_g\rangle\ ,
 \end{array}
 \label{12.2}
\end{equation}
 \begin{equation}
  \displaystyle \hat\psi''_{{\bf k}\sigma}+(d-1)\frac{a'}{a}\hat\psi'_{{\bf k}\sigma}+
  k^2\hat\psi_{{\bf k}\sigma}=0\ .
   \label{12.3}
\end{equation}
Here $\varkappa_d$ is the Einstein gravitational constant in
$D$--dimensional spacetime. (Dimension
$[\varkappa_d\hbar]=[l]^{D-2}$.) The left hand sides of equations
(\ref{12.2}) satisfy the Bianchi identity:
\begin{equation}
 \begin{array}{c}
\displaystyle
 \frac1{2\varkappa_d}d(d-1)\left[a^{d-3}{a'}^2\right]'-
 \frac1{2\varkappa_d}d(d-1)\frac{a'}{a}\left[2a^{d-2}{a''}+(d-3)a^{d-3}{a'}^2\right]\equiv 0\ .
 \end{array}
 \label{12.4}
\end{equation}
In the right hand side of equations (\ref{12.2}), the identity
(\ref{12.4}) generates condition of the graviton EMT  conservation
that satisfies if the equations of motion (\ref{12.3}) are taken
into account. Regarding the origin of the system of equations
(\ref{12.2}) and (\ref{12.3}), we should  make the following
comment. In this case it is inappropriate to invoke the reference
to the path integral and factorization of its measures because the
path integral inevitably leads to the theory of ghosts interacting
with the macroscopic gravity. We can only mention a heuristic
recipe: one should refer to the density of Einstein equations with
mixed indices, define the exponential parameterization of the
metric, and expand the equations into a series of metric
fluctuations with an accuracy of the second--order terms.
Deviations from this recipe (for example, linear parameterization
$\hat g_{ik}=g_{ik}+\hat h_{ik}$) lead to a system of inconsistent
classical and quantum equations. To remove this sort of
inconsistency, one is forced to use artificial transactions
outside the formalism of the theory (see, for example,
\cite{42}).

While working with the system of equations (\ref{12.2}),
(\ref{12.3}), we face with two mathematical problems. The first
problem is that in the framework of that system of equations,
except in very special cases, it is impossible to formulate the
dynamics of operators on a given background that is to get the
solution of the equation (\ref{12.3}) as an accurate operator
function of time. This is due to the fact that formulae of
(\ref{12.2}) in reality are not yet specific equations. They are
only a layout of Einstein equations with radiation corrections.
These equations can only be obtained after regularization and
renormalizations of the ultraviolet divergences. In addition, the
functional form of equations depends on which quantum
gravitational effects are to be taken into account outside the
sector of vacuum (i.e. zero) fluctuations of the graviton field.
The only possible way to study the system of equations
(\ref{12.2}) and (\ref{12.3}) is (i) to obtain the solution of
operator equation (\ref{12.3}) in a form of a functional of the
scale factor without specifying the dependence on $a(\eta)$ {\it
with a clear emphasis on zero fluctuations in this functional},
(ii) to substitute the obtained functional in (\ref{12.2}) under
certain assumptions about the state vector; (iii) to regularize
and renormalize and finally (iv) to solve the macroscopic Einstein
equations, obtained after these operations. Implementation of the
program, an essential element of which is the allocation of zero
fluctuations generating ultraviolet divergences, is possible only
when using the method of asymptotic expansions of solutions of
operator equation in the square of wavelength of the graviton
modes. Thus, the problem of the lack of macroscopic Einstein
equations in the original formulation of this theory with
divergences limits the methods of this theory to the short--wave
approach. Note that this fact was clearly indicated by DeWitt
\cite{7}.

The second problem is related to the infrared instability of the
theory, with the object of the theory being a conformal
non--invariant massless quantum field. The problem is due to the
fact that not every representation of the asymptotic series can be
substituted into energy--momentum tensor to perform the summation
over the wave numbers. For example, if in the explicit form, a
term in the asymptotic series contains a large parameter $k^{2n}$
in the denominator, then starting from $n=2$ in the integration
over the wave numbers the infrared divergences will appear. Such
an asymptotic series can not be used even for the renormalization
of ultraviolet divergences, because when it is used in the space
of the physical dimension $d=3$, the logarithmic divergences arise
simultaneously at the ultraviolet and the infrared limits. In the
method of dimensional regularization the problem is reduced to the
fact that it is impossible to choose an interim dimension $d$ in a
way such that the integral exists at both limits.

Formally, the technical problem described above is partly solved
by reformatting the asymptotic series. In particular, the
following method will be used, in which parameter of the
asymptotic expansion is the effective frequency
 \begin{equation}
 \begin{array}{c}
 \displaystyle \omega_k^2=k^2+\rho\ ,
 \qquad
\rho=\frac{d-1}{4d}a^2R_{(d)}=-\frac{d-1}{4}\left[
2\frac{a''}{a}+(d-3)\frac{a^{'2}}{a^2}\right]\ .
\end{array}
 \label{12.5}
\end{equation}
In this method, the integrals over the wave numbers can be defined
in terms of the principal value. Contributions of the poles at
$k=\sqrt{-\rho}$ can not be mathematically verified if only
because there are such contributions from each term of the
infinite asymptotic series. The inability to describe infrared
effects is the principal disadvantage of a theory with
divergences, which uses only asymptotic expansions with respect to
the wavelength. Meanwhile, as general considerations and the
results of this work show, in the physics of conformal
non--invariant massless field the most interesting and innovative
effects occur in the infrared spectrum. The method of describing
these effects, based on the exact BBGKY chain, can not be used in
the theory with divergences, because a method regularizing the
infinite chain of moments of the spectral function does not exist.

The above problems automatically reduces the interest toward the
theory with divergences. However, given that all previous works in
this area have been implemented in the framework of regularization
and renormalization, let us conduct our analysis to the end. In
calculations, it is enough to consider the equation for the
convolution. After identity transformations, using the equation of
motion (\ref{12.3}), we get
\begin{equation}
 \begin{array}{c}
\displaystyle
\frac1{2\varkappa_d}d(d-1)\left[2a^{d-2}{a''}+(d-3)a^{d-3}{a'}^2\right]=
 -\frac{d-1}{16\varkappa_d}\sum_{{{\bf k}\sigma}}\left(W'_{{\bf k}\sigma}a^{d-1}\right)'\ ,
 \end{array}
 \label{12.6}
\end{equation}
where
\[
\displaystyle W_{{\bf k}\sigma}=
\langle\Psi_g|\hat \psi_{{\bf k}\sigma}^+\hat \psi_{{\bf k}\sigma}|\Psi_g\rangle
\]
is the spectral function of gravitons. The calculation of the
spectral function by the method of asymptotic expansion with
respect to the square of wavelength was described in Section
\ref{swg}. Now we need to repeat this calculation excluding the
ghosts, but with input from zero fluctuations in the spacetime of
dimension $D=d+1$. The relevant calculations do not require
additional comments. A spectral function is represented as:
\begin{equation}
\displaystyle W_{{\bf k}\sigma}=W^{(vac)}_{{\bf k}\sigma}+W^{(exc)}_{{\bf k}\sigma}\ ,
\label{12.7}
\end{equation}
where $W^{(vac)}_{{\bf k}\sigma}$ is the vacuum component of the
spectral function and $W^{(exc)}_{{\bf k}\sigma}$ is the spectral
function of excitations. After passage to the limit $d\to 3$, the
contribution of $W^{(exc)}_{{\bf k}\sigma}$ to the EMT of short
gravitons is exactly the same as (\ref{4.9}), (\ref{4.10}). In the
future, we discuss only the contribution from vacuum components of
the spectral function. In the calculations, we must keep in mind
that in the $d$--dimensional space the number of internal degrees
of freedom of transverse gravitons is $w_g=(d+1)(d-2)/2$. The
solution for the vacuum spectral function is expressed in terms of
the functional (\ref{4.4}):
\begin{equation}
\begin{array}{c}
\displaystyle \sum_\sigma W^{(vac)}_{{\bf k}\sigma}=
\frac{4\varkappa_d\hbar}{a^{d-1}}\cdot\frac{(d+1)(d-2)}{4\epsilon_k}=
\frac{4\varkappa_d\hbar}{a^{d-1}}\cdot
\frac{(d+1)(d-2)}{4\omega_k}\sum_{s=0}^{\infty}(-1)^s\hat J_k^s\cdot 1\ .
\end{array}
\label{12.8}
\end{equation}
The powers of operator $\hat J_k^s\cdot 1$ are defined by formulas
(\ref{4.5}), in which $\omega_k^2$ has the form (\ref{12.5}).
After substitution of (\ref{12.8}) into (\ref{12.6}), the
zero--term in the asymptotic expansion creates an integral,
calculated by the rules of dimensional regularization:
\begin{equation}
\begin{array}{c}
  \displaystyle \sum_{{\bf
  k}}\frac{1}{\omega_k}=
  \frac{1}{(2\pi)^d}\frac{2\pi^{d/2}}{\Gamma(d/2)}
  \int\limits_0^{\infty}\frac{k^{d-1}dk}{(k^2+\rho)^{1/2}}=
    \frac{\displaystyle \Gamma\left[(3-d)/2\right]}
  {\displaystyle 2^{d-1}\pi^{(d+1)/2}(1-d)}\rho^{(d-1)/2}\ .
  \end{array}
  \label{12.9}
\end{equation}
The $\Gamma-$function in (\ref{12.9}) diverges for $d\to 3$.
Therefore, calculation of the integral (\ref{12.9}) and
transformation of expressions with $\Gamma-$functions are carried
out with those values of $d$ which provide the existence of the
integral and $\Gamma-$functions. At the final stage, the result of
these calculations is analytically continued to the vicinity
$d=3$. All other terms of the asymptotic expansion (\ref{12.8})
generate finite integrals and do not require a dimensional
regularization. For reasons of heuristic rather than mathematical
nature, it is considered that these terms are negligible compared
to the contribution of the principal term of the asymptotic
expansion (see below the effective Lagrangian (\ref{12.19})).
Convolution of $D$--dimensional Einstein's equations (\ref{12.6}),
containing the main term of the vacuum EMT of gravitons, has the
form:
\begin{equation}
 \begin{array}{c}
\displaystyle
\frac1{2\varkappa_d}d(d-1)\left[2a^{d-2}{a''}+(d-3)a^{d-3}{a'}^2\right]=
\frac{\hbar
(d+1)(d-2)}{2^{d+3}\pi^{(d+1)/2}}\Gamma\left(\frac{3-d}{2}\right)
\left[\left(\frac{\rho^{(d-1)/2}}{a^{d-1}}\right)'a^{d-1}\right]'\
.
 \end{array}
 \label{12.10}
\end{equation}
Other Einstein equations can be obtained using the Bianchi
identities. A complete system of Einstein vacuum equations is
written in $D$--covariant form:
\begin{equation}
 \begin{array}{c}
\displaystyle R_{i(d)}^k-\frac12\delta_i^kR_{(d)}+
\\[5mm]
\displaystyle +
\frac{\varkappa_d\hbar(d+1)(d-2)(d-1)^{\frac{d-1}{2}}}{2^{2d+2}(d\pi)^{\frac{d+1}{2}}}
\Gamma\left(\frac{3-d}{2}\right)\times
\left[\left(R_{(d)}^{\frac{d-1}{2}}\right)^{;k}_{;i}-\delta_i^k\left(R_{(d)}^{\frac{d-1}{2}}\right)^{;l}_{;l}-
\left(R_{i(d)}^k-\frac1{d+1}\delta_i^kR_{(d)}\right)R_{(d)}^{\frac{d-1}{2}}\right]=0\ .
 \end{array}
 \label{12.11}
\end{equation}
Equation (\ref{12.11}) are obtained by the variation of action
\begin{equation}
 \begin{array}{c}
\displaystyle S_{vac}=\int\sqrt{|g_{(d)}|}d^{\scriptscriptstyle D}x
\left[-\frac{1}{2\varkappa_d}R_{(d)}+
\frac{\hbar(d-2)(d-1)^{\frac{d-1}{2}}}{2^{2d+2}(d\pi)^{\frac{d+1}{2}}}
\Gamma\left(\frac{3-d}{2}\right)
R_{(d)}^{\frac{d+1}{2}}\right]\ .
 \end{array}
 \label{12.12}
\end{equation}
It is obvious from (\ref{12.11}), (\ref{12.12}) that the method of
dimensional regularization retains overall covariance of the
theory. Of course, quantum corrections, appearing in
(\ref{12.11}), satisfy the condition of conservation.

Renormalization and removal of regularization (limit $d\to 3$) are
held at the level of action. A parameter with the dimension of
length, which will eventually acquire the status of
renormalization scale, is contained within the theory. This
parameter, referred to as $L_g$, is appears in the
$D$--dimensional constant of gravity:
\begin{equation}
\displaystyle \varkappa_d=\varkappa\cdot L_g^{d-3}\ .
\label{12.13}
\end{equation}
The technique of removal the regularization assumes conservation
of dimensionality for those objects in which the limit operation
is performed. There are two such objects: the measure of
integration $d\mu$ and the density of the Lagrangian
$\mathcal{L}$. As can be seen from (\ref{12.12}), (\ref{12.13}),
the first (Einstein) term of the action is written down as
\begin{equation}
\begin{array}{c}
\displaystyle S^{(1)}_{vac}=\int\mathcal{L}^{(1)}d\mu\ ,
\\[5mm]
\displaystyle \mathcal{L}^{(1)}=-\frac{1}{2\varkappa}R_{(d)}\ ,\qquad d\mu=\sqrt{|g_{(d)}|}L_g^{4-D}d^Dx\ ,
 \end{array}
 \label{12.14}
\end{equation}
where $D$--dimensional objects $\mathcal{L}$ and $d\mu$ have the
same dimensions as the corresponding 4--dimensional objects. In
this sector of the theory the limit transition is trivial:
$R_{(d)}\to R$, $d\mu\to \sqrt{-g}d^4x$. In the sector of quantum
corrections to the Einstein theory, we introduce the same measure
and obtain the density of the Lagrangian:
\begin{equation}
\displaystyle
\mathcal{L}^{(2)}=
\frac{\hbar L_g^{d-3}(d-2)(d-1)^{\frac{d-1}{2}}}{2^{2d+2}(d\pi)^{\frac{d+1}{2}}}
\Gamma\left(\frac{3-d}{2}\right)
R_{(d)}^{\frac{d+1}{2}}
\label{12.15}
\end{equation}
It is necessary to emphasize that  {\it the operations of
renormalizations and removal of regularization have to be
mathematically well--defined and generally--covariant.} The
condition of mathematical certainty assumes that {\it the
renormalization is conducted before the lifting of
regularization.} At the same time, the general--covariance of the
procedure is automatically fulfilled if the counter--terms imposed
in the Lagrangian are the $D$--dimensional invariants. Note also
that if the mathematical value is finite at $d=3$, then the above
formulated conditions do not prevent the expansion of this
quantity in a Taylor series over the parameter $(3-d)/2$. In
particular, we can write:
\begin{equation}
 \begin{array}{c}
\displaystyle L_g^{d-3}R_{(d)}^{\frac{d+1}{2}}\equiv R_{(d)}^2\left(L_g^2R_{(d)}\right)^{\frac{d-3}{2}}=
R_{(d)}^2\left(1+\frac{3-d}{2}\ln\frac{\mu_g^2}{R_{(d)}}+...\right)\ ,
 \end{array}
 \label{12.16}
\end{equation}
where $\mu_g=1/L_g$; ellipsis designate the terms which do not
contribute to the final result. The substitution (\ref{12.16}) in
(\ref{12.15}) provides:
\begin{equation}
\begin{array}{c}
\displaystyle
\mathcal{L}^{(2)}=\frac{\hbar (d-2)(d-1)^{\frac{d-1}{2}}}{2^{2d+2}(d\pi)^{\frac{d+1}{2}}}
\Gamma\left(\frac{3-d}{2}\right)R_{(d)}^2+
\frac{\hbar (d-2)(d-1)^{\frac{d-1}{2}}}{2^{2d+2}(d\pi)^{\frac{d+1}{2}}}
\Gamma\left(\frac{5-d}{2}\right)R_{(d)}^2\ln\frac{\mu_g^2}{R_{(d)}}+...\ .
\end{array}
\label{12.17}
\end{equation}
According to (\ref{12.17}), the source Lagrangian of the theory
requires a $D$--invariant counter--term, which removes the
contribution proportional to the diverging $\Gamma$--function:
\begin{equation}
\displaystyle
\mathcal{L}_0^{(2)}=-\frac{\hbar (d-2)(d-1)^{\frac{d-1}{2}}}{2^{2d+2}(d\pi)^{\frac{d+1}{2}}}
\Gamma\left(\frac{3-d}{2}\right)R_{(d)}^2+\frac{\hbar}{4f^2}R_{(d)}^2\ .
\label{12.18}
\end{equation}
In (\ref{12.18}), there is a new finite constant of the theory of
gravity  $1/f^2$. The removal of the regularization in the
renormalized Lagrangian is conducted by the regular transition:
\begin{equation}
\begin{array}{c}
\displaystyle
\mathcal{L}_{ren}=\lim_{d\to 3}\left(\mathcal{L}^{(1)}+\mathcal{L}^{(2)}+
\mathcal{L}^{(2)}_0\right)=
-\frac{1}{2\varkappa}R+\frac{\hbar}{4f^2}R^2+\frac{\hbar}{1152\pi^2}
R^2\ln\frac{\mu_g^2}{R}=
-\frac{1}{2\varkappa}R+\frac{\hbar}{1152\pi^2}
R^2\ln\frac{\lambda_g^2}{R}\ ,
\end{array}
\label{12.19}
\end{equation}
where
\[
\displaystyle \lambda_g^2=\mu_g^2\exp\frac{288\pi^2}{f^2}
\]
is the renorm--invariant scale. There is a heuristic argument
allowing to use the obtained expression: quantum corrections in
the Lagrangian (\ ref (12.19)) dominate over all other neglected
terms of the asymptotic series over The logarithmic parameter
$\ln(\lambda_g^2/R)\gg 1$.

The renormalized Einstein vacuum equations with quantum
corrections obtained from the Lagrangian (\ref{12.19}) are as
follows:
\begin{equation}
 \begin{array}{c}
\displaystyle R_i^k-\frac12\delta_i^kR+
\frac{\varkappa\hbar}{288\pi^2}\left\{
 \left[R\ln\frac{\lambda_g^2}{R}\right]_{;i}^{;k}-
 \delta_i^k\left[R\ln\frac{\lambda_g^2}{R}\right]_{;l}^{;l}-\left(RR_i^k-\frac14\delta_i^kR^2\right)\ln\frac{\lambda_g^2}{R}-
 \frac18\delta_i^kR^2\right\}=0\ .
 \end{array}
 \label{12.20}
\end{equation}
Note that exactly the same equations are obtained from
$D$--dimensional equations (\ref{12.11}), provided that the
operations are performed in the same sequence: {\it first a
renormalization with the introduction of $D$--covariant
counter--terms is conducted, and then a limit transition to the
physical dimension is performed.}

\subsection{Intrinsic Contradiction of Theory with no Ghosts:
Impossibility of One-Loop Renormalization }\label{nonren}

We are still discussing a formal model --- self--consistent theory
of gravitons with no ghosts. In the previous section it was shown
that the renormalization of divergences, that inevitably arise in
this model, requires the imposition of an additional term
quadratic in the curvature in the Lagrangian. It is now necessary
to draw attention to two mathematical facts: (i) the need for a
modification of Einstein theory is caused by quantum effects
contained in the Lifshitz operator equation (\ref{12.3}); (ii) the
original Lagrangian and operator equations of the modified theory
have the form:
\begin{equation}
 \begin{array}{c}
\displaystyle \mathcal{L}=\int\left(-\frac{1}{2\varkappa}\hat R+
\frac{\hbar}{4f^2}\hat R^2\right)\sqrt{-\hat g}d^4x\ ,
 \end{array}
 \label{12.21}
\end{equation}
\begin{equation}
 \begin{array}{c}
\displaystyle
\sqrt{-\hat g}\left[\frac{1}{\varkappa}\left(\hat R_i^k-
\frac12\delta_i^k\hat R\right)+\frac{\hbar}{f^2}\left(
 \hat D_i\hat D^k\hat R-
 \delta_i^k\hat D_l\hat D^l\hat R-
 \hat R\hat R_i^k+\frac14\delta_i^k\hat R^2\right)\right]=0\ ,
 \end{array}
 \label{12.22}
\end{equation}
where $\hat D_i$ is a covariant derivative in a space with the
operator metric $\hat g_{ik}$. It is quite obvious that these
facts contradict each other: the quantum effects in the Lifshitz
equation lead to a theoretical model that contradicts the Lifshitz
equation. Let us demonstrate that the contradiction is a direct
consequence of the non--renormalizability of the model
(\ref{12.21}) off the graviton mass shell.

Equations (\ref{12.22}), after their linearization describe
quantized waves of two types --- tensor and scalar. It makes sense
to discuss the problem of the scalar modes only in the event that
at least preliminary criteria for consistency of modified theory
will be obtained. Therefore, first of all, we should reveal
properties of the tensor modes. Here is an expression for the
Lagrangian of a system consisting of self--consistent cosmological
field and tensor gravitons:
\begin{equation}
\begin{array}{c}
 \displaystyle
 S=\int dtNa^3\left\{-\frac{3}{\varkappa N^2}\frac{{\dot a}^2}{a^2}+
 \frac{9\hbar}{f^2N^4}\left(\frac{\ddot a}{a}-
 \frac{\dot N}{N}\frac{\dot a}{a}+\frac{{\dot a}^2}{a^2}\right)^2+\right.
  \\[5mm]
 \displaystyle \left.
 +\frac{1}{8}\left[\frac{1}{\varkappa}+
 \frac{6\hbar}{N^2f^2}\left(\frac{\ddot a}{a}-
 \frac{\dot N}{N}\frac{\dot a}{a}+
 \frac{{\dot a}^2}{a^2}\right)\right]\sum_{{{\bf k}\sigma}}\left(\frac{1}{N^2}
 \frac{d{\hat \psi}_{{\bf k}\sigma}^+}{dt}\frac{d{\hat \psi}_{{\bf k}\sigma}}{dt}-
 \frac{k^2}{a^2}\hat \psi_{{\bf k}\sigma}^+\hat \psi_{{\bf k}\sigma}\right)\right\}\ .
 \end{array}
\label{12.23}
 \end{equation}
The equation for gravitons is produced either by the linearization
of the equation (\ref{12.22}), or from (\ref{12.23}) by the
variation procedure:
\begin{equation}
\begin{array}{c}
 \displaystyle \left(1-\frac{\varkappa\hbar}{f^2}R\right)\left(\hat \psi''_{{\bf k}\sigma}
 +2\frac{a'}{a}\hat \psi'_{{\bf k}\sigma} +k^2\hat \psi_{{\bf k}\sigma}\right)-
 \frac{\varkappa\hbar}{f^2}R'\hat \psi'_{{\bf k}\sigma}=0\ .
 \end{array}
\label{12.24}
 \end{equation}
Please note that the last term in (\ref{12.24}) makes it
impossible to retain the Lifshitz equation. After the
transformation
\[
 \displaystyle \hat \psi_{{\bf k}\sigma}=a^{-1}
 \left(1-\varkappa\hbar R/f^2\right)^{-1/2}\hat\varphi_{{\bf k}\sigma}
\]
equation (\ref{12.24}) has a form
\begin{equation}
\begin{array}{c}
 \displaystyle \hat\varphi''_{{\bf k}\sigma} +
 \left[k^2+a^2\left(\frac{R}{6}+P\right)\right]\hat\varphi_{{\bf k}\sigma}=0\ .
 \end{array}
\label{12.25}
 \end{equation}
In (\ref{12.25}), the deviation from the Lifshitz equation is
manifested in the effective frequency of gravitons --- the latter
contains an additional function of curvature's derivatives
 \begin{equation}
\begin{array}{c}
\displaystyle P=-\frac12\left[\ln\left(1-\varkappa\hbar R/f^2\right)\right]^{;l}_{;l}-
-\frac14
\left[\ln\left(1-\varkappa\hbar R/f^2\right)\right]^{;l}
\left[\ln\left(1-\varkappa\hbar R/f^2\right)\right]_{;l}\ .
 \end{array}
\label{12.26}
 \end{equation}
When calculating quantum corrections to the macroscopic equations,
the modification of the effective frequency leads to additional
divergences. Averaged vacuum equations (\ref{12.22}), after their
polynomial expansion in powers of curvature, look as follows
(finite logarithmic corrections are omitted):
\begin{equation}
 \begin{array}{c}
\displaystyle R_i^k-\frac12\delta_i^kR+
\varkappa\hbar\left(\frac{\Gamma(\varepsilon)}{288\pi^2}+
\frac{1}{f_0^2}\right)\left(
 R_{;i}^{;k}-
 \delta_i^kR_{;l}^{;l}-
 RR_i^k+\frac14\delta_i^kR^2\right)+
 \\[5mm]
\displaystyle +(\varkappa\hbar)^2\frac{\Gamma(\varepsilon)}{48\pi^2f_0^2}\left(
 R_{\; ;l;i}^{;l\; \;\; ;k}-
 \delta_i^kR_{\; ;l;m}^{;l\; \;\; ;m}-
 R_i^kR_{;l}^{;l}+
 \frac12R_{;i}R^{;k}-\frac14\delta_i^kR_{;l}R^{;l}\right)=0\ .
 \end{array}
 \label{12.27}
\end{equation}
Here $\Gamma(\varepsilon)\sim 1/\varepsilon$ is a divergent
$\Gamma$--function obtained by dimensional regularization;
$1/f_0^2$ is a seed constant of a theory with quadratic invariant.
The complete quantum Lagrangian corresponding to equations
(\ref{12.27}) has the form:
\begin{equation}
 \begin{array}{c}
\displaystyle \mathcal{L}=\int\left[-\frac{1}{2\varkappa}\hat R+
\hbar\left(\frac{\Gamma(\varepsilon)}{1152\pi^2}+\frac{1}{4f_0^2}\right)\hat R^2+
\varkappa\hbar^2\frac{\Gamma(\varepsilon)}{192\pi^2f_0^2}
\hat R^{;l}\hat R_{;l}\right]\sqrt{-\hat g}d^4x\ .
 \end{array}
 \label{12.28}
\end{equation}
Renormalization of the second term in (\ref{12.28}) is performed
by selecting the seed constant:
\[
\displaystyle \frac{1}{f_0^2}=-\frac{\Gamma(\varepsilon)}{288\pi^2}+\frac{1}{f^2}\ .
\]
However, a divergent coefficient forms before the third term. To
overcome this divergence, it is necessary to introduce a new
seeding "fundamental"\ constant of the modified theory of gravity
$1/h_0^2$ with a renormalization rule:
\[
\displaystyle \displaystyle \frac{1}{h_0^2}=\frac{\Gamma(\varepsilon)}{48\pi^2}
\left(\frac{\Gamma(\varepsilon)}{288\pi^2}-\frac{1}{f^2}\right)+\frac{1}{h^2}\ .
\]

Further actions are obvious and pointless: Lifshitz equation is
the subject of the next modification; quantum corrections generate
another new divergence; to renormalize the new divergence a new
theory of gravity is introduced, etc. The only conclusion to be
drawn from this procedure is that {\it based on the criteria of
quantum field theory, the one--loop self--consistent theory of
gravitons in the isotropic Universe, and not possessing the
property of one--loop finiteness outside of  mass shell, does not
exists as a mathematical model. In such a theory it is impossible
to quantitatively analyze any physical effect.} The theory of
gravitons without ghosts is non--renormalizable even in the
one--loop approximation. It is also important to stress that the
correct alternative to a {\it non--renormalizable} theory is only
a {\it finite} theory with the graviton--ghost compensation of
divergences.

In the future, from our perspective, the method of regularization
and renormalization in general will be excluded from the arsenal
of quantum theory of gravity, including one from the theory of
one--loop quantum effects involving matter fields. Correct
alternatives to existing methods of analysis of these effects to
be found in extended supergravities, finite at least in one--loop
approximation.

The situation prevailing in the scientific literature is a
paradoxical one. On the one hand, inadequate nature of the
regularization and renormalization methods in the quantum theory
of gravity should be obvious from the latest development trends in
the theories of supergravity and superstrings. On the other hand,
however, in all works we know on cosmological applications of
one--loop quantum gravity theoretical models are used, which,
according to the criteria of quantum field theory, do not exist.
We cannot comment on the specific results obtained in these models
by the reasons clear from the content of this Section. Once again
we should emphasize that {\it the self--consistent theory of
gravitons, if it exists as a theoretical model, must be finite
outside the mass shell of gravitons.} Effects arising in the
finite theory are described in the main text of this work.

\[
\]
{\bf ACKNOWLEDGMENT.}
We are deeply grateful to Ludwig D. Faddeev of the Steklov Mathematical Institute and Mikhail Shifman and Arkady Vainshtein of the University of Minnesota for discussions of the structure and content of the theory.  Also, we would like to express our deep appreciation to our friend and colleague Walter Sadowski for invaluable advise and help in the preparation of the manuscript.

\end{document}